\documentclass[a4paper,11pt]{article}
\makeatletter
\gdef\@fpheader{}

\newif\ifnotoc\notocfalse
\newif\ifemailadd\emailaddfalse
\newif\iftoccontinuous\toccontinuousfalse
\newif\ifnatbibsort\natbibsorttrue

\relax

\RequirePackage{amsmath}
\RequirePackage{amssymb}
\RequirePackage{epsfig}
\RequirePackage{graphicx}
\ifnatbibsort\RequirePackage[numbers,sort&compress]{natbib}\else\RequirePackage[numbers,compress]{natbib}\fi
\RequirePackage[colorlinks=true
  ,urlcolor=blue
  ,anchorcolor=blue
  ,citecolor=blue
  ,filecolor=blue
  ,linkcolor=blue
  ,menucolor=blue
  ,linktocpage=true
  ,pdfproducer=medialab
  ,pdfa=true
]{hyperref}

\def\@subheader{\@empty}
\def\@keywords{\@empty}
\def\@abstract{\@empty}
\def\@xtum{\@empty}
\def\@dedicated{\@empty}
\def\@arxivnumber{\@empty}
\def\@collaboration{\@empty}
\def\@collaborationImg{\@empty}
\def\@proceeding{\@empty}
\def\@preprint{\@empty}

\newcommand{\subheader}[1]{\gdef\@subheader{#1}}
\newcommand{\keywords}[1]{\if!\@keywords!\gdef\@keywords{#1}\else%
\PackageWarningNoLine{\jname}{Keywords already defined.\MessageBreak Ignoring last definition.}\fi}
\renewcommand{\abstract}[1]{\gdef\@abstract{#1}}
\newcommand{\dedicated}[1]{\gdef\@dedicated{#1}}
\newcommand{\arxivnumber}[1]{\gdef\@arxivnumber{#1}}
\newcommand{\proceeding}[1]{\gdef\@proceeding{#1}}
\newcommand{\xtumfont}[1]{\textsc{#1}}
\newcommand{\correctionref}[3]{\gdef\@xtum{\xtumfont{#1} \href{#2}{#3}}}

\newcommand\preprint[1]{\gdef\@preprint{\hfill #1}}

\newcommand\note[2][]{%
\if!#1!%
\stepcounter{footnote}\footnotetext{#2}%
\else%
{\renewcommand\thefootnote{#1}%
\footnotetext{#2}}%
\fi}

\newtoks\auth@toks
\renewcommand{\author}[2][]{%
  \if!#1!%
    \auth@toks=\expandafter{\the\auth@toks#2\ }%
  \else
    \auth@toks=\expandafter{\the\auth@toks#2$^{#1}$\ }%
  \fi
}

\newtoks\affil@toks\newif\ifaffil\affilfalse
\newcommand{\affiliation}[2][]{%
\affiltrue
  \if!#1!%
    \affil@toks=\expandafter{\the\affil@toks{\item[]#2}}%
  \else
    \affil@toks=\expandafter{\the\affil@toks{\item[$^{#1}$]#2}}%
  \fi
}

\newtoks\email@toks\newcounter{email@counter}%
\newcommand{\emailAdd}[1]{%
\emailaddtrue%
\ifnum\theemail@counter>0\email@toks=\expandafter{\the\email@toks, \@email{#1}}%
\else\email@toks=\expandafter{\the\email@toks\@email{#1}}%
\fi\stepcounter{email@counter}}
\newcommand{\@email}[1]{\href{mailto:#1}{\tt #1}}

\newcommand*\collaboration[1]{\gdef\@collaboration{#1}}
\newcommand*\collaborationImg[2][]{\gdef\@collaborationImg{#2}}

\newcommand\afterLogoSpace{\smallskip}
\newcommand\afterSubheaderSpace{\vskip3pt plus 2pt minus 1pt}
\newcommand\afterProceedingsSpace{\vskip21pt plus0.4fil minus15pt}
\newcommand\afterTitleSpace{\vskip23pt plus0.06fil minus13pt}
\newcommand\afterRuleSpace{\vskip23pt plus0.06fil minus13pt}
\newcommand\afterCollaborationSpace{\vskip3pt plus 2pt minus 1pt}
\newcommand\afterCollaborationImgSpace{\vskip3pt plus 2pt minus 1pt}
\newcommand\afterAuthorSpace{\vskip5pt plus4pt minus4pt}
\newcommand\afterAffiliationSpace{\vskip3pt plus3pt}
\newcommand\afterEmailSpace{\vskip16pt plus9pt minus10pt\filbreak}
\newcommand\afterXtumSpace{\par\bigskip}
\newcommand\afterAbstractSpace{\vskip16pt plus9pt minus13pt}
\newcommand\afterKeywordsSpace{\vskip16pt plus9pt minus13pt}
\newcommand\afterArxivSpace{\vskip3pt plus0.01fil minus10pt}
\newcommand\afterDedicatedSpace{\vskip0pt plus0.01fil}
\newcommand\afterTocSpace{\bigskip\medskip}
\newcommand\afterTocRuleSpace{\bigskip\bigskip}
\newlength{\affiliationsSep}
\newcommand\beforetochook{\pagestyle{myplain}\pagenumbering{roman}}

\DeclareFixedFont\trfont{OT1}{phv}{b}{sc}{11}

\renewcommand\maketitle{
\pagestyle{empty}
\thispagestyle{titlepage}
\setcounter{page}{0}
\noindent{\small\scshape\@fpheader}\@preprint\par
\afterLogoSpace

\if!\@subheader!\else\noindent{\trfont{\@subheader}}\fi
\afterSubheaderSpace

\if!\@proceeding!\else\noindent{\sc\@proceeding}\fi
\afterProceedingsSpace

{\LARGE\flushleft\sffamily\bfseries\@title\par}
\afterTitleSpace

\hrule height 1.5\p@%
\afterRuleSpace

\if!\@collaboration!\else
{\Large\bfseries\sffamily\raggedright\@collaboration}\par
\afterCollaborationSpace
\fi
\if!\@collaborationImg!\else
{\normalsize\bfseries\sffamily\raggedright\@collaborationImg}\par
\afterCollaborationImgSpace
\fi

{\bfseries\raggedright\sffamily\the\auth@toks\par}
\afterAuthorSpace

\ifaffil\begin{list}{}{%
\setlength{\leftmargin}{0.28cm}%
\setlength{\labelsep}{0pt}%
\setlength{\itemsep}{\affiliationsSep}%
\setlength{\topsep}{-\parskip}}
\itshape\small%
\the\affil@toks
\end{list}\fi
\afterAffiliationSpace

\ifemailadd 
\noindent\hspace{0.28cm}\begin{minipage}[l]{.9\textwidth}
\begin{flushleft}
\textit{E-mail:} \the\email@toks
\end{flushleft}
\end{minipage}
\else
\PackageWarningNoLine{\jname}{E-mails are missing.\MessageBreak Plese use \protect\emailAdd\space macro to provide e-mails.}
\fi
\afterEmailSpace
\if!\@xtum!\else\noindent{\@xtum}\afterXtumSpace\fi
\if!\@abstract!\else\noindent{\renewcommand\baselinestretch{.9}\textsc{Abstract:}}\ \@abstract\afterAbstractSpace\fi
\if!\@keywords!\else\noindent{\textsc{Keywords:}} \@keywords\afterKeywordsSpace\fi
\if!\@arxivnumber!\else\noindent{\textsc{ArXiv ePrint:}} \href{https://arxiv.org/abs/\@arxivnumber}{\@arxivnumber}\afterArxivSpace\fi
\if!\@dedicated!\else\vbox{\small\it\raggedleft\@dedicated}\afterDedicatedSpace\fi
\ifnotoc\else
\iftoccontinuous\else\newpage\fi
\beforetochook\hrule
\tableofcontents
\afterTocSpace
\hrule
\afterTocRuleSpace
\fi
\setcounter{footnote}{0}
\pagestyle{myplain}\pagenumbering{arabic}
}

\renewcommand{\baselinestretch}{1.1}\normalsize
\divide\@tempdima\baselineskip
\@tempcnta=\@tempdima
\skip\@mpfootins = \skip\footins
\renewcommand{\@dotsep}{10000}

\newcommand\ps@myplain{
\pagenumbering{arabic}
\renewcommand\@oddfoot{\hfill-- \thepage\ --\hfill}
\renewcommand\@oddhead{}}
\let\ps@plain=\ps@myplain

\newcommand\ps@titlepage{\renewcommand\@oddfoot{}\renewcommand\@oddhead{}}

\numberwithin{equation}{section}

\renewcommand\section{\@startsection{section}{1}{\z@}%
                                   {-3.5ex \@plus -1.3ex \@minus -.7ex}%
                                   {2.3ex \@plus.4ex \@minus .4ex}%
                                   {\normalfont\large\bfseries}}
\renewcommand\subsection{\@startsection{subsection}{2}{\z@}%
                                   {-2.3ex\@plus -1ex \@minus -.5ex}%
                                   {1.2ex \@plus .3ex \@minus .3ex}%
                                   {\normalfont\normalsize\bfseries}}
\renewcommand\subsubsection{\@startsection{subsubsection}{3}{\z@}%
                                   {-2.3ex\@plus -1ex \@minus -.5ex}%
                                   {1ex \@plus .2ex \@minus .2ex}%
                                   {\normalfont\normalsize\bfseries}}
\renewcommand\paragraph{\@startsection{paragraph}{4}{\z@}%
                                   {1.75ex \@plus1ex \@minus.2ex}%
                                   {-1em}%
                                   {\normalfont\normalsize\bfseries}}
\renewcommand\subparagraph{\@startsection{subparagraph}{5}{\z@}%
                                   {1.75ex \@plus1ex \@minus .2ex}%
                                   {-1em}%
                                   {\normalfont\normalsize\itshape}}

\def\fnum@figure{\textbf{\figurename\nobreakspace\thefigure}}
\def\fnum@table{\textbf{\tablename\nobreakspace\thetable}}

\long\def\@makecaption#1#2{%
  \vskip\abovecaptionskip
  \sbox\@tempboxa{\small #1. #2}%
  \ifdim \wd\@tempboxa >\hsize
    \small #1. #2\par
  \else
    \global \@minipagefalse
    \hb@xt@\hsize{\hfil\box\@tempboxa\hfil}%
  \fi
  \vskip\belowcaptionskip}

\renewenvironment{thebibliography}[1]{%
\begin{oldthebibliography}{#1}%
\small%
\raggedright%
\setlength{\itemsep}{5pt plus 0.2ex minus 0.05ex}%
}%
{%
\end{oldthebibliography}%
}
\makeatother

\usepackage{enumitem}
\title{\boldmath   Topological Control of Quantum Chaos Diagnostics: OTOCs, Spectral Statistics, and Information Scrambling in Ising Model }
\author[a,b]{Reza Pirmoradian,}
\author[c,d]{Soheir Rouhani,}
\author[a,e,f]{M. Reza Tanhayi}
\affiliation[a]{School of Quantum Physics and Matter
	Institute for Research in Fundamental Sciences (IPM),
	P.O. Box 19395-5531, Tehran, Iran}
\affiliation[b]{Ershad Damavand, Institute of Higher Education (EDI)\\
P.O. Box 14168-34311, Tehran, Iran}
\affiliation[c]{Department of Mathematics and Computer Science, Amirkabir University of Technology, Tehran, Iran}
\affiliation[d]{Department of Mathematics, Faculty of Mathematical Sciences, Alzahra University,
Tehran, Iran}
\affiliation[e]{Department of Physics, CT.C, Islamic Azad University, Tehran, Iran\\ P.O. Box 14676-86831,
	Tehran, Iran}
\affiliation[f]{Institute of Biosocial and Quantum Science and Technologies, CT.C, Islamic Azad University, Tehran, Iran}
\emailAdd{rezapirmoradian@ipm.ir}
\emailAdd{So.rouhani@aut.ac.ir}
\emailAdd{mtanhayi@ipm.ir}

\begin{abstract} 
{We investigate the integrability-to-chaos transition and information scrambling in Ising spin networks via a graph-theoretic formulation. Modeling spins as vertices and interactions as edges encoded by adjacency matrices across path, Erdos-Renyi, and Watts-Strogatz topologies, we demonstrate that long-range couplings and heterogeneous degree distributions markedly accelerate quantum information propagation. The Hamiltonian comprises local and normalized non-local interactions; tuning the non-local coupling and field heterogeneity drives integrability breaking. To quantify scrambling, we employ bipartite mutual and tripartite information. Increasing non-local interactions drives tripartite information to large negative values, signaling deep information scrambling. The squared commutators constructed from out-of-time-order correlators (OTOCs) exhibit early-time exponential growth; equivalently, the OTOC four-point functions themselves decay exponentially, yielding quantum Lyapunov exponents that scale systematically with parameters governing the chaotic regime. Complementing this, Krylov complexity reveals rapid operator growth in the chaotic phase, synchronizing with OTOC and mutual information dynamics. Spectrally, the transition manifests as a shift from Poissonian to Wigner-Dyson level spacing statistics. The spectral form factor (SFF) exhibits the characteristic slope-dip-ramp-plateau structure, enabling the extraction of Thouless and Heisenberg times. Crucially, a reduced Thouless time strongly correlates with accelerated information and operator scrambling. Ultimately, this work establishes a unified framework bridging network topology with information-theoretic, operator, and spectral diagnostics, offering insights into thermalization and non-equilibrium dynamics in quantum many-body systems.

}
\end{abstract} 
\begin{document} 
\maketitle
\flushbottom

\section{Introduction}
Over the past two decades, isolated non-equilibrium quantum systems have emerged as a central theme across both theoretical and experimental physics. Much of the early progress in statistical mechanics rested on exactly solvable models---the Ising model~\cite{Ising:1925em,PhysRev.65.117}, the Bethe ansatz~\cite{Bethe:1931hc}, and the Jordan-Wigner transformation~\cite{Jordan:1928wi}---and these very frameworks still serve as indispensable benchmarks when one tackles far-from-equilibrium phenomena. More recently, such systems have become a natural arena in which to study information scrambling, entanglement growth, energy diffusion, and the onset of quantum chaos~\cite{Calabrese:2006rx,Eisert:2014jea,DAlessio:2015qtq}.

Closed quantum dynamics poses a deep conceptual puzzle. In classical mechanics, nonlinear equations of motion together with deterministic chaos drive irreversible behavior. Isolated quantum systems, on the other hand, obey the unitary, time-reversible Schr\"odinger equation:
\begin{equation}
    i\hbar\frac{\partial }{\partial t}\left| \psi(t) \right\rangle=\hat H
    \left| \psi(t) \right\rangle,
\end{equation}
where \(|\psi(t)\rangle\) is the state vector and \(\hat{H}\) is the Hamiltonian. One is then led to ask: how can macroscopic irreversibility, subsystem thermalization, and chaotic behavior arise from microscopic laws that are, at their core, reversible? The question traces back to Boltzmann's ergodic hypothesis and remains among the most pressing in many-body physics. Within the setting of isolated quantum systems, the Eigenstate Thermalization Hypothesis (ETH) provides the most widely accepted resolution. The central tenet of ETH is that thermal equilibrium is not merely a feature of statistical ensembles; it is already encoded in individual energy eigenstates. In this sense, ETH bridges unitary microscopic dynamics and the thermodynamic irreversibility one observes at macroscopic scales~\cite{RevModPhys.83.863,Deutsch:1991,Srednicki:1994mfb}.

To probe these non-equilibrium regimes both empirically and theoretically, the quantum quench protocol has become a standard tool. One prepares a many-body system in a chosen initial state---often the ground state of a pre-quench Hamiltonian---and then abruptly changes a control parameter such as an external magnetic field or an interaction strength. After this sudden change the system is driven far from equilibrium and evolves unitarily under the new post-quench Hamiltonian~\(\hat{H}'\). The key objective is to track the late-time relaxation of local observables and the growth of entanglement entropy, thereby assessing whether the system equilibrates to a state describable by conventional statistical ensembles. This protocol has proved valuable not only for theoretical modeling but also for experiments carried out with ultracold atoms, trapped ions, and photonic platforms~\cite{Rigol:2008,Rigol:2009,Calabrese:2007rg,Blatt:2012chk,Bloch:2012uep,Cirac:2012}.

Quench dynamics, however insightful, is not the only route to understanding thermalization. The present study pursues an alternative, complementary approach that brings together the physics of interacting spin networks and the mathematical machinery of graph theory. Prototypical interacting Hamiltonians provide highly controllable settings in which to analyze collective behavior and correlation spreading. Adopting a graph-theoretical viewpoint allows one to bring topological and structural metrics to bear on the problem, quantifying rigorously the complexity of the underlying interaction network. Such a hybrid framework makes it possible to explore systematically the crossover between integrable and chaotic regimes and to show explicitly how the connectivity of the interaction graph governs the onset of quantum chaos and eigenstate thermalization. Merging these analytical tools, we believe, opens a productive path toward unraveling the microscopic mechanisms behind thermalization and toward drawing a sharp boundary between integrable and non-integrable quantum dynamics~\cite{Grabarits:2024snh}.

The long-time fate of a quantum system depends crucially on whether its Hamiltonian is integrable. In strictly integrable models, an extensive set of independent, quasi-local conserved charges~\(I_j\) prevents complete thermalization. The long-time steady state is therefore not described by the standard Gibbs ensemble but by the Generalized Gibbs Ensemble (GGE):

\begin{equation}
    \rho_{\text{GGE}}=\frac{1}{\mathcal{Z}}\exp\left(-\sum_{j}\lambda_j I_j\right),
\end{equation}

where the Lagrange multipliers~\(\lambda_j\) are fixed uniquely by the initial expectation values of the conserved quantities~\cite{Rigol:2007ti,Tasaki:1998,Goldstein:2005aib,Pirmoradian:2023uvt}. Non-integrable (chaotic) systems, by contrast, lack such macroscopic conservation laws. They typically undergo genuine thermalization and relax to an equilibrium state governed by the canonical Gibbs ensemble:
\begin{equation}
    \rho=\frac{1}{\mathcal{Z}}e^{-\beta \hat{H}}.
\end{equation}

This dichotomy underpins the physical content of ETH. In chaotic systems that satisfy ETH, the matrix elements of local observables in the energy eigenbasis are smooth functions of energy, while off-diagonal fluctuations are suppressed exponentially by the thermodynamic entropy. It is precisely this mathematical structure that guarantees individual highly excited eigenstates already carry thermal properties~\cite{Huse:2014tqa,Nandkishore:2014kca}.

A particularly clean theoretical laboratory for studying the crossover between these two dynamical regimes is the one-dimensional Ising spin chain in the presence of both transverse and longitudinal magnetic fields. With the longitudinal field set to zero the model is integrable and can be mapped exactly onto non-interacting fermions through the Jordan-Wigner transformation. Switching on a finite longitudinal field, however, explicitly breaks integrability by destroying the infinite tower of conservation laws. The system is then driven into a chaotic, non-integrable regime with robust thermalizing dynamics, which makes it an ideal testbed for the ETH~\cite{Pfeuty:1970qrn}. In these spin chains the propagation of information and correlations is bounded by fundamental velocity limits set by the Lieb-Robinson bounds~\cite{Lieb:1972wy,Nachtergaele:2006,Hastings:2005pr}, and the generation of diagonal entropy gives additional microscopic insight into the thermalization process~\cite{Barankov:2008qq}.

Entanglement entropy is one of the most widely used diagnostics for non-equilibrium dynamics and the approach to thermalization. In integrable systems the von Neumann entropy of a spatial subsystem grows ballistically---linearly in time---at early stages, with a spreading rate bounded by the maximum quasiparticle velocity, a constraint tied directly to the Lieb-Robinson bounds~\cite{Abanin:2017}. Chaotic, non-integrable systems behave differently: operator spreading and information scrambling are rapid, and although the entanglement entropy initially follows a similar linear rise, it eventually saturates at late times to a volume-law value. That saturation value agrees with the thermodynamic entropy and, for random pure states, closely approaches the Page value---a hallmark of ETH-driven thermalization~\cite{Page:1993df,Bennett:1995tk,Affleck:1986bv}.

Going beyond single-subsystem measures, mutual information offers a rigorous way to track how quantum correlations spread in space. The bipartite mutual information quantifies the total shared information between two subsystems, and its time evolution reveals how correlations build up and are subsequently scrambled. In integrable regimes, where quasiparticles propagate coherently, information transfer respects a clear light-cone structure; the mutual information between distant regions therefore grows with a delayed, quasi-linear profile and often shows pronounced revivals. In chaotic systems governed by ETH the picture changes: mutual information spreads more rapidly and diffusely, then decays to small values as information becomes globally scrambled across the full Hilbert space.

When one wants to capture the delocalization of information among several partitions, tripartite mutual information becomes a sensitive probe. Both the magnitude and, importantly, the sign of this quantity carry diagnostic weight. In chaotic systems, where information is dispersed quickly over all available degrees of freedom, the tripartite information typically takes on negative values. The negativity reflects the monogamy of mutual information and signals the extensive fragmentation that accompanies thermalizing dynamics. In integrable models, on the other hand, the tripartite information stays small in magnitude or can even remain positive, consistent with the persistence of stable quasiparticle excitations and local conservation laws. Taken together, the time profiles of bipartite and tripartite information complement entanglement entropy and provide a coherent set of diagnostics for distinguishing chaotic thermalization from integrable dynamics.

A further, and by now standard, tool for characterizing quantum chaos and microscopic information scrambling is the OTOC. Physically, the OTOC quantifies how much two initially commuting local operators fail to commute once one of them is evolved forward in time in the Heisenberg picture. In weakly interacting or integrable systems the squared commutator stays parametrically small. In genuinely chaotic many-body systems it grows rapidly, signalling that local information has been delocalized throughout the Hilbert space. During the scrambling regime the decay of the OTOC---and the accompanying exponential growth of the squared commutator---proceeds at a rate set by the quantum Lyapunov exponent~\(\lambda_L\). This exponent is the quantum counterpart of the classical Lyapunov exponent and measures the sensitivity of the system to initial perturbations. A key result obtained from holographic duality places a universal upper bound on~\(\lambda_L\), one that scales linearly with the temperature of the system~\cite{Shenker:2013pqa,Maldacena:2015waa}. Saturation of this bound signals maximal chaos, a property exhibited most famously by black-hole horizons and by certain strongly coupled conformal field theories~\cite{Hayden:2007cs,Pirmoradian:2021wvo}.

Complementary to dynamical correlators, the SFF probes the spectral correlations and level statistics that are intrinsic to quantum chaos:
\begin{equation}
    \text{SFF}(\beta,t)=|Z(\beta+it)|^2,
\end{equation}
where \(Z(\beta)=\text{Tr}[ e^{-\beta \hat{H}} ]\) is the analytically continued thermal partition function~\cite{Altland:1997,Verbaarschot:2000dy}. Closely related to the SFF is the distribution of nearest-neighbor energy-level spacings, which remains one of the most robust spectral diagnostics for telling chaotic systems apart from integrable ones. In chaotic systems that obey ETH, level spacings are strongly correlated and follow the Wigner-Dyson distribution---a direct manifestation of level repulsion, whereby adjacent energy levels actively avoid crossing. Integrable systems, by contrast, possess uncorrelated energy levels drawn from a Poisson distribution, reflecting the absence of level repulsion that follows from the extensive set of conserved quantities. This statistical distinction is a cornerstone of random-matrix-theory applications in quantum chaos and is routinely analyzed in parallel with the SFF.

Krylov complexity, introduced more recently, offers a different angle on operator growth and scrambling. Under Heisenberg evolution an initial local operator~\(\hat{\mathcal{O}}_0 \equiv \hat{O}\) generates a Krylov subspace through repeated application of the Liouvillian superoperator~\(\hat{\mathcal{L}}(\cdot) = [\hat{H},\cdot]\). The resulting sequence of operators is then orthogonalized via the Lanczos algorithm, yielding a set of orthonormal basis states~\(|\mathcal{O}_n\rangle\). Krylov complexity,~\(K(t)\), is defined as the expectation value of the index~\(n\) in this basis:
\begin{equation}
    K(t)=\sum_{n}n|a_n(t)|^2.
\end{equation}
In chaotic systems~\(K(t)\) grows exponentially, capturing the rapid scrambling of operators within the Krylov space. In integrable models, on the other hand, it typically shows bounded or at most polynomial growth, often accompanied by oscillatory behavior~\cite{Parker:2018yvk,Balasubramanian:2022tpr}. The contrast is sharp enough to make Krylov complexity a sensitive diagnostic of the dynamical approach to thermalization and of the structural validity of ETH~\cite{Alishahiha:2024rwm}.

Taken together, these diagnostics---entanglement entropy, mutual information, the OTOC, the SFF, and Krylov complexity---form a broad and mutually reinforcing toolkit for non-equilibrium quantum systems. They shed light on thermalization itself and, at the same time, draw connections between many-body physics, quantum information theory, and holographic gravity. Within this landscape, minimal models such as the Ising chain remain essential theoretical laboratories: despite their mathematical simplicity they expose the mechanisms that govern complex quantum dynamics and continue to inform the design of quantum simulators and other emerging quantum technologies~\cite{Essler:2004ht,Gogolin:2015gts,Schollwock:2011,Vidal:2003lvx}.

\section{The Ising Model and the Analysis of Integrability and Chaos}\label{sec:Ising}

The Ising model holds a special place in statistical mechanics and quantum many-body physics. Its microscopic definition is elementary---binary spins coupled to their neighbors---yet the phenomenology it supports is strikingly rich: cooperative behavior, spontaneous symmetry breaking, and universal criticality all emerge from this minimal setup. In the quantum one-dimensional version, where a transverse field competes with the spin--spin coupling, the model furnishes the simplest analytically tractable realization of a continuous quantum phase transition~\cite{Pfeuty:1970qrn}. Under the Jordan--Wigner transformation~\cite{Jordan:1928wi} the chain maps exactly onto free fermions, so that its spectrum, correlations, and quench dynamics can all be written down in closed form; the classical two-dimensional counterpart, meanwhile, is solved exactly by Onsager's construction~\cite{PhysRev.65.117}. It is this combination of simplicity and exact solvability that makes the Ising model a conceptual laboratory in which fundamental notions---order parameters, scale invariance, universality---can be tested under full theoretical control, not merely a convenient computational tool.

The same laboratory turns out to be well suited to questions on the other side of many-body physics: thermalization and chaos~\cite{RevModPhys.83.863,DAlessio:2015qtq}. Strictly integrable models cannot thermalize in the conventional sense, so probing the onset of chaos requires deforming the Hamiltonian away from the free-fermion point. The deformation we adopt here supplements the local Ising model with a longitudinal field and an infinite-range coupling,
\begin{equation}
    \hat{H}=\hat{H}_{\mathrm{local}}-\frac{g}{\sqrt{N}}\sum_{i<j}\hat{\sigma}_{i}^{z}\hat{\sigma}_{j}^{z},
    \label{eq:H}
\end{equation}
where
\begin{equation}
    \hat{H}_{\mathrm{local}}=-J\sum_{(i,j)\in \mathbb{E}}\hat{\sigma}_{i}^{z}\hat{\sigma}_{j}^{z}
    -h_{x}\sum_{i=1}^{N}\hat{\sigma}_{i}^{x}-h_{z}\sum_{i=1}^{N}\hat{\sigma}_{i}^{z},
    \label{eq:Hlocal}
\end{equation}
with $J$ setting the pairwise coupling along the edges $\mathbb{E}$ of the interaction graph, $h_{x}$ and $h_{z}$ the transverse and longitudinal fields, and $g$ the strength of the all-to-all term. For the path graph, $\mathbb{E}=\{(i,i+1)\}$ and Eq.~(\ref{eq:Hlocal}) reduces to the familiar chain form. Either the longitudinal field or the infinite-range interaction alone is sufficient to destroy the free-fermion structure; together they place the model firmly in the non-integrable regime. The prefactor $1/\sqrt{N}$ keeps the collective interaction under control as the number of spins grows, in the spirit of infinite-range models \`a la Sachdev--Ye~\cite{Sachdev:1993}, while remaining strong enough, at the system sizes accessible to exact diagonalization, to push the spectral statistics from Poisson to Wigner--Dyson form---the standard fingerprint of quantum chaos, thermalization, and the ETH~\cite{Jurcevic:2014,Pachos:2004tqe,Oganesyan:2007,Haake:2010fgh}.

\begin{figure}[tbp]
\centering
\includegraphics[width=.495\textwidth,origin=c]{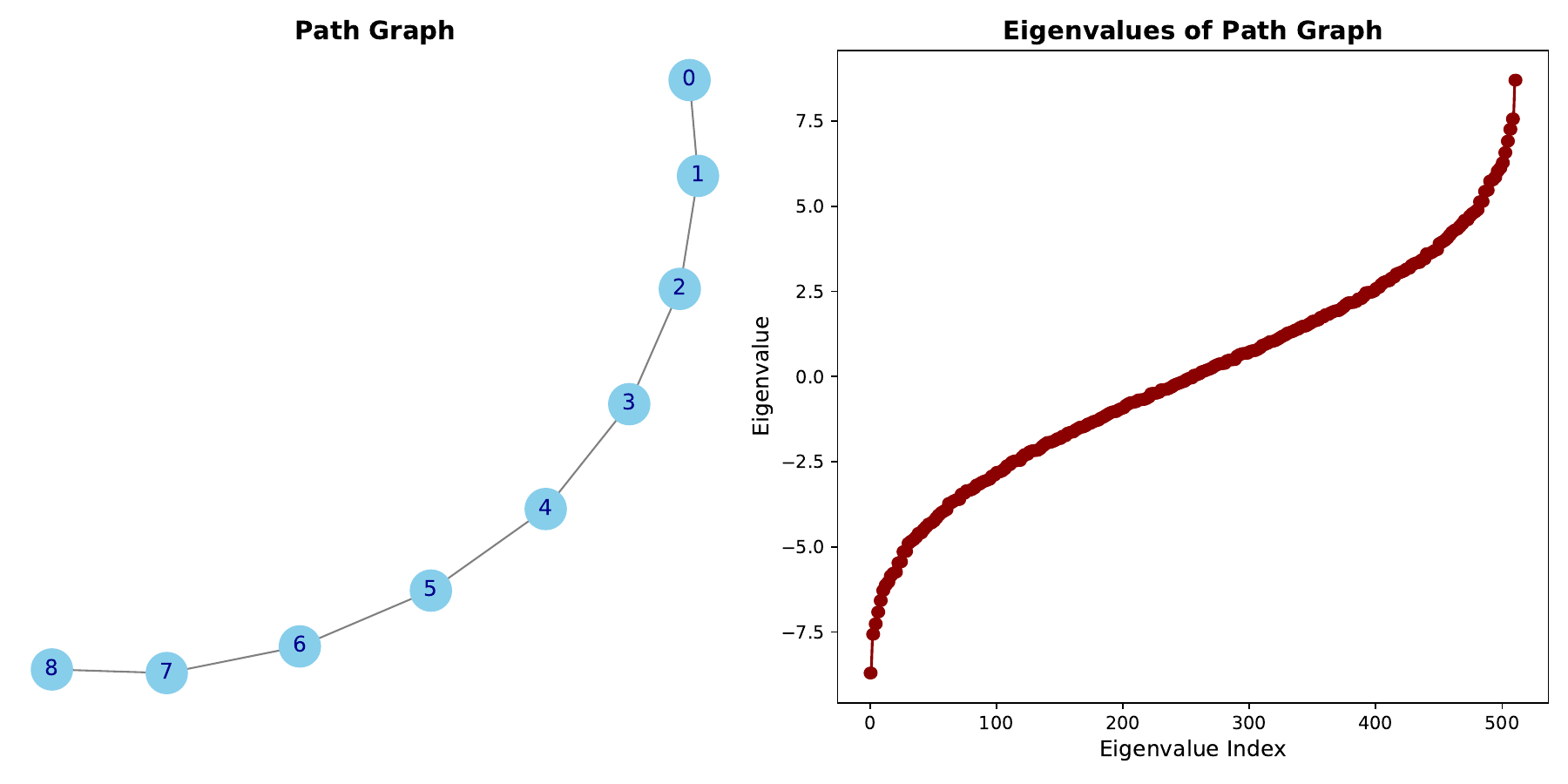}
\includegraphics[width=.495\textwidth,origin=c]{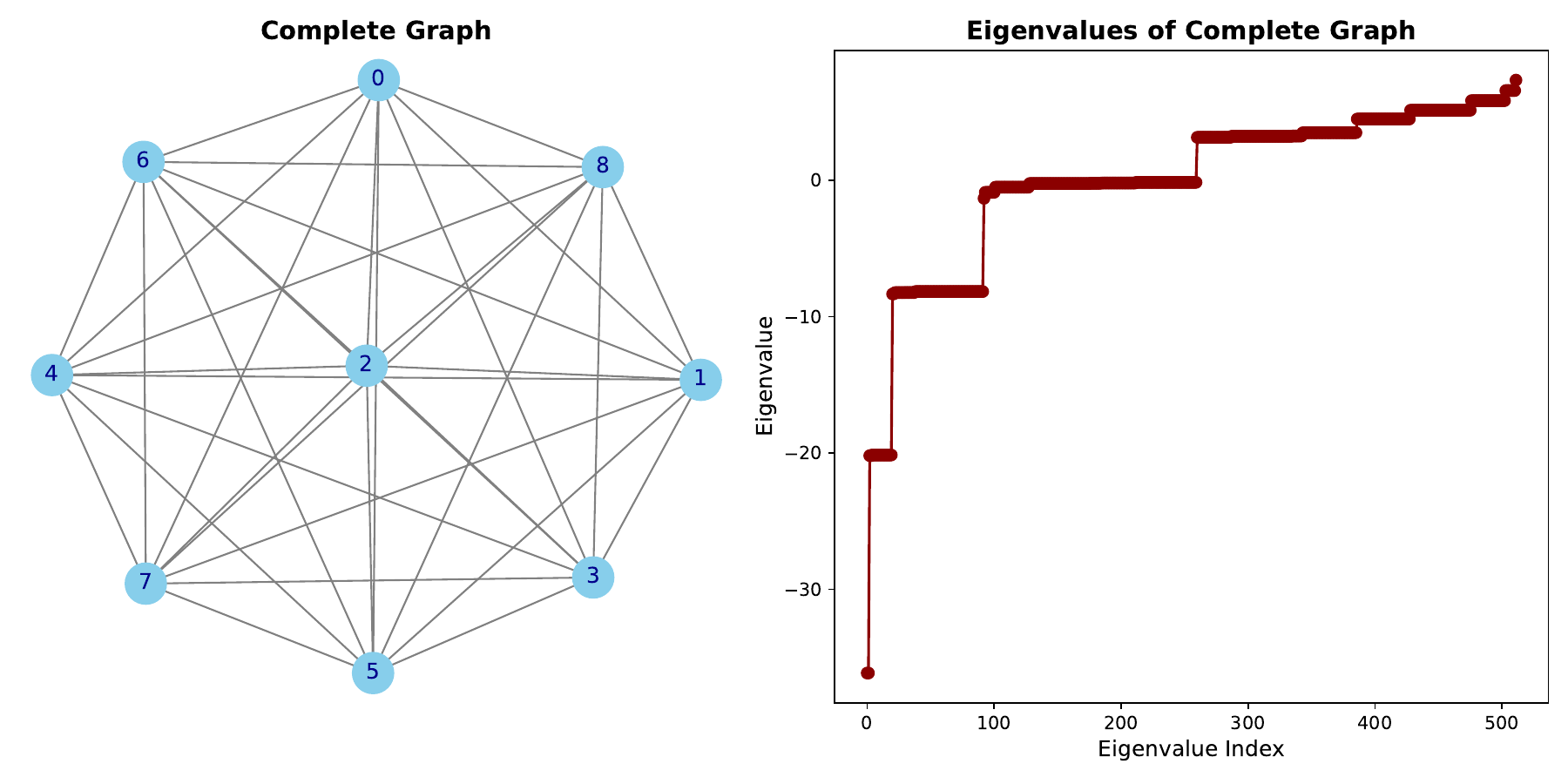}
\includegraphics[width=.495\textwidth,origin=c]{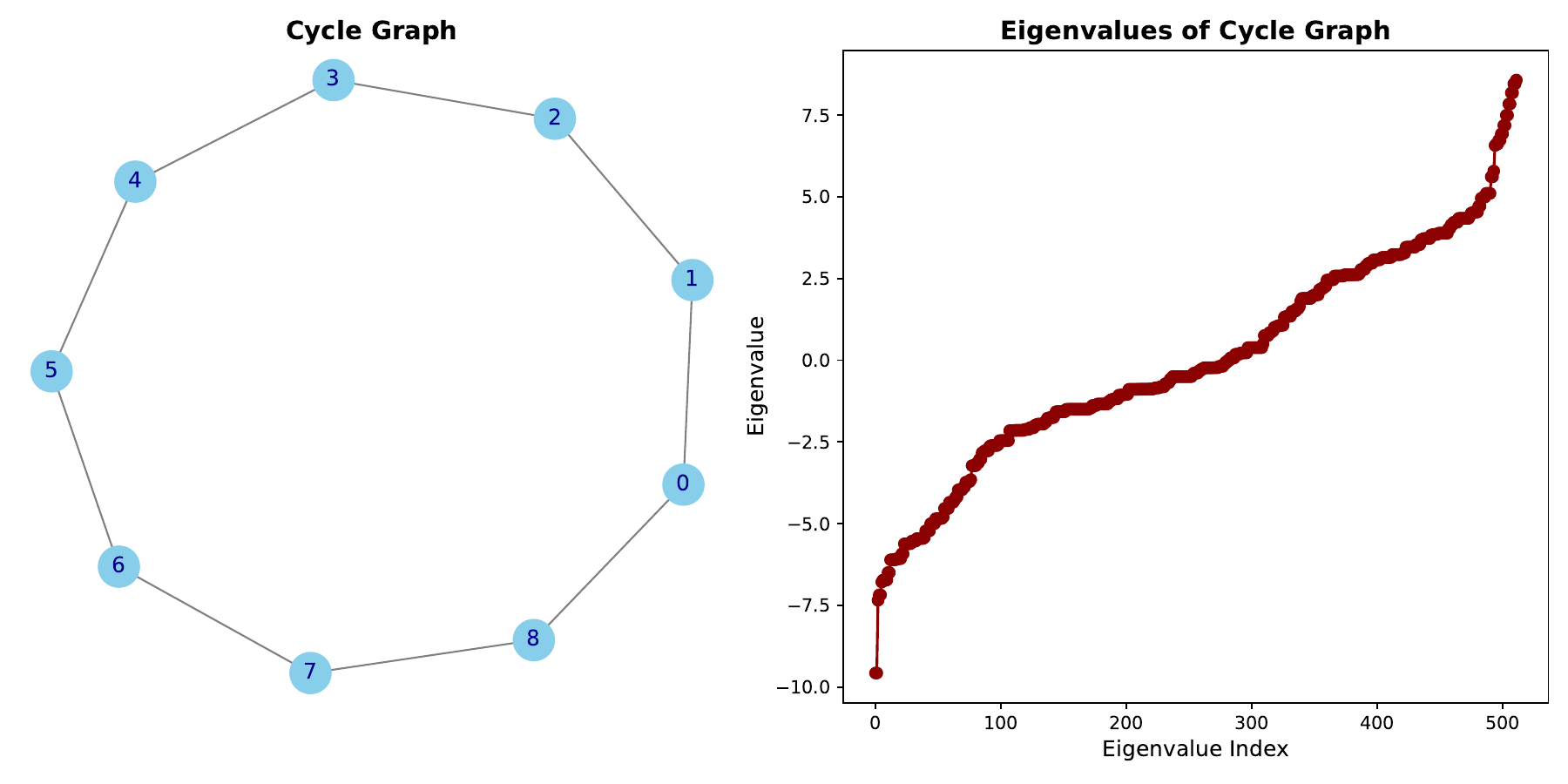}
\includegraphics[width=.495\textwidth,origin=c]{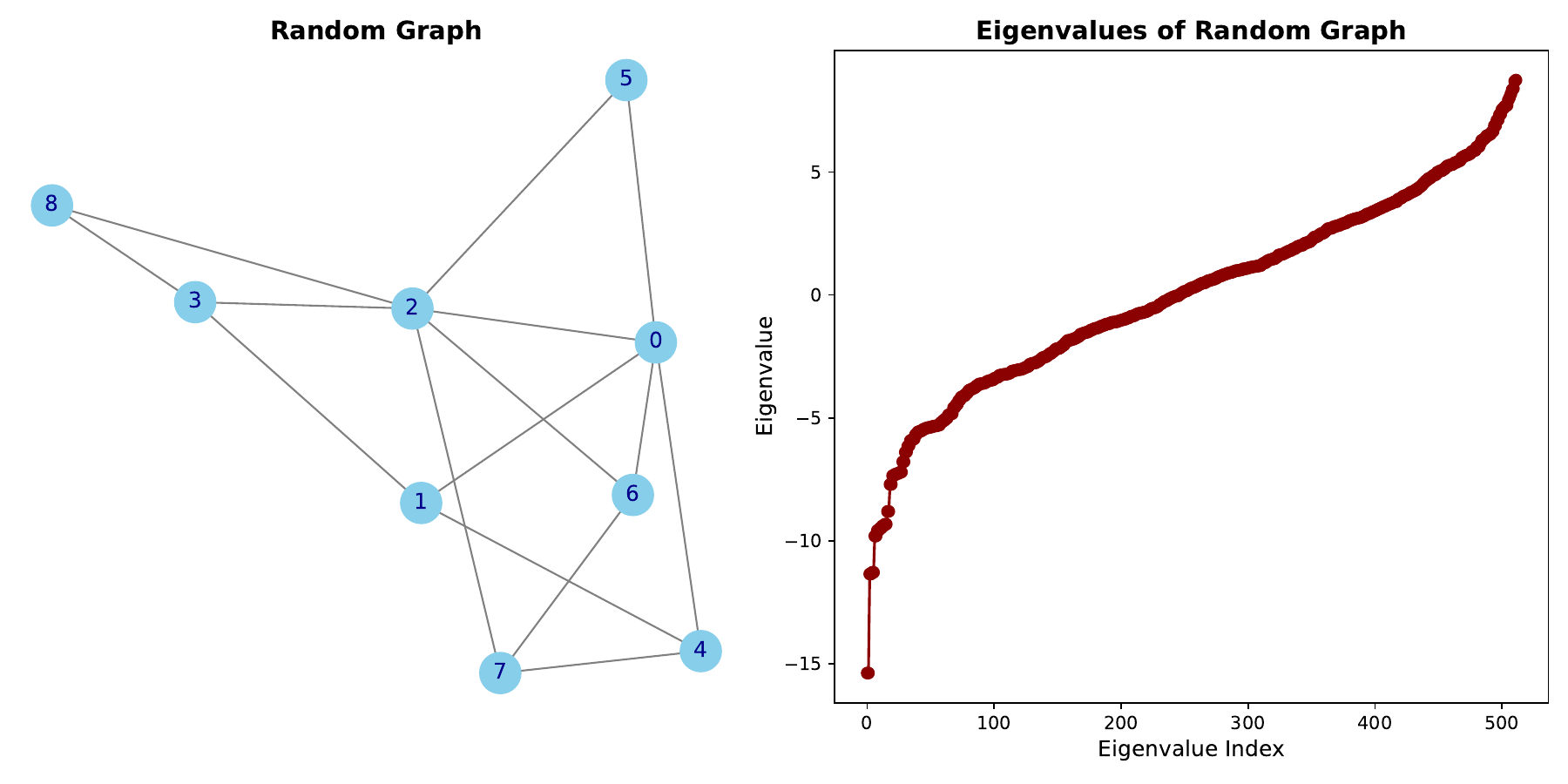}
\caption{Ordered energy eigenvalues of the Ising Hamiltonian defined on four interaction topologies: path (top left), complete (top right), cycle (bottom left) and random (bottom right) graphs.}
\label{Graph_eigenvalues}
\end{figure}

A complementary handle on the physics, one that we exploit throughout this work, is the topology of the interaction network. Graph theory supplies a natural language for this purpose~\cite{Grabarits:2024yhi}: each vertex of a graph carries a spin-$1/2$ degree of freedom whose local basis is formed by the eigenstates of $\hat{\sigma}^{z}$ with eigenvalues $\pm1$, and each edge specifies a pairwise coupling between the corresponding spins. Choosing the underlying graph---path, cycle, complete, Erd\H{o}s--R\'enyi, Watts--Strogatz, and so on---amounts to engineering the coupling structure of the Hamiltonian and, with it, the organization of the spectrum. Fig.~\ref{Graph_eigenvalues} illustrates the point by displaying the ordered energy eigenvalues for four representative topologies: path, complete, cycle, and random graphs.

The imprint of topology on the spectrum is immediate. Path and cycle graphs encode one-dimensional geometries with open and periodic boundary conditions, respectively. The complete graph, owing to its permutation symmetry, reduces to a highly symmetric collective model whose regularity shields it from generic chaotic behavior. Disordered topologies such as random graphs, by contrast, destroy the free-fermion mappings available on regular lattices and open up a natural setting for non-integrability and rich spectral behavior~\cite{Grabarits:2024yhi}.

It is useful to begin by recalling the integrable point. For the chain topology, setting $g=0$ and $h_{z}=0$, the Jordan--Wigner transformation maps the model onto non-interacting fermions. The resulting system carries an extensive family of mutually commuting conserved charges and, after a quench, relaxes not to a standard Gibbs ensemble but to a GGE built from those charges~\cite{Rigol:2016itf,Ilievski:2015jhc}. Dynamically, integrability reveals itself through entanglement growth governed by ballistic quasiparticle propagation---growth that saturates below the thermal value---and through long-lived oscillations of local observables~\cite{Eisler:2007}.

\begin{figure}[tbp]
\centering
\includegraphics[width=.495\textwidth,origin=c]{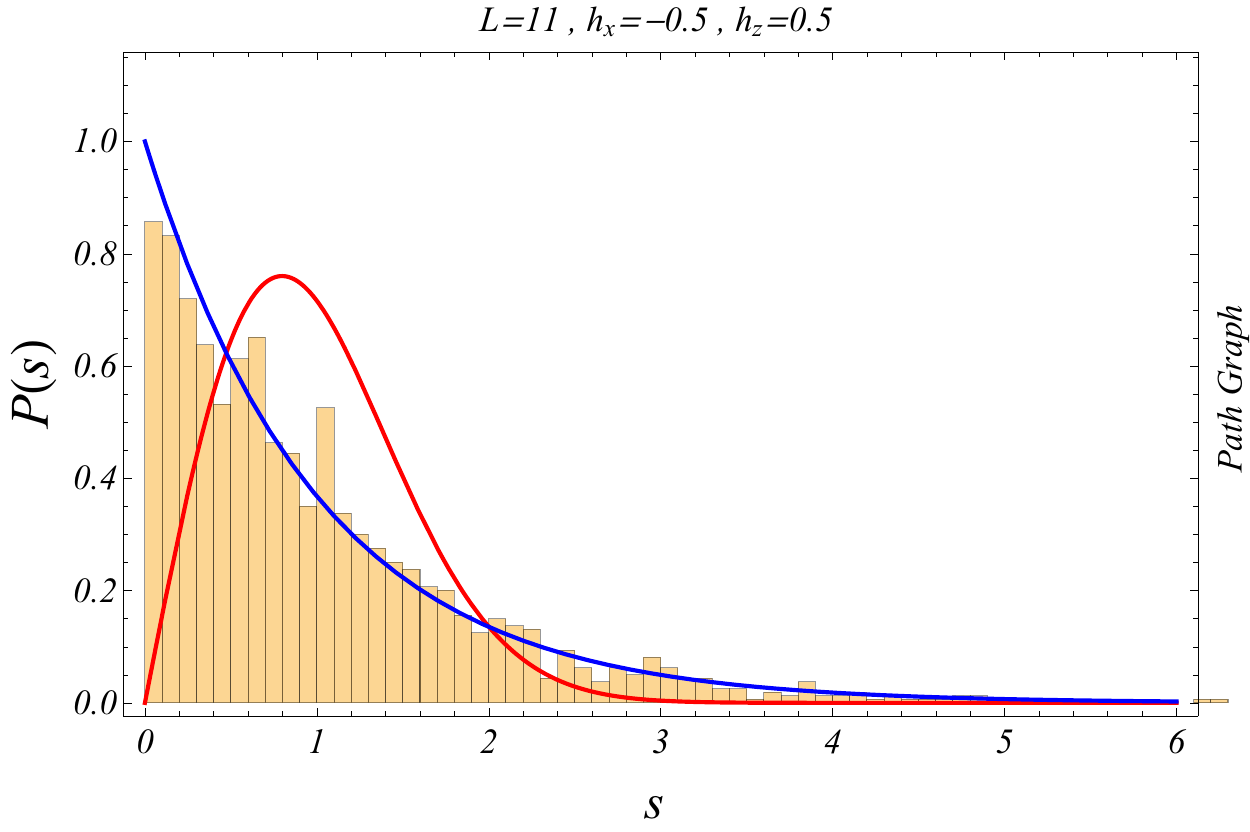}
\includegraphics[width=.495\textwidth,origin=c]{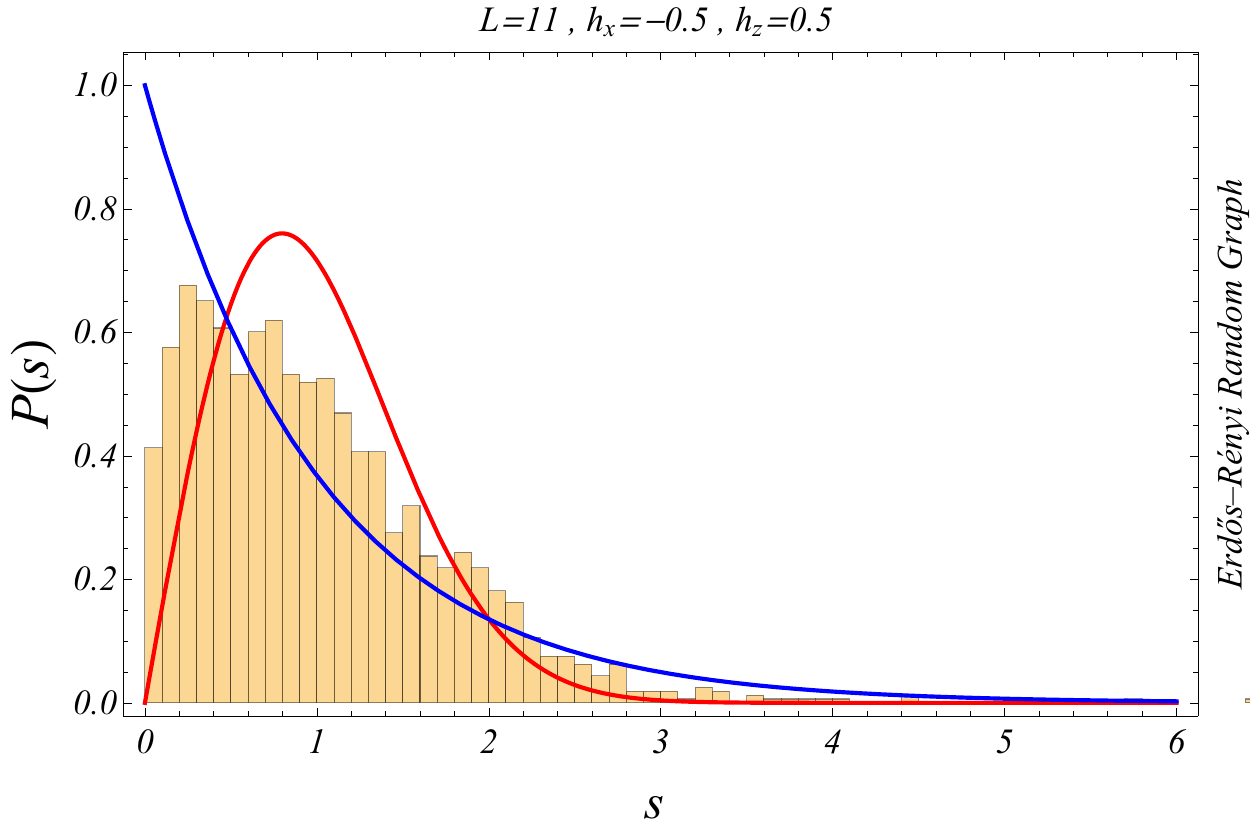}
\includegraphics[width=.495\textwidth,origin=c]{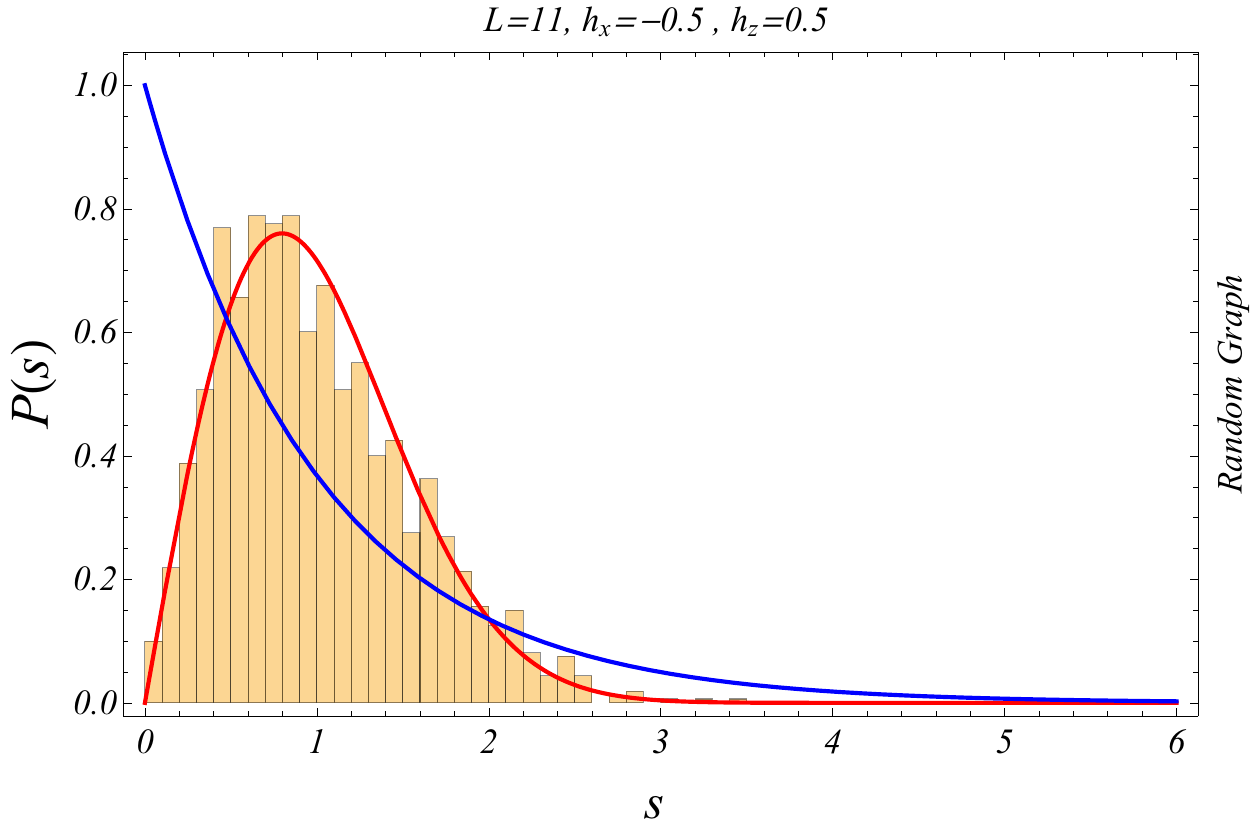}
\includegraphics[width=.495\textwidth,origin=c]{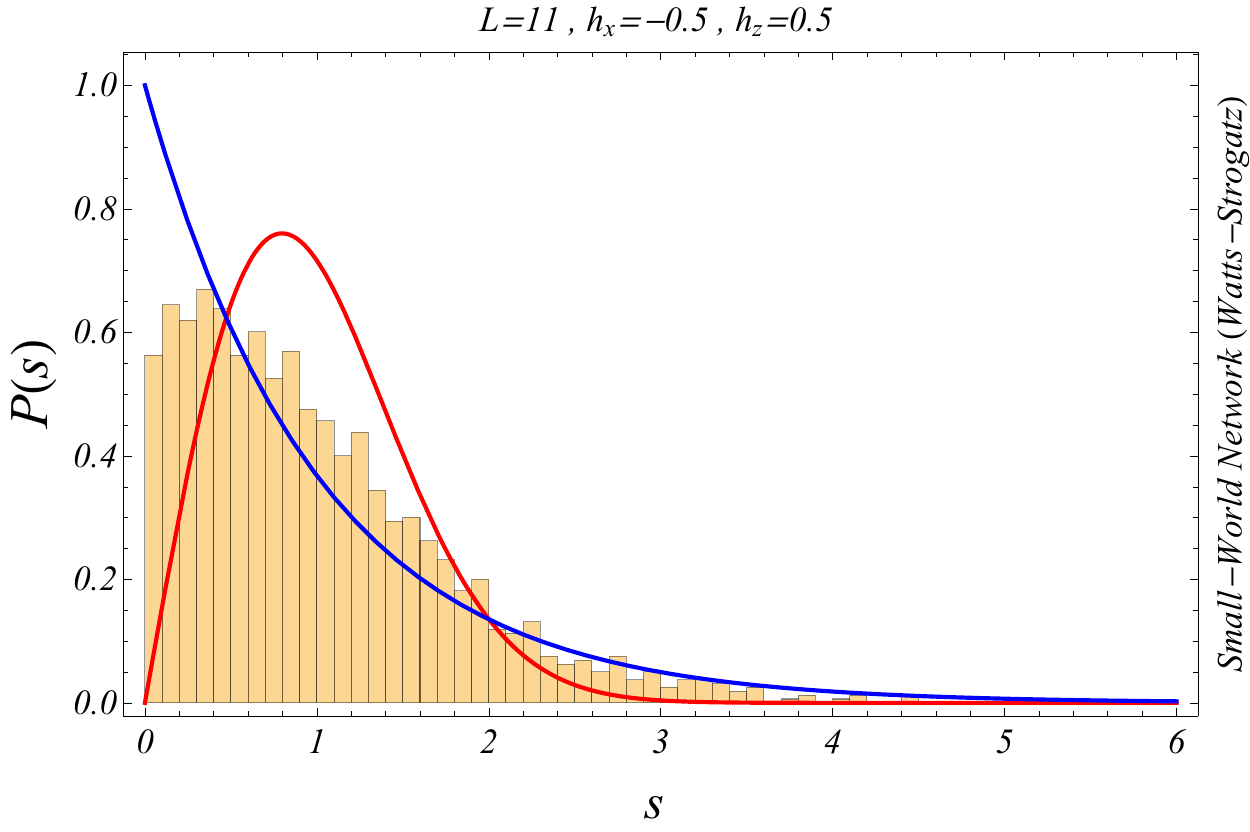}
\caption{Level-spacing distribution of the Ising spectrum in the non-integrable regime for four interaction topologies: path graph (top left), Erd\H{o}s--R\'enyi graph (top right), random graph (bottom left) and Watts--Strogatz small-world network (bottom right), with $h_{x}=-0.5$, $h_{z}=0.5$ and $L=11$. Blue and red curves denote the Poisson distribution and the GOE Wigner surmise, respectively.}
\label{hisGraph}
\end{figure}

Switching on $g\neq0$ and/or $h_{z}\neq0$ dissolves this tower of conservation laws and drives the model into the chaotic regime. The most immediate diagnostic of this transition is the statistics of level spacings. As chaos sets in, the Poissonian statistics of the integrable limit gives way, after unfolding, to the Wigner surmise of the Gaussian orthogonal ensemble (GOE),
\begin{equation}
    P(s)=\frac{\pi}{2}\,s\,\exp\!\left(-\frac{\pi}{4}s^{2}\right),
    \label{eq:WD}
\end{equation}
which encodes level repulsion and the absence of any spectral structure beyond the symmetries of the Hamiltonian~\cite{Wigner:1951}. (The GOE is the appropriate ensemble here because the Hamiltonian is real symmetric and therefore time-reversal invariant.) Fig.~\ref{hisGraph} tracks this crossover numerically: for the path, Erd\H{o}s--R\'enyi, random, and Watts--Strogatz small-world topologies, the level-spacing histograms in the non-integrable regime are compared against the Poisson law (blue curves) and the Wigner--Dyson surmise (red curves). The depletion of small spacings, signaling level repulsion, is clearly developed---most prominently for the disordered topologies---while residual finite-size effects at $L=11$ keep the histograms within the crossover region. The same random-matrix structure underlies the rapid growth of entanglement entropy seen in numerical simulations and quantum-simulator experiments alike~\cite{Islam:2015mom,Bernien:2017ubn}.

Beyond spectroscopy, the chaotic regime is characterized by thermalization in the sense of the ETH~\cite{Deutsch:2018ulr}: individual energy eigenstates reproduce microcanonical predictions for local observables, so that at long times the system retains no memory of the initial state other than the conserved energy. Where an integrable model remembers its initial conditions through its extensive charge algebra and equilibrates to a GGE, a chaotic one possesses no local integrals of motion beyond the global charges; its long-time behavior is indistinguishable from that of a standard Gibbs ensemble~\cite{Reimann:2008}. This restructuring of the eigenstates goes hand in hand with fast entanglement growth saturating at the thermal, Page-like value characteristic of random states, with the scrambling of local information~\cite{Hosur:2015ylk}, and with a rapid increase of operator complexity.

On a practical level, the infinite-range coupling also accelerates the spreading of correlations toward thermal equilibrium, as one can see from the fast development of multipartite entanglement and the exponential growth of operator size~\cite{Nahum:2017yvy}. These observations, taken together, establish the local and non-local Ising models as a controlled setting in which the diagnostics of chaos and thermalization, in particular OTOCs and Krylov complexity, analyzed in the subsequent sections can be examined on equal footing.

\section{Entanglement Dynamics and Entropy Growth}\label{sec:EntDyn}

A quantitative understanding of how entanglement and entropy grow in nonequilibrium quantum systems is central to the problem of quantum information spreading, the establishment of long-range correlations, and the build-up of operator complexity. Beyond the characterization of transient, far-from-equilibrium behaviour, these processes underpin the modern diagnostics of many-body chaos, notably OTOCs and Krylov complexity, which quantify the sensitivity of the dynamics to local perturbations and the efficiency of scrambling \cite{Swingle:2016var,Garcia-Mata:2022voo,Nandy:2024evd}.

On a practical level, the natural setting for such an analysis is provided by spin models, and in particular by the local and nonlocal Ising Hamiltonians introduced in Sec.~\ref{sec:Ising} and their non-integrable deformations. Within this family, the time evolution of entanglement is controlled not only by the microscopic structure of the interactions but also by the integrability of the model and the validity of the ETH. In integrable systems the extensive set of conserved charges constrains relaxation to a GGE rather than to a thermal ensemble, and the entanglement entropy, although growing ballistically, retains a memory of the initial state. In chaotic, ETH-compliant systems, by contrast, the dynamics acquires an effectively stochastic character described by Random matrix theory (RMT): information is scrambled rapidly, operator complexity grows exponentially at early times \cite{Parker:2018yvk}, and entanglement saturates at thermal values consistent with the microcanonical ensemble.

A detailed investigation of these dynamics therefore does more than characterize a particular model: it connects quantum chaos, statistical mechanics, critical field theory and quantum information theory within a single framework, and one-dimensional Ising models together with their graph-based generalizations serve as ideal conceptual laboratories in which the interplay between localization, thermalization and complexity can be dissected under full control.

\begin{figure}[tbp]
\centering
\includegraphics[width=.495\textwidth,origin=c]{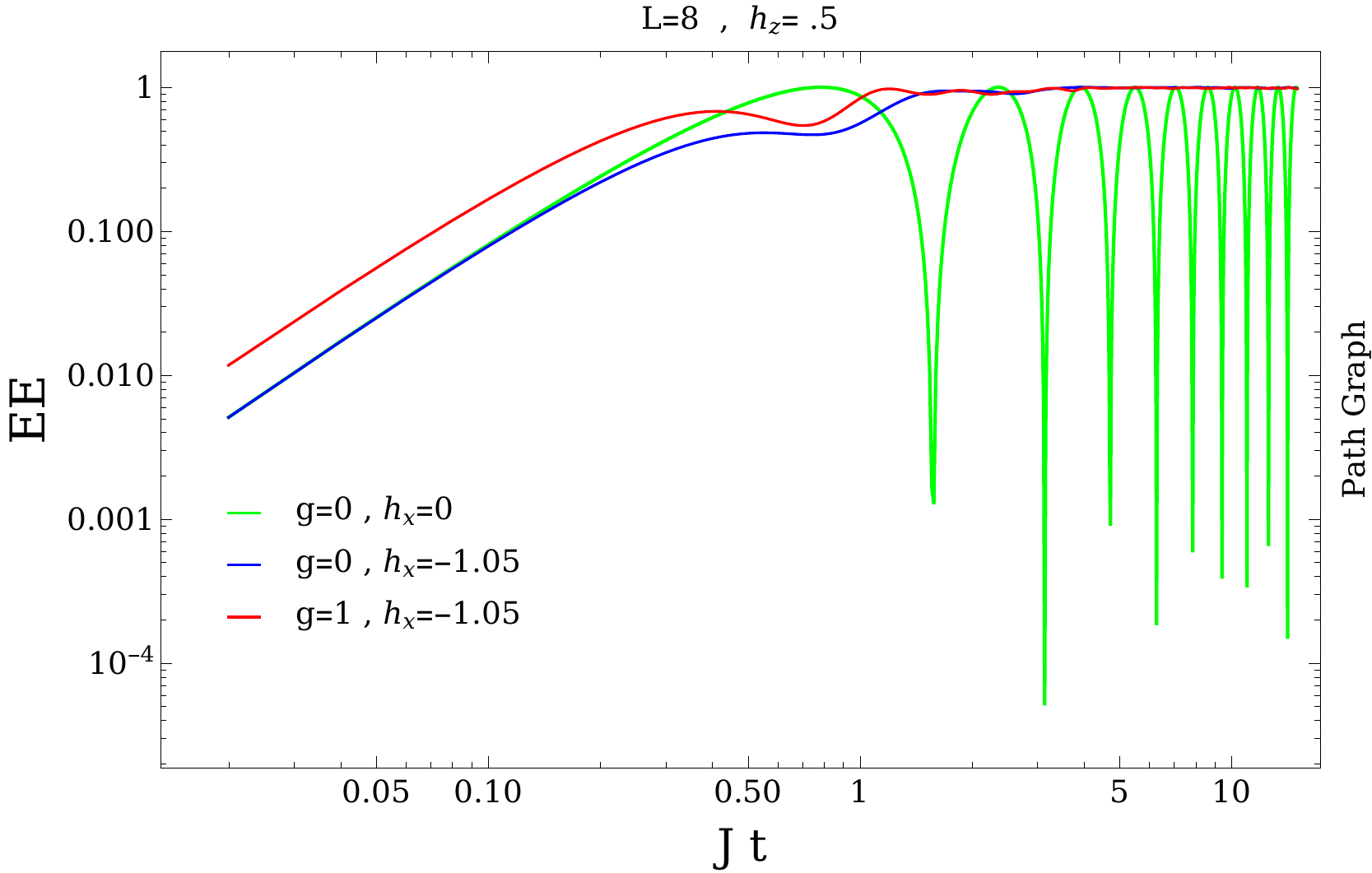}
\includegraphics[width=.495\textwidth,origin=c]{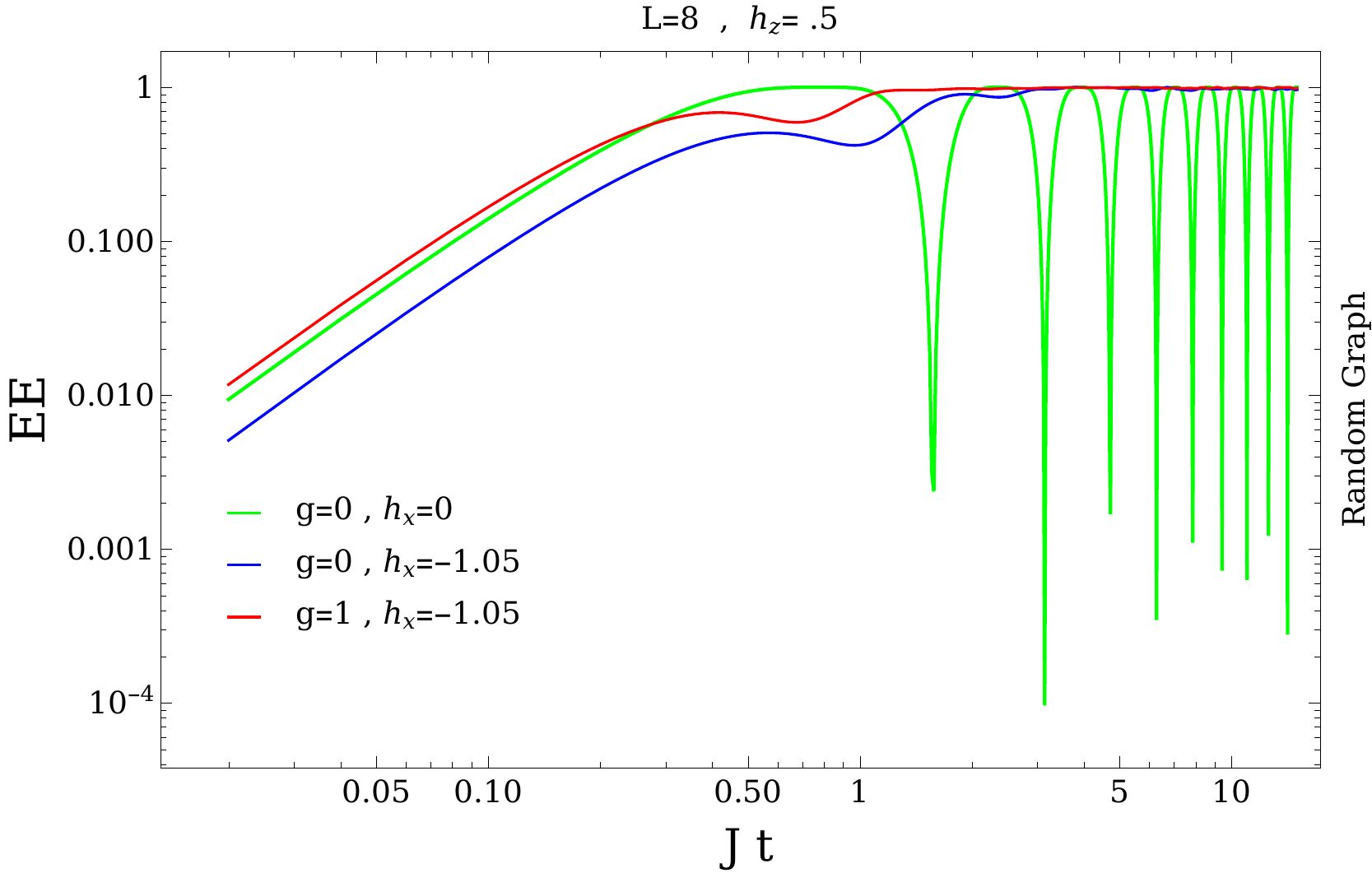}
\caption{Time evolution of the single-site (first qubit) entanglement entropy for the Ising model defined on a path graph (left) and a random graph (right), with $h_z=0.5$ and $L=8$, averaged over $180$ random initial product states. The three curves correspond to $g=0,\ h_x=0$ (green), $g=0,\ h_x=-1.05$ (blue) and $g=1,\ h_x=-1.05$ (red).}
\label{plEE}
\end{figure}

Fig.~\ref{plEE} presents the time evolution of the entanglement entropy of a single spin for the path and random graphs, comparing three representative parameter sets. This observable is a standard diagnostic of quantum thermalization: non-ergodic or localized phases host area-law entangled eigenstates, whereas thermalizing, ETH-compliant dynamics exhibits volume-law entanglement, and the local entropy saturates at the thermal value \cite{Hastings:2007iok}. To suppress sample-to-sample fluctuations and ensure robustness against the choice of initial conditions, all results are averaged over $180$ realizations of random product states. Each realization is a product state over the Bloch sphere,
\begin{equation}
    |\theta,\phi\rangle=\prod_{i=1}^{N}\left(\cos\frac{\theta}{2}\,|Z_{+}^{i}\rangle+e^{i\phi}\sin\frac{\theta}{2}\,|Z_{-}^{i}\rangle\right),
    \label{eq:productstate}
\end{equation}
where $\theta$ and $\phi$ are the spherical coordinates of the Bloch vector and $|Z_{\pm}^{i}\rangle$ denote the eigenstates of $\hat{\sigma}_i^z$. Within a single realization the angles are taken identical on all sites, i.e.\ the system is prepared in a spatially uniform product state, while between realizations $\theta$ and $\phi$ are sampled randomly so as to cover different energy densities across the spectrum.

For a system evolving from such a state into $|\psi(t)\rangle$, the reduced density matrix of a subsystem $A$,
\begin{equation}
    \rho_A(t)=\mathrm{Tr}_{B}\left[|\psi(t)\rangle\langle\psi(t)|\right],
\end{equation}
with $B$ the complement of $A$, forms the basis for all entanglement measures considered here \cite{Ghasemi:2021jiy,Pirmoradian:2023uvt} and encodes the loss of local information due to scrambling. The degree of entanglement between $A$ and $B$ is quantified by the von Neumann entropy
\begin{equation}
    \mathcal{S}_A(t)=-\mathrm{Tr}\left[\rho_A(t)\ln\rho_A(t)\right].
\end{equation}

The three curves in Fig.~\ref{plEE} illustrate how ergodicity, rather than integrability alone, governs this dynamics. For $h_x=0$ the Hamiltonian is diagonal in the $\hat{\sigma}^z$ basis and every local magnetization is conserved; the evolution reduces to pure dephasing, and the entropy displays pronounced oscillations and long-time revivals without approaching a thermal plateau. Switching on the transverse field ($h_x=-1.05$) restores ergodicity: the entropy rises rapidly and saturates smoothly, and the additional all-to-all coupling ($g=1$) further increases the entanglement velocity and shortens the saturation time. The pronounced late-time revivals visible in the dephasing case ($h_x=0$), particularly for the path graph, reflect coherent recurrences of a non-ergodic dynamics and weaken as the connectivity of the graph increases, whereas for the thermalizing curves only weak finite-size recurrences remain.

It is instructive to contrast this with the strictly integrable limit $g=0$, $h_z=0$. There, the entanglement entropy grows linearly in time, with the ballistic spreading of correlations governed by the quasiparticle picture and bounded by the Lieb--Robinson velocity \cite{Kim:2013etb}. The extensive set of conserved charges nevertheless prevents true thermalization: the steady-state entropy remains below the thermal value and is described by a GGE, and local observables retain persistent oscillations instead of relaxing to Gibbs predictions \cite{Rigol:2016itf,Ilievski:2015jhc}. In the chaotic regimes with $g\neq0$ and/or $h_z\neq0$, by contrast, information is scrambled efficiently, the entanglement entropy approaches volume-law thermal values, and local observables relax as dictated by the ETH \cite{Cardy:2015xaa}. This behaviour can be summarized by the phenomenological form
\begin{equation}
    \mathcal{S}_A(t)\approx
    \begin{cases}
        v_E\, t, & t<t_{\mathrm{sat}},\\[2pt]
        \mathcal{S}_{\mathrm{thermal}}, & t\geq t_{\mathrm{sat}},
    \end{cases}
\end{equation}
where $\mathcal{S}_{\mathrm{thermal}}$ is the entropy of the thermal state \cite{Reimann:2008}. In the nonlocal models both quantities respond monotonically to the strength of the deformation: increasing $g$ or $h_z$ enhances the entanglement velocity $v_E$ while reducing $t_{\mathrm{sat}}$, as seen in Fig.~\ref{plEE}. This ties the entanglement dynamics directly to the spectral analysis of Sec.~\ref{sec:Ising}---the level-spacing statistics and the absence of conserved charges---and underscores the role of nonlocal interactions and non-integrability in shaping the relaxation \cite{Bertini:2016tmj}.

Entanglement growth thus provides a dynamical diagnostic fully parallel to the OTOC and Krylov-complexity analyses developed in the subsequent sections: rapid spreading of entanglement and rapid growth of operators are two faces of the same scrambling mechanism, and together they allow integrable and chaotic dynamics to be distinguished quantitatively.

\begin{figure}[tbp]
\centering
\includegraphics[width=.495\textwidth,origin=c]{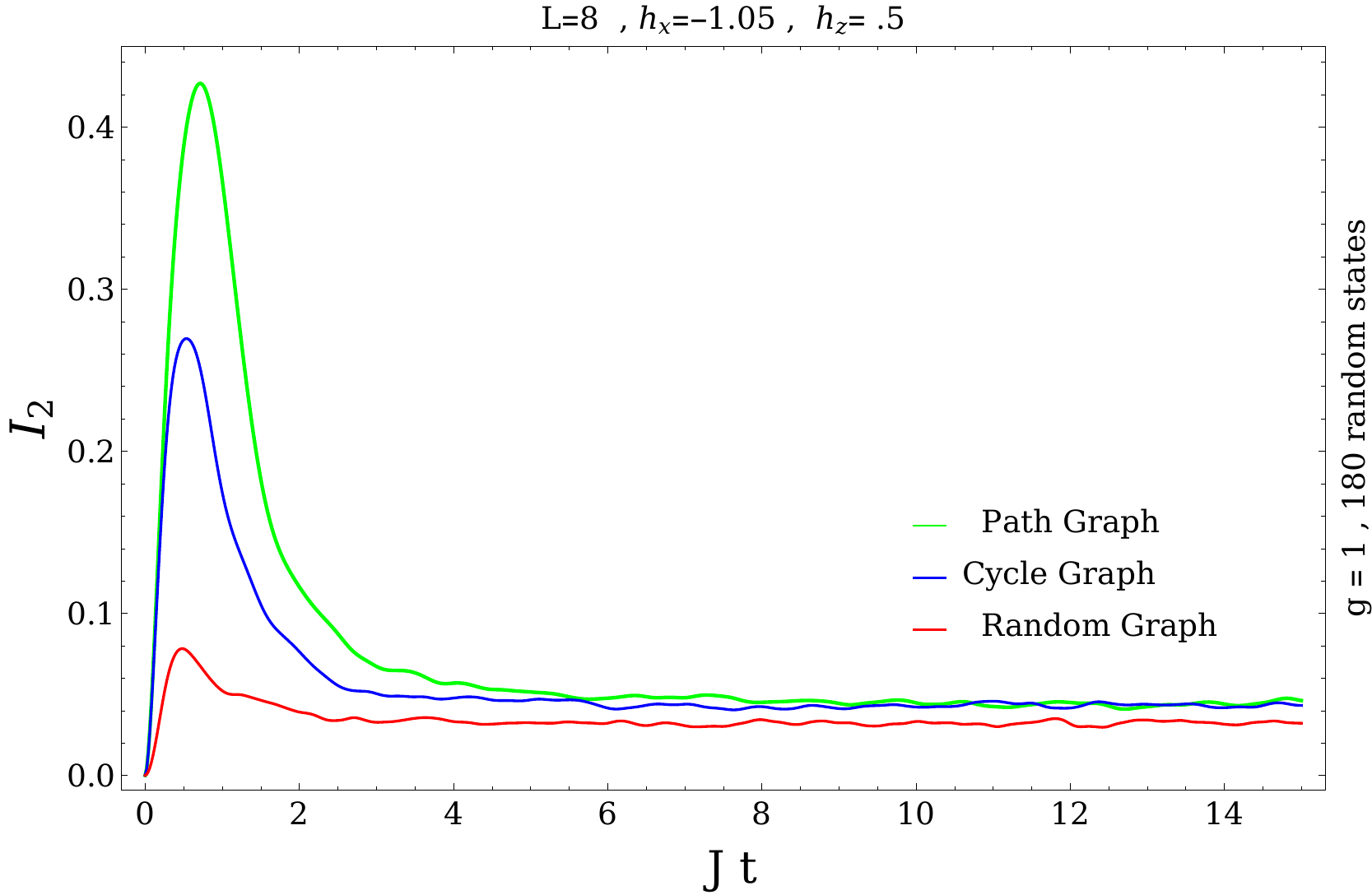}
\includegraphics[width=.495\textwidth,origin=c]{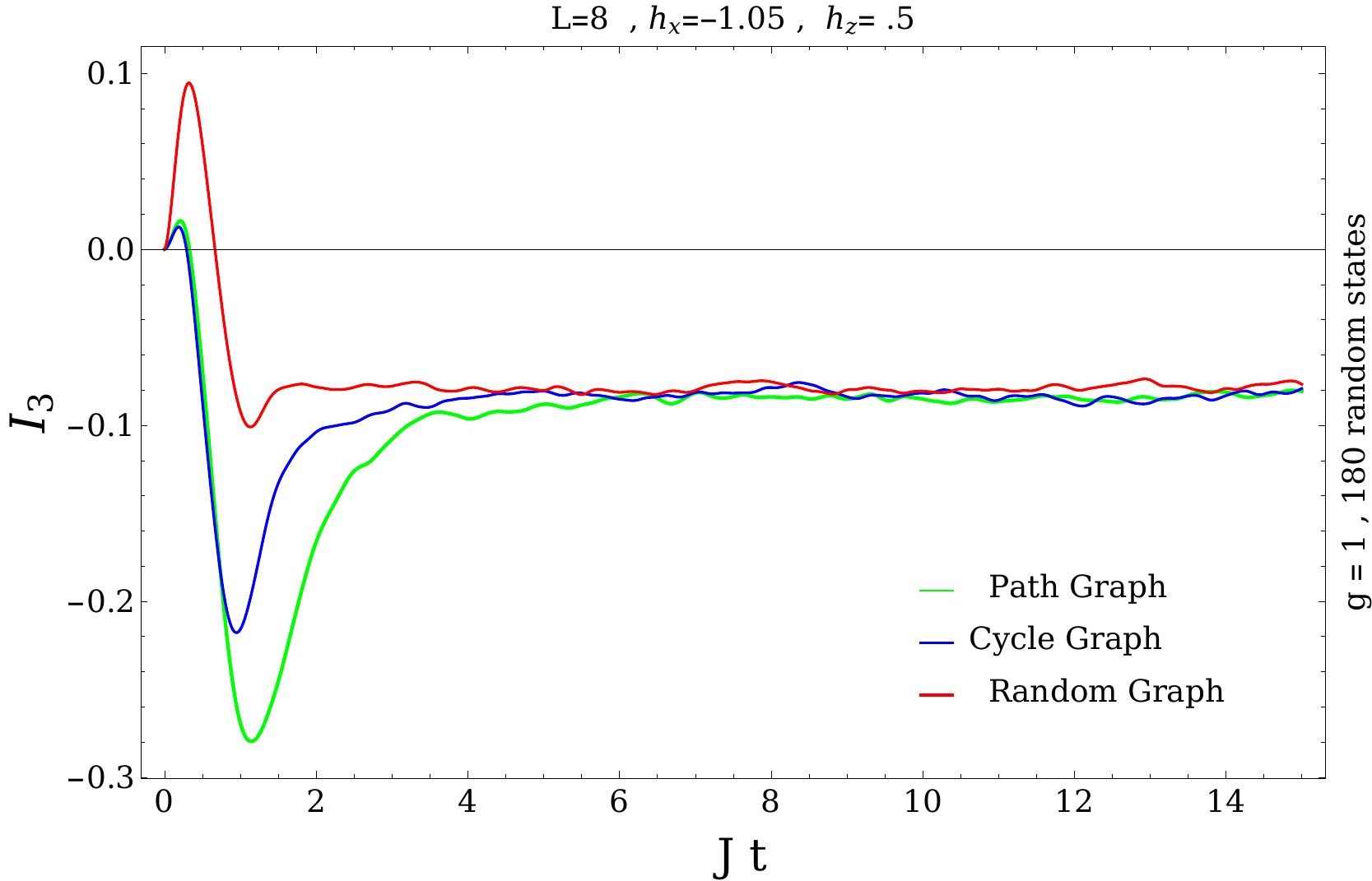}
\caption{Bipartite (left) and tripartite (right) mutual information under the Ising Hamiltonian with $g=1$, $h_x=-1.05$, $h_z=0.5$ and $L=8$, for a path graph, a cycle graph and a random graph; the results were averaged over $180$ random initial states and are shown as a function of $Jt$.}
\label{plI}
\end{figure}

Beyond the entropy of a single subsystem, mutual information between disjoint regions offers a sharper probe of how quantum information disperses. The bipartite mutual information, $I_2(A:B)=S_A+S_B-S_{AB}$, captures the total (classical and quantum) correlations shared by two subsystems and resolves the formation and decay of local correlations in time. When quasiparticles are mobile and the Lieb--Robinson velocity enforces an effective light cone \cite{Daley:2012xhf}, the mutual information between spatially separated regions grows only after a delay set by their separation and then increases approximately linearly, often accompanied by oscillations and transient revivals---a pattern characteristic of integrable systems, in which information remains carried by stable ballistic quasiparticle modes. By contrast, in chaotic systems governed by the ETH, information is rapidly scrambled across the entire many-body spectrum: after an initial rise, the mutual information between small local subsystems decays on short timescales, signalling that local information has been delocalized into global, highly non-local degrees of freedom, with no pronounced revivals.

An even more sensitive diagnostic is the tripartite mutual information, $I_3(A:B:C)=I_2(A:B)+I_2(A:C)-I_2(A:BC)$, whose sign and magnitude directly diagnose scrambling \cite{Hosur:2015ylk,Hayden:2007cs}. When information is dispersed globally, $I_3$ becomes large and negative, reflecting the monogamy of mutual information and the fact that the information initially localized in one subsystem is recoverable only from the full system. Close to the integrable limit, by contrast, $|I_3|$ is suppressed or $I_3$ turns positive, consistent with long-lived quasiparticles and with revivals of locally retrievable information, a regime in which the monogamy constraint on mutual information need not hold.

Fig.~\ref{plI} illustrates these diagnostics for the nonlocal Ising model ($g=1$) on the path, cycle and random graphs. In all three topologies the bipartite mutual information exhibits an early-time peak followed by relaxation to a small stationary value, while the tripartite mutual information develops a negative dip at intermediate times before settling onto a small negative plateau---the hallmark of information delocalization in the chaotic regime. The topology mainly affects the transient dynamics: along the sequence path $\to$ cycle $\to$ random, i.e.\ as the effective connectivity increases, the early-time peak of $I_2$ and the depth of the $I_3$ dip both decrease, since highly connected networks distribute information globally on a faster timescale. Together with the entropy growth discussed above, these time-resolved correlation measures provide a consistent picture of the route to thermal equilibrium and of the distinctions between chaotic and integrable behaviour.

\section{OTOC and chaos diagnostics}\label{sec:OTOC}

How quantum information propagates, and under which conditions chaotic behavior emerges, are central questions in the nonequilibrium dynamics of many-body systems. A particularly powerful probe of both issues is the OTOC. Given two Hermitian operators $\hat{W}$ and $\hat{V}$, it is customary to monitor the squared commutator
\begin{equation}
    C(t)=-\langle[\hat{W}(t),\hat{V}(0)]^{2}\rangle,
\label{sqc}
\end{equation}
where $\hat{W}(t)=e^{i\hat{H}t}\hat{W}e^{-i\hat{H}t}$ denotes the Heisenberg-evolved operator and $\langle\cdot\rangle$ stands for the expectation value in the initial state or in a thermal ensemble \cite{Pirmoradian:2025xxl}. The qualifier ``out-of-time-order'' refers to the fact that, upon expanding the commutator, one encounters the four-point function $f_{WV}(t)\equiv\langle\hat{W}(t)\hat{V}\hat{W}(t)\hat{V}\rangle$, in which the operators appear in a non-chronological order. Physically, $C(t)$ tracks how rapidly an initially local perturbation ceases to commute with a separate probe; its space-time dependence therefore encodes the butterfly velocity at which local disturbances spread and quantum information becomes scrambled \cite{Jaiswal:2024fotoc,Roberts:2014isa,Khemani:2018sdn}.

When the two operators are additionally unitary ($\hat{W}^{2}=\hat{V}^{2}=\hat{I}$, as for Pauli strings), the squared commutator collapses to
\begin{equation}
    C_{WV}(t)=2\big(1-\mathrm{Re}\,f_{WV}(t)\big),
\label{sqcf}
\end{equation}
so that the growth of $C_{WV}$ is entirely governed by the decay of $f_{WV}$.

In chaotic systems the early-time behavior of the squared commutator is exponential,
\begin{equation}
    C(t)\sim\epsilon\,e^{\lambda_{L}t},
\end{equation}
where $\lambda_{L}$ is the quantum Lyapunov exponent---bounded, for thermal states at temperature $T$, by the Maldacena--Shenker--Stanford limit $\lambda_{L}\le 2\pi T$ (with $\hbar=k_{B}=1$) \cite{Maldacena:2015waa}---and $\epsilon$ sets the size of the initial perturbation. The associated Lyapunov time $t_{L}\sim 1/\lambda_{L}$ estimates the interval over which a small disturbance grows to system-wide proportions \cite{Swingle:2016var,Yao:2016ayk,Khorasani:2023usq}. The presence of this exponential regime, followed by saturation of $f_{WV}$ at a small value, is the standard signature of operator scrambling and sharply distinguishes chaotic from integrable dynamics.

Integrable models, by contrast, exhibit at most polynomial growth of $C(t)$, or bounded oscillations, since an extensive set of conserved quantities constrains the propagation of information \cite{Garcia-Mata:2022voo,Daley:2012xhf}. This dichotomy is fully consistent with the spectral analysis of Sec.~\ref{sec:Ising} and the entanglement and entropy analysis of Sec.~\ref{sec:EntDyn}, where chaotic and integrable geometries were accompanied, respectively, by fast and slow growth of these quantities.

The OTOC also admits a natural interpretation in Krylov space. Within the universal operator growth hypothesis, the exponential rise of $C(t)$ is the dynamical counterpart of a linear growth of the Lanczos coefficients, $b_{n}\sim\alpha n$, which in turn drives the rapid increase of Krylov complexity and the delocalization of the operator wave function over the Krylov basis \cite{Parker:2018yvk,Roberts:2015,Swingle:2016jdj}.

For the graph-based Ising models considered in this work, switching on nonlocal couplings together with a longitudinal field increases $\lambda_{L}$ (equivalently, decreases $t_{L}$). Scrambling is therefore accelerated and entanglement builds up more rapidly, in agreement with the entropy production reported in the previous section \cite{Hosur:2015ylk,Fan:2017}.

Finally, from a spectral viewpoint, $C(t)$ is tied to the structure of the many-body spectrum: in nonintegrable systems obeying the ETH, the level statistics cross over from Poisson to Wigner--Dyson form, and the resulting level repulsion is accompanied by a larger Lyapunov rate and faster operator spreading, again in parallel with the rapid entropy growth discussed above.

Taken together, these observations show that the OTOC provides a unified framework for diagnosing integrable and chaotic nonequilibrium dynamics, and they set the stage for the Krylov-complexity and spectral analysis of the following sections.

\begin{figure}[tbp]
\centering
\includegraphics[width=.495\textwidth,origin=c]{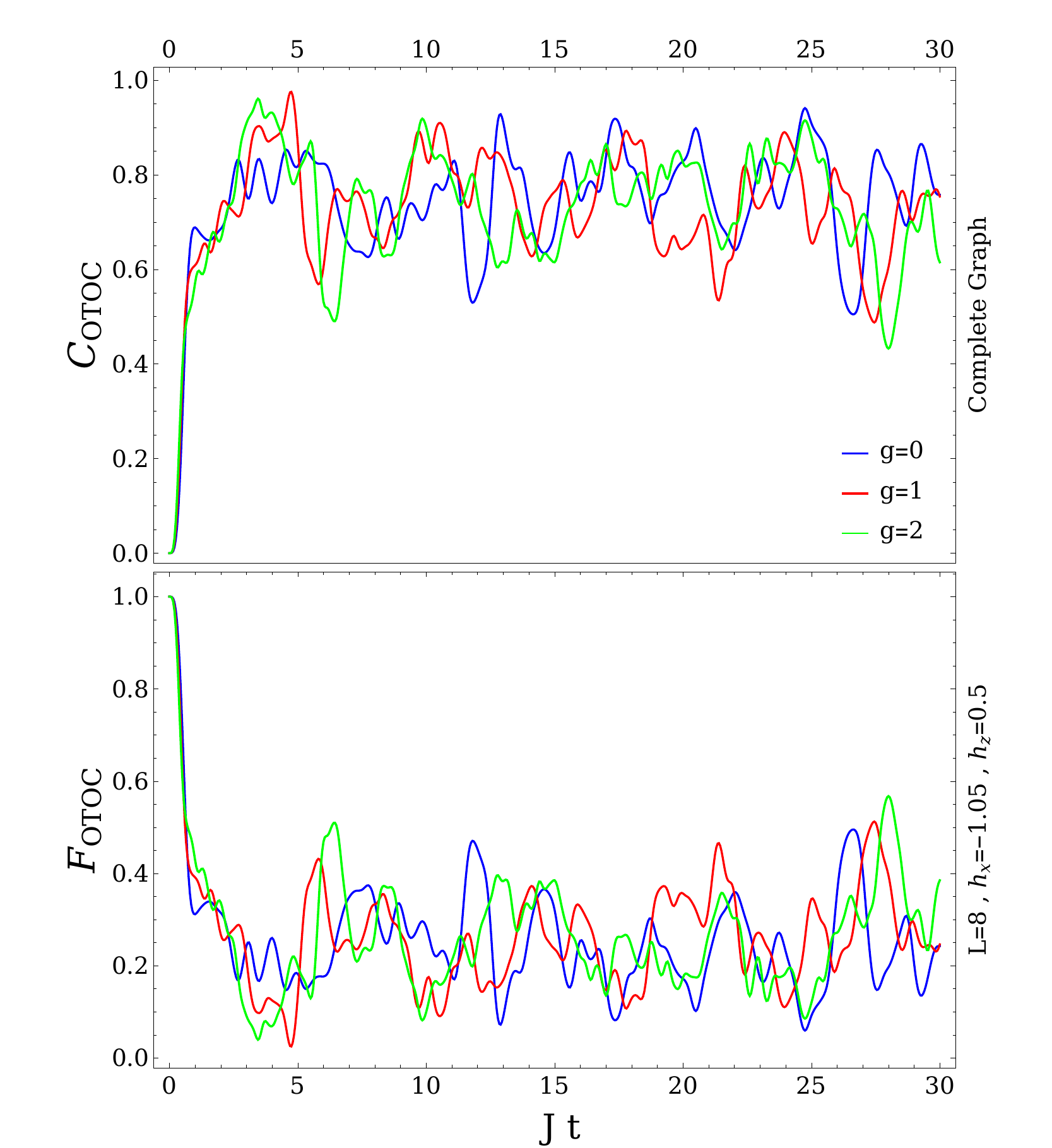}
\includegraphics[width=.495\textwidth,origin=c]{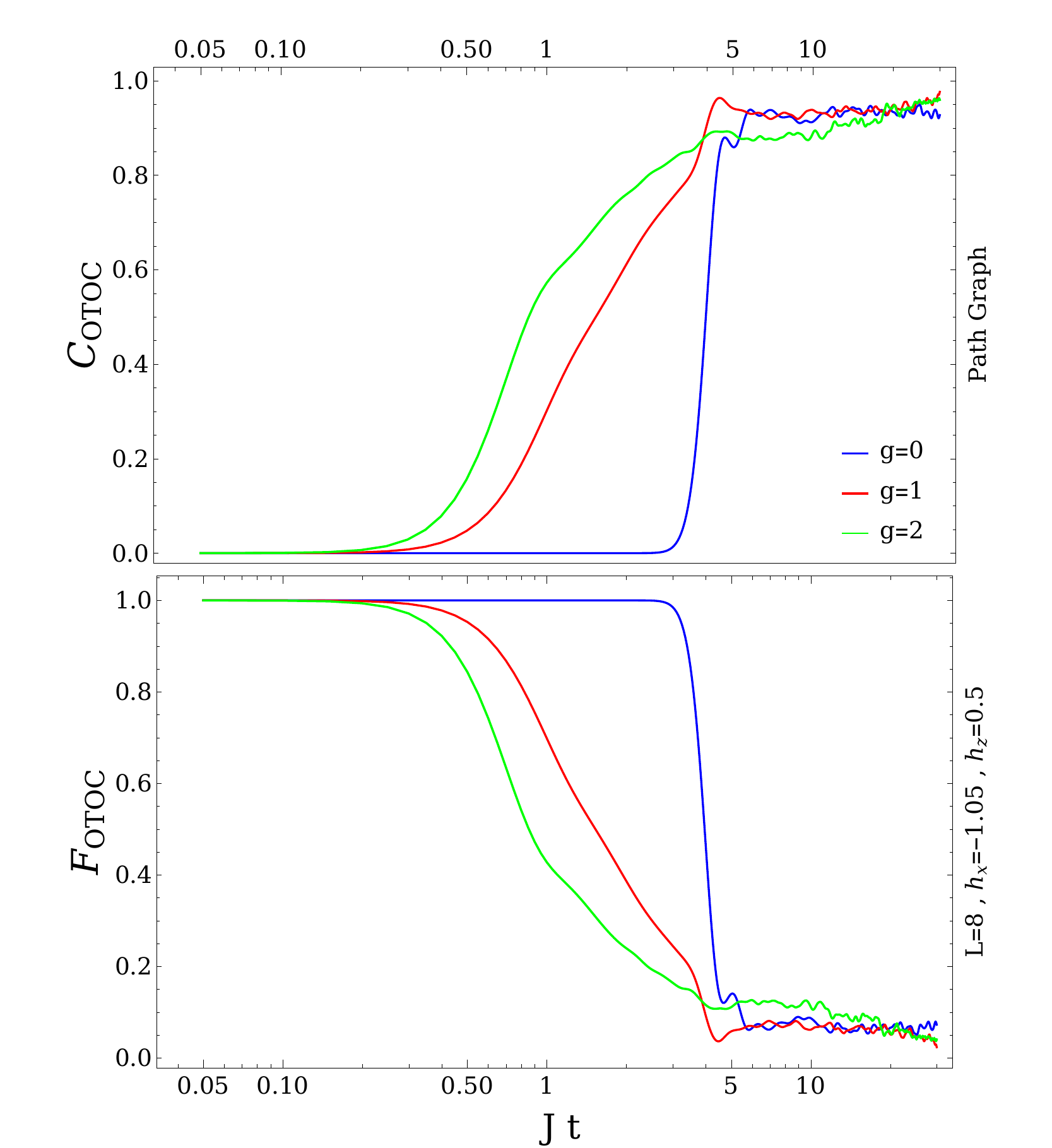}
\includegraphics[width=.495\textwidth,origin=c]{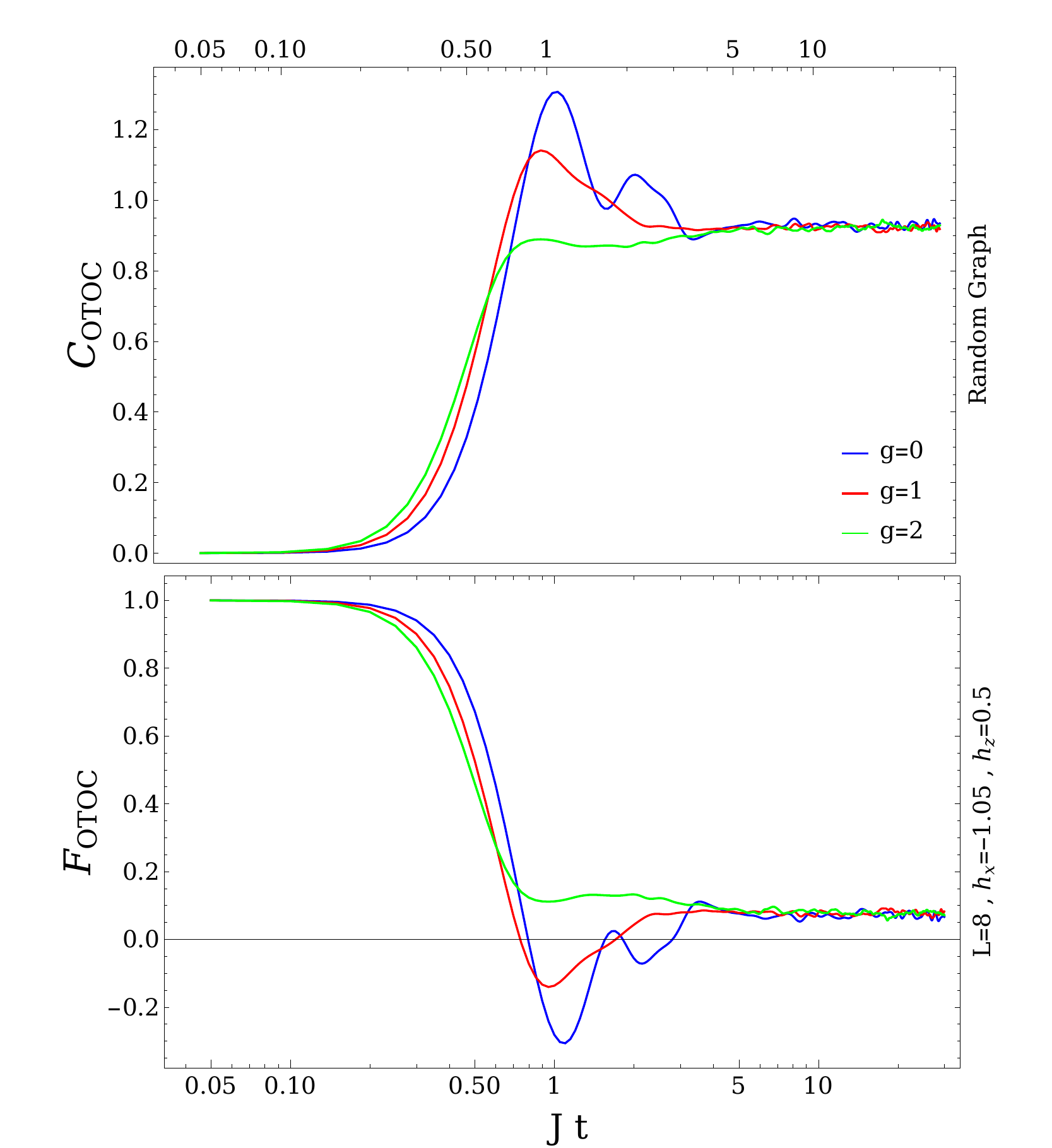}
\caption{Time evolution of the normalized squared-commutator OTOC
$C_{\mathrm{OTOC}}\equiv 1-\mathrm{Re}\,f_{WV}$ (upper panels) and of the
out-of-time-order four-point function $F_{\mathrm{OTOC}}\equiv f_{WV}$ (lower
panels) for the Ising model defined on the complete graph (top left), the path
graph (top right), and the random graph (bottom). Note that $C_{\mathrm{OTOC}}$
is plotted in units of half the squared commutator of Eqs.~(\ref{sqc}), (\ref{sqcf}),
$C_{\mathrm{OTOC}}=\tfrac{1}{2}C_{WV}=1-\mathrm{Re}\,F_{\mathrm{OTOC}}$; this
normalization does not affect the extracted Lyapunov exponents. The curves
correspond to $g=0,1,2$ as indicated in the legends; all data were obtained for
$L=8$, $h_{x}=-1.05$, and $h_{z}=0.5$.}
\label{fig:OTOC}
\label{OTOCgraphbeta0}
\end{figure}

As Fig.~\ref{OTOCgraphbeta0} illustrates, the path and random graphs---precisely the geometries that exhibit chaotic signatures---show a sharp initial rise of $C_{\mathrm{OTOC}}$, consistent with exponential growth, followed by relaxation toward a nearly constant plateau beyond the scrambling time. Late-time oscillations are strongly suppressed in these cases, indicating that the operators have effectively scrambled. The complete graph behaves differently: $C_{\mathrm{OTOC}}$ remains large and oscillatory at all accessible times, with no clean saturation, as expected for an infinite-range, integrable model.

\section{Krylov complexity, spectral diagnostics of chaos, and their connection to entanglement}\label{sec:Krylov}

Krylov complexity offers a systematic and basis-independent way to quantify operator growth in nonequilibrium quantum systems \cite{Nandy:2024evd,Bhattacharya:2022gbz}. Given an initial operator $\hat{\mathcal{O}}_{0}$---for instance, a local Pauli string $\hat{W}$---repeated action of the Liouvillian $\hat{L}=[\hat{H},\cdot]$ under the Lanczos algorithm generates an orthonormal Krylov basis $\{|\mathcal{O}_{n})\}$ in which the Heisenberg-evolved operator $|\mathcal{O}(t))=e^{i\hat{L}t}|\mathcal{O}_{0})$ expands as
\begin{equation}
    |\mathcal{O}(t))=\sum_{n=0}^{\infty}\phi_{n}(t)|\mathcal{O}_{n}),\qquad \sum_{n}|\phi_{n}(t)|^{2}=1.
\end{equation}
Krylov complexity is then the first moment of this distribution,
\begin{equation}
    K(t)=\sum_{n=0}^{\infty}n\,|\phi_{n}(t)|^{2},
\end{equation}
i.e., the expectation value of the ``position'' operator $\hat{N}=\sum_{n}n|\mathcal{O}_{n})(\mathcal{O}_{n}|$ in Krylov space; its growth measures how far the operator has spread along the Krylov chain \cite{Alishahiha:2022nhe,Roberts:2016hpo,Brown:2017jil,Doroudiani:2019llj,Pirmoradian:2020}. An entirely parallel construction applies to states: running the Lanczos algorithm on $\hat{H}$ with an initial state $|\psi_{0}\rangle$ produces a basis in which the time-evolved state delocalizes, and the corresponding moment is known as spread, or state, complexity \cite{Balasubramanian:2022tpr}. The numerical results reported below are of this latter type, with the initial state $|Y+\rangle\equiv\bigotimes_{i=1}^{L}|+y\rangle_{i}$, i.e., the product state in which every spin points along the $+y$ direction.

\begin{figure}[tbp]
\centering
\includegraphics[width=.495\textwidth,origin=c]{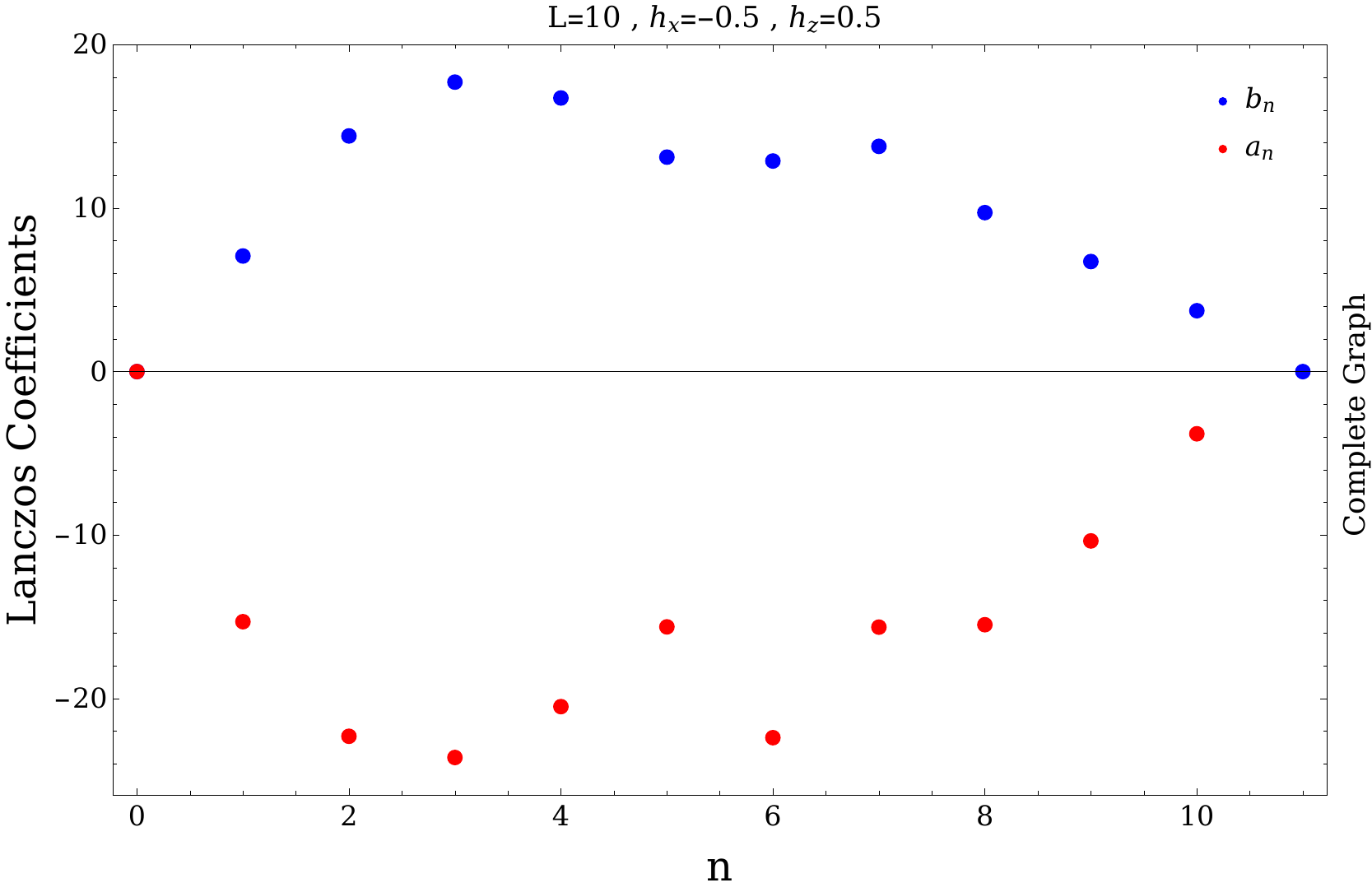}
\includegraphics[width=.495\textwidth,origin=c]{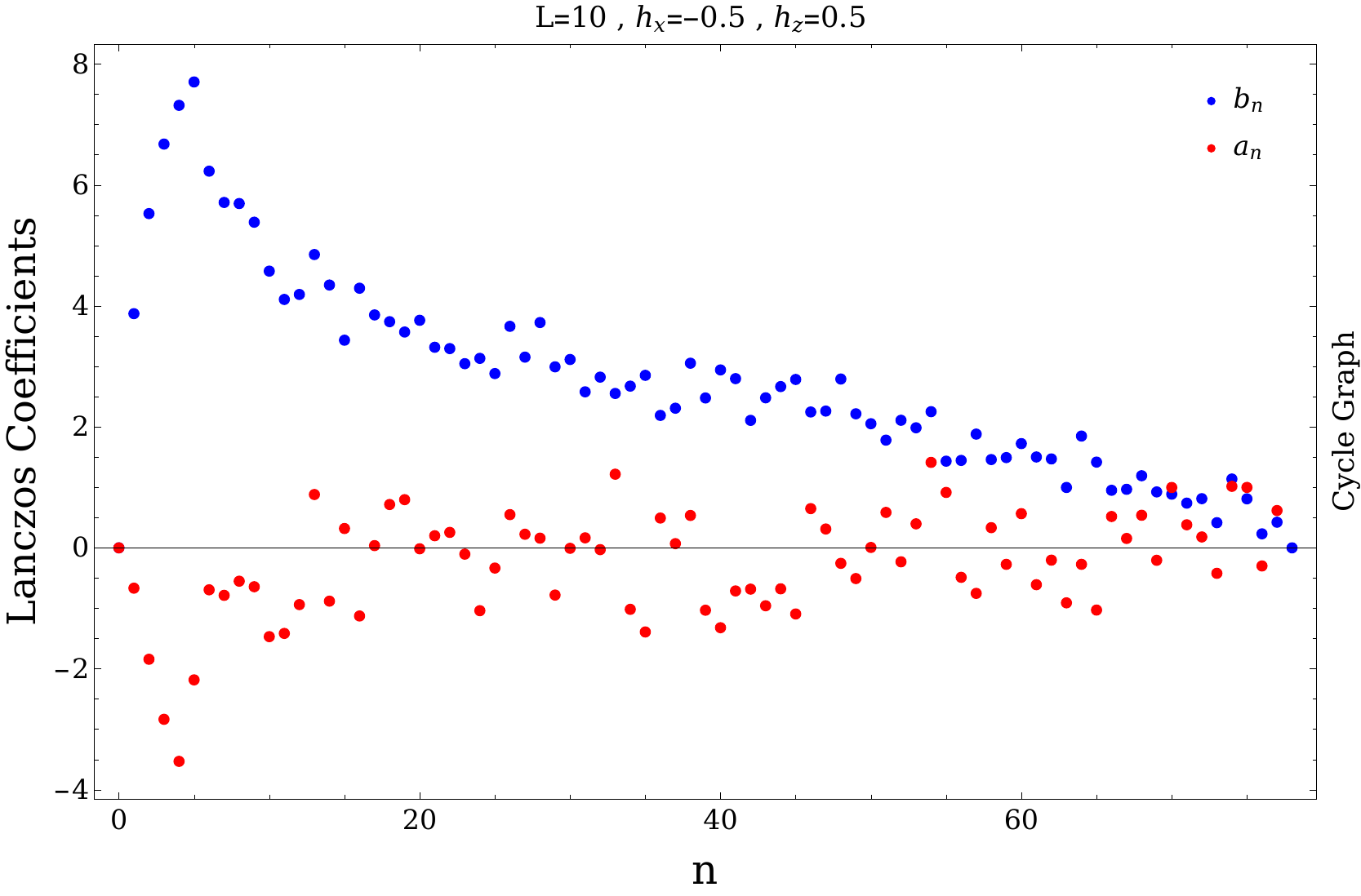}
\includegraphics[width=.495\textwidth,origin=c]{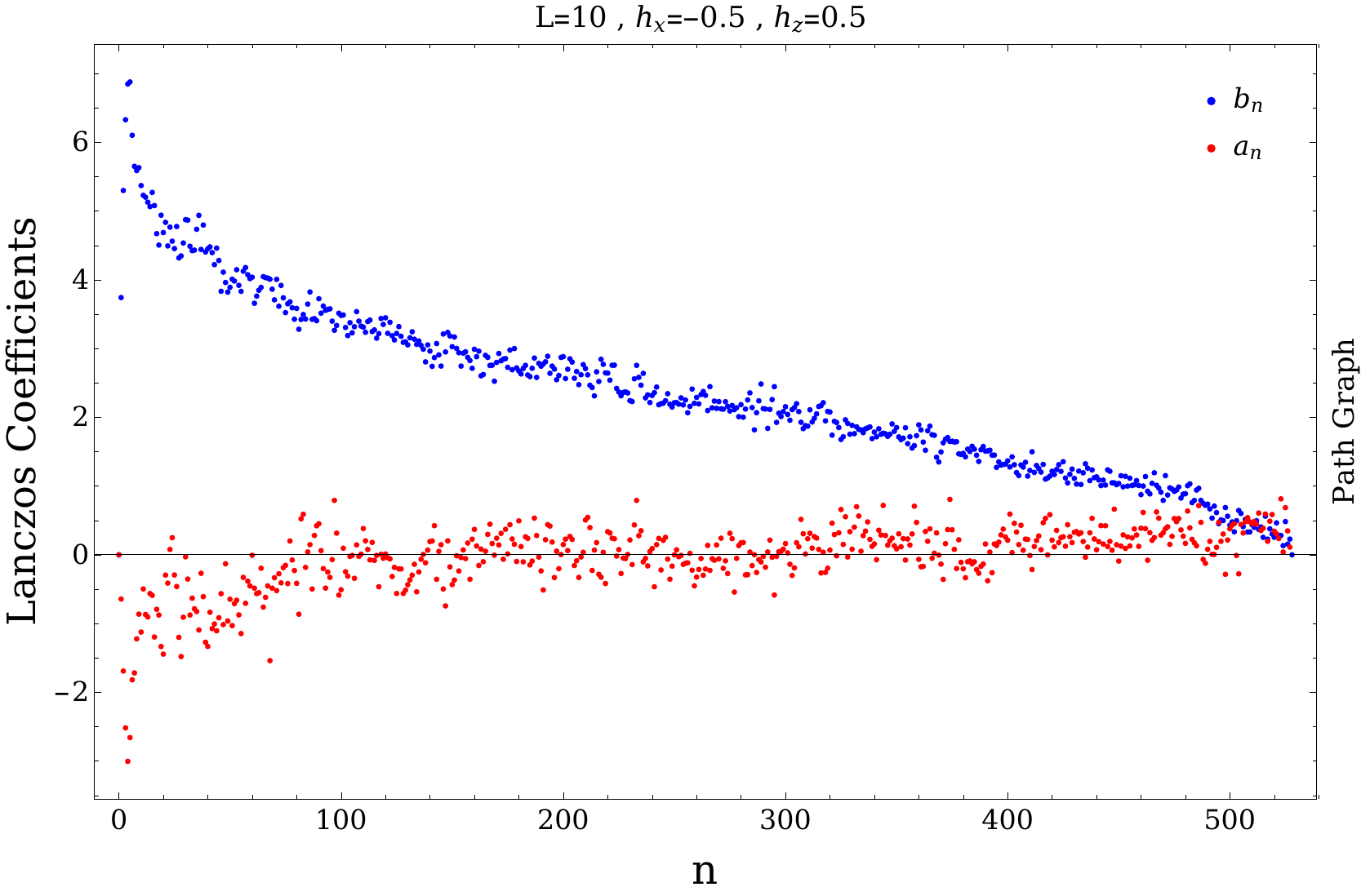}
\includegraphics[width=.495\textwidth,origin=c]{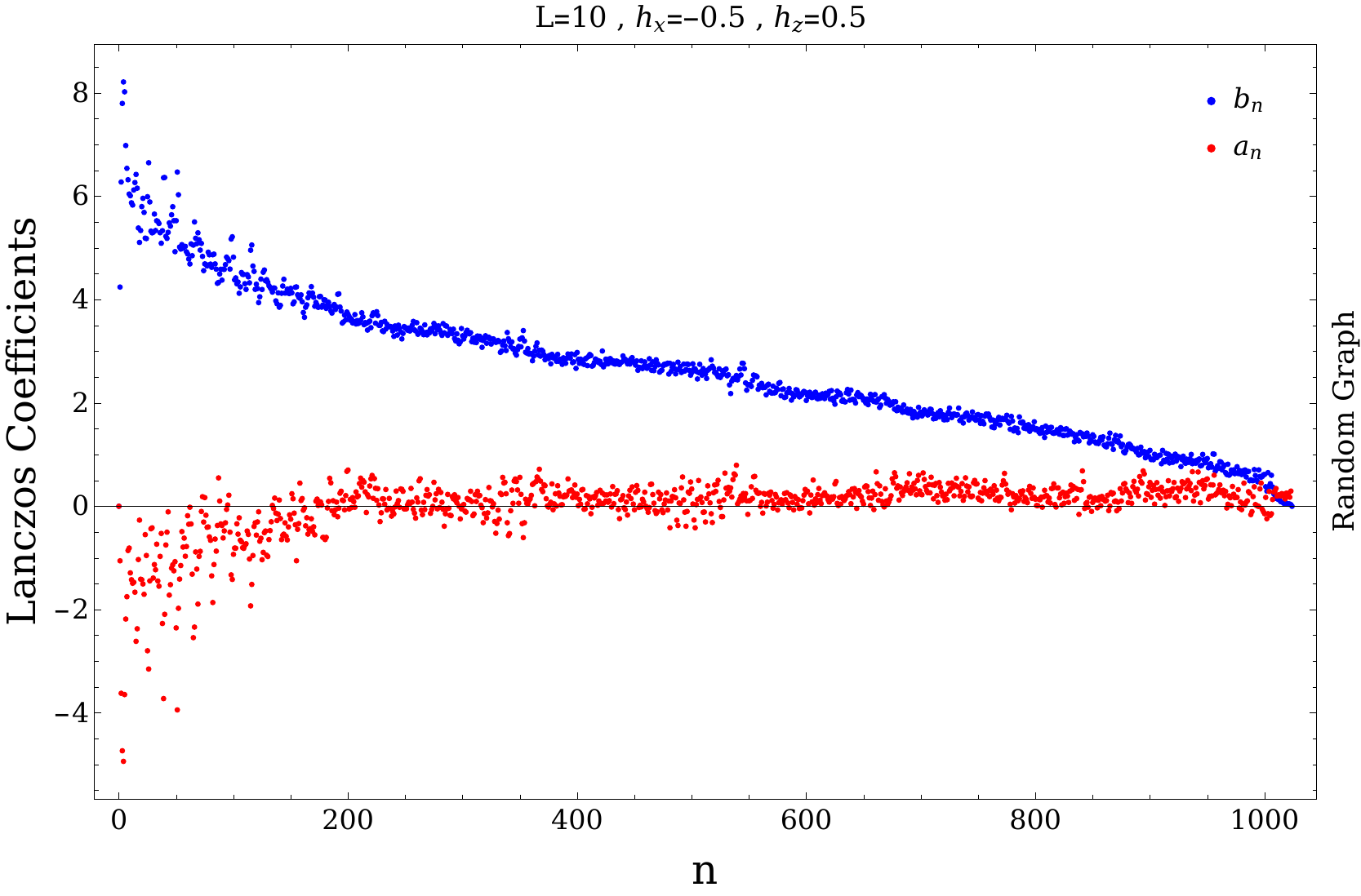}
\caption{Diagonal ($a_{n}$, red) and off-diagonal ($b_{n}$, blue) Lanczos coefficients versus the Krylov index $n$ for the graph-based Ising model with $h_{x}=-0.5$, $h_{z}=0.5$, and $L=10$, initialized in the state $|Y+\rangle$. Panels correspond to the complete (top left), cycle (top right), path (bottom left), and random (bottom right) graphs. The length of the sequence reflects the dimension of the Krylov subspace explored by the time evolution.}
\label{bnanYP0graph}
\end{figure}

The behavior of $K(t)$ is controlled by the Lanczos coefficients, which tridiagonalize the generator of time evolution (the Liouvillian $\hat{L}$ for operators, or $\hat{H}$ for states) in the Krylov basis,
\begin{equation}
    \hat{L}|\mathcal{O}_{n})=a_{n}|\mathcal{O}_{n})+b_{n+1}|\mathcal{O}_{n+1})+b_{n}|\mathcal{O}_{n-1}),\quad n\ge0,\quad b_{0}=0,
\end{equation}
where $a_{n}$ and $b_{n}$ denote the diagonal and off-diagonal coefficients, respectively \cite{Bhattacharya:2023zqt,Bhattacharya:2022gbz,Caputa:2024vrn}. Both sequences extracted from our numerics are displayed in Fig.~\ref{bnanYP0graph}. Within the universal operator growth hypothesis, chaotic dynamics in the thermodynamic limit is signalled by an asymptotically linear growth $b_{n}\simeq\alpha n+\gamma$, which drives an exponential early-time rise of $K(t)$ \cite{Parker:2018yvk}; integrable models, by contrast, exhibit slow or bounded $b_{n}$ and at most polynomial growth with pronounced revivals. In the finite systems studied here the Krylov chain necessarily terminates at the dimension of the accessible Hilbert space, so $b_{n}$ eventually decreases and $K(t)$ crosses over to saturation; the chaotic and integrable regimes are then distinguished by the length of the chain over which $b_{n}$ remains appreciable and by the early-time growth rate of $K(t)$, in line with ETH expectations \cite{Camargo:2024deu,Baggioli:2024wbz,Nahum:2017yvy,Khorasani:2021zus,Vasli:2023syq}.

\begin{figure}[tbp]
\centering
\includegraphics[width=.495\textwidth,origin=c]{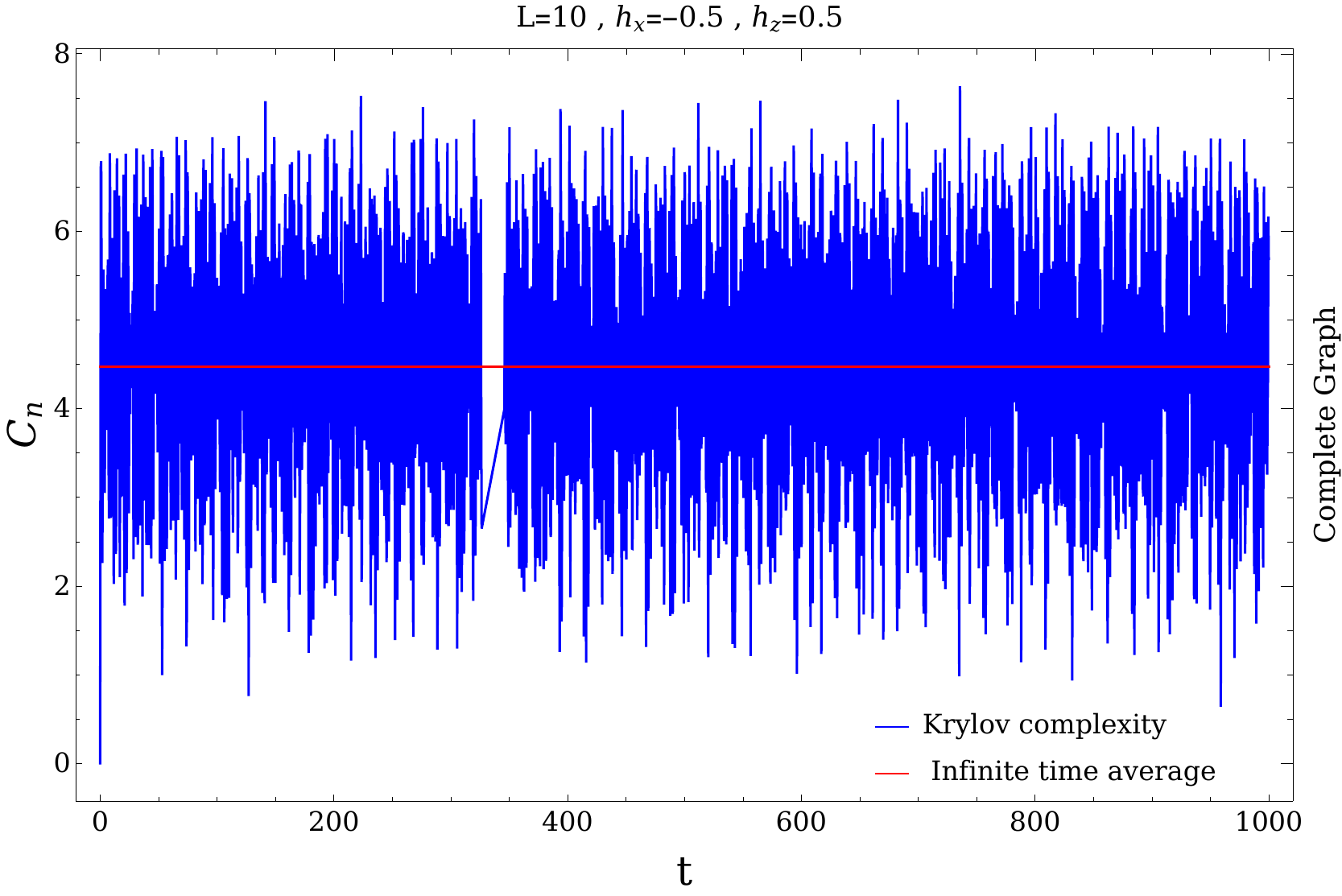}
\includegraphics[width=.495\textwidth,origin=c]{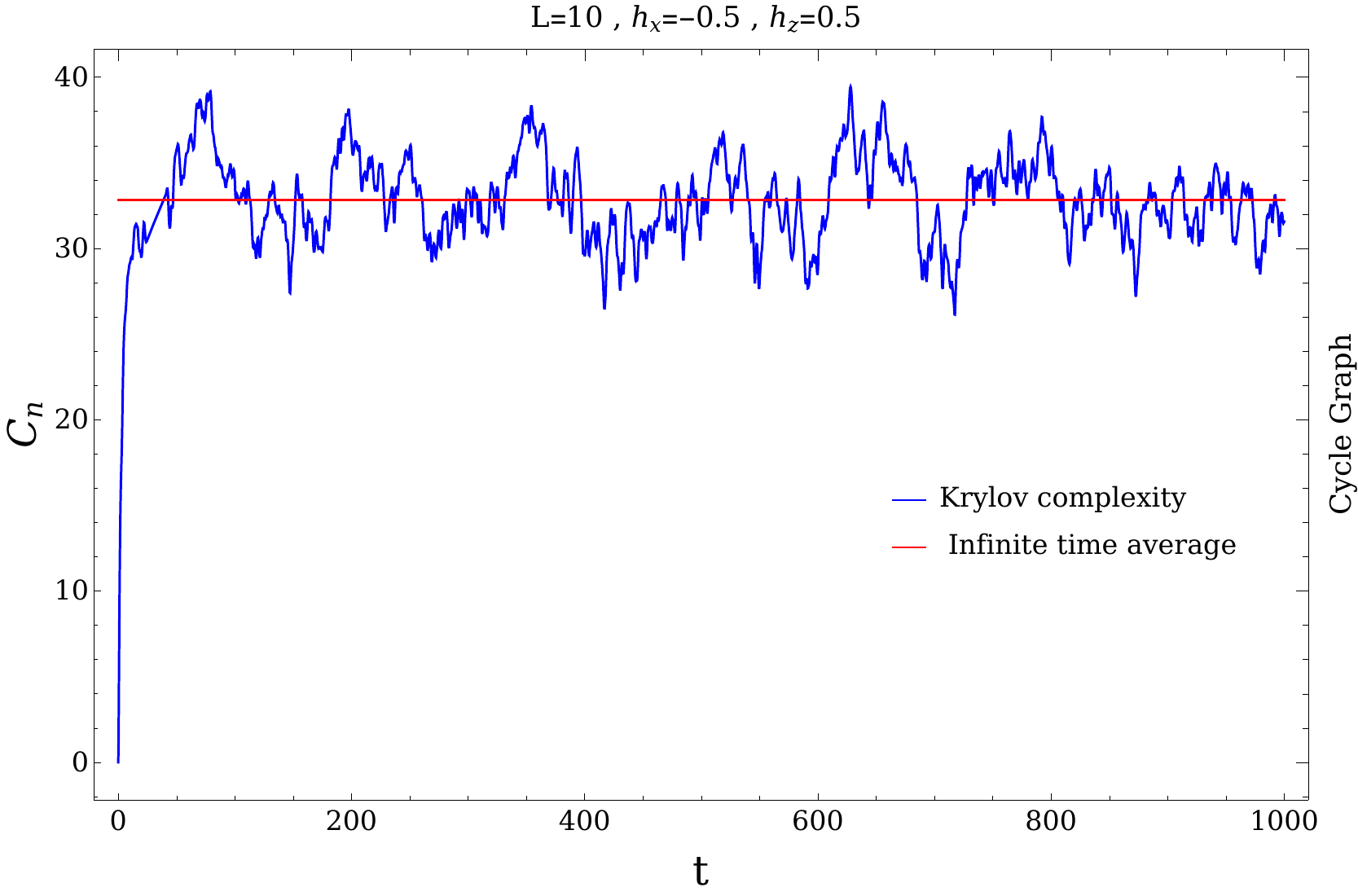}
\includegraphics[width=.495\textwidth,origin=c]{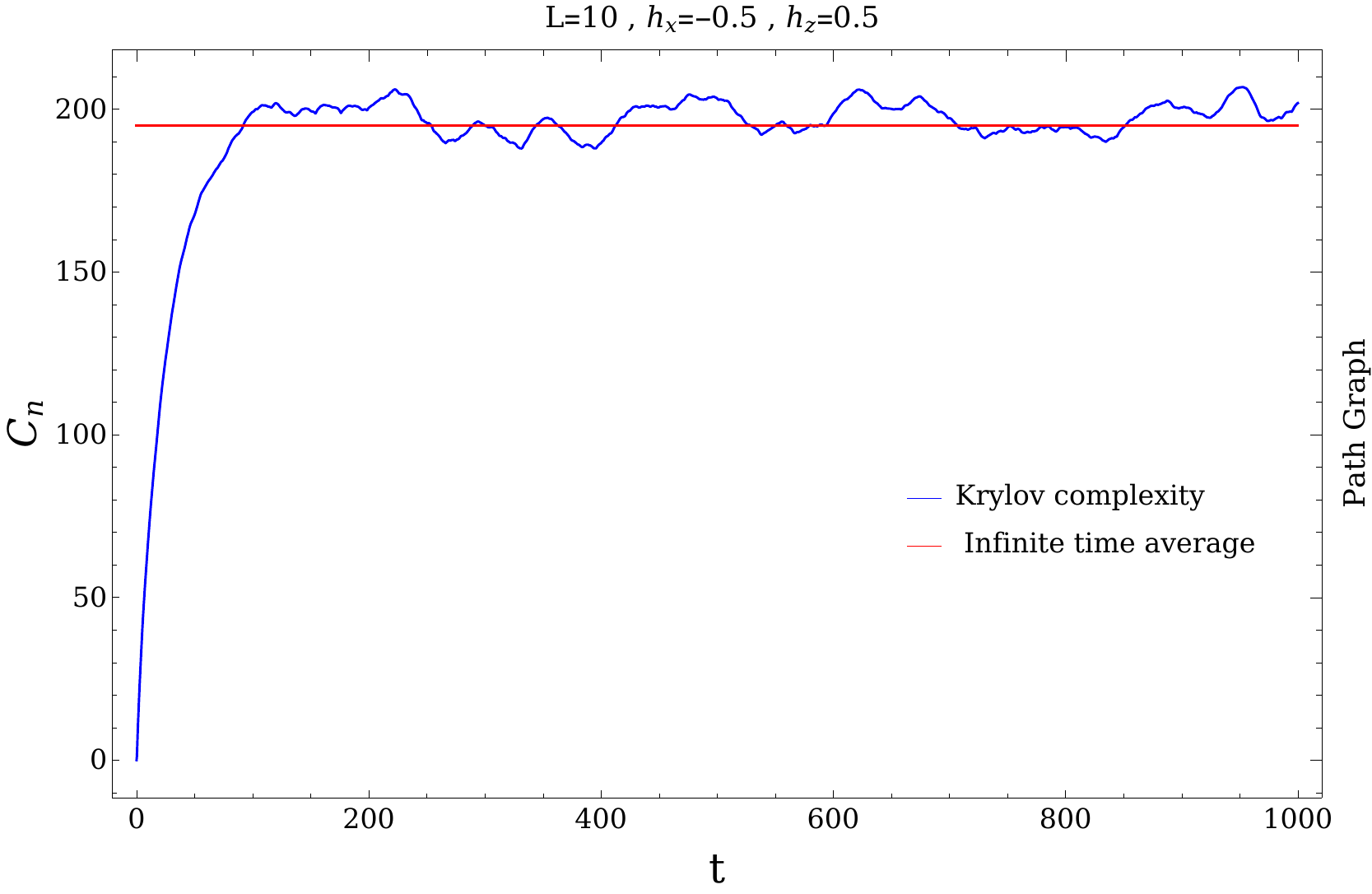}
\includegraphics[width=.495\textwidth,origin=c]{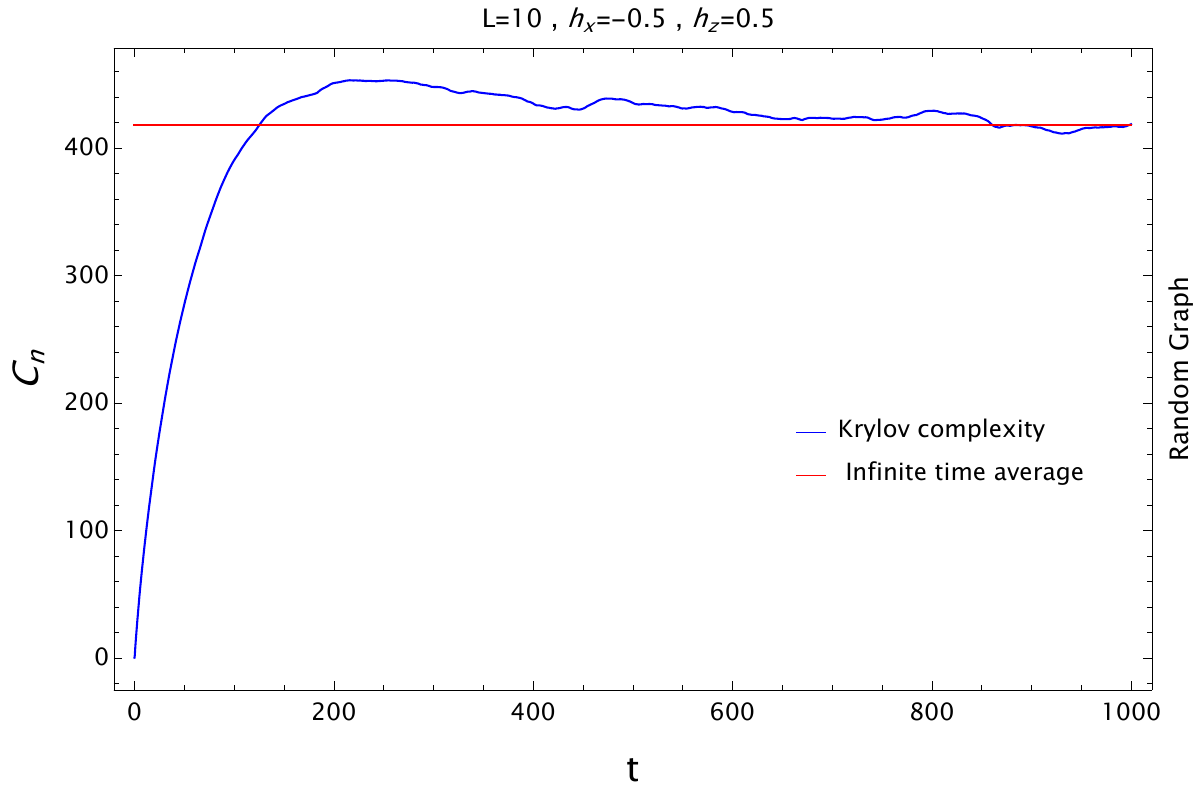}
\caption{Growth of the Krylov (spread) complexity $K(t)$ (plotted as $C_{n}$ in the panels) for the same setups as in Fig.~\ref{bnanYP0graph}. The horizontal red line marks the infinite-time average $\bar{K}=\mathrm{Tr}(\rho_{\mathrm{DE}}\hat{N})$. The random graph (bottom right) exhibits a rapid rise followed by a pronounced overshoot and a slow relaxation toward $\bar{K}$, the signature of chaotic scrambling; the complete graph (top left), in contrast, displays persistent large-amplitude oscillations about the mean with no genuine saturation.}
\label{SPComYP0graph}
\end{figure}

The geometry of the underlying graph leaves a clear imprint on this structure (Figs.~\ref{bnanYP0graph} and \ref{SPComYP0graph}). The length of the Lanczos sequence effectively explored by the dynamics reflects the dimension of the accessible Krylov subspace. On the complete graph the evolution is confined to a parametrically small subspace, $n\sim L$, as expected for an infinite-range model whose large symmetry algebra restricts the available phase space; notably, the diagonal coefficients $a_{n}$ attain large, irregular values in this case, reflecting the highly structured mean-field spectrum. On the random graph, by contrast, the sequence reaches dimensions of order $2^{L}$, with $b_{n}$ decaying only slowly and remaining nonzero across the entire chain: the initial state $|Y+\rangle$ delocalizes over essentially the full Hilbert space, the hallmark of chaotic scrambling. The cycle and path graphs realize an intermediate situation, in which residual lattice symmetries keep the explored subspace below the full dimension while still permitting scrambling-like growth of the correlators.

The connection between Krylov complexity and entanglement is equally direct. For a bipartition of the system into $A$ and $B$, the von Neumann entropy of $A$,
\begin{equation}
    \mathcal{S}_{\mathrm{vN}}(\rho_{A})=-\mathrm{Tr}(\rho_{A}\ln\rho_{A}),
\end{equation}
grows rapidly whenever the state (or a local operator) explores the Hilbert space ergodically, distributing information across the bipartition \cite{Roberts:2014isa,Khemani:2018sdn,Calabrese:2007rg}. In ETH-obeying systems the rise of $K(t)$ and of $\mathcal{S}_{\mathrm{vN}}$ are therefore concomitant, and Krylov complexity complements the OTOC and entropy growth as a thermalization diagnostic.

This picture is reinforced by spectral considerations. Nonintegrable systems with Wigner--Dyson statistics support long Krylov chains with slowly decaying Lanczos coefficients and, consequently, rapid early-time growth of both $K(t)$ and $\mathcal{S}_{\mathrm{vN}}$, whereas integrable systems with Poissonian spectra exhibit short chains and slow, bounded complexity growth \cite{Rakovszky:2017qit,Cotler:2017jue}. Moreover, the Krylov-space probability distribution $|\phi_{n}(t)|^{2}$ is tied to the autocorrelation function of the initial operator (or state),
\begin{equation}
    \phi_{0}(t)=(\mathcal{O}_{0}|\mathcal{O}(t)),
\end{equation}
whose decay provides another proxy for scrambling \cite{Roberts:2016hpo,Khemani:2018sdn}.

Turning to the nonlocal Ising models studied here, the combination of long-range couplings and a longitudinal field enhances $\lambda_{L}$, prolongs the window over which $b_{n}$ remains appreciable, and amplifies both $K(t)$ and $\mathcal{S}_{\mathrm{vN}}$. In our numerics, the early-time growth rate of the Krylov complexity tracks the rate of entanglement production, $\mathrm{d}K/\mathrm{d}t\propto\mathrm{d}\mathcal{S}_{\mathrm{vN}}/\mathrm{d}t$, in the scrambling regime, indicating that $K(t)$ can serve as a quantitative proxy for entanglement growth and thermalization.

Further insight follows from the spectral decomposition of the complexity. Expanding the initial state in the energy eigenbasis, $|\psi_{0}\rangle=\sum_{j}c_{j}|E_{j}\rangle$, one finds
\begin{equation}\label{kcomplexityeq}
    K(t)=\mathrm{Tr}(\rho_{\mathrm{DE}}\hat{N})+\sum_{j\ne k}e^{i\omega_{jk}t}c_{j}^{*}c_{k}N_{jk},
\end{equation}
with $\omega_{jk}=E_{j}-E_{k}$, $N_{jk}=\langle E_{j}|\hat{N}|E_{k}\rangle$, and the diagonal-ensemble density matrix
\begin{equation}
    \rho_{\mathrm{DE}}=\sum_{j}|c_{j}|^{2}|E_{j}\rangle\langle E_{j}|.
\end{equation}
In practice, whenever the Hamiltonian possesses a global symmetry and the initial state lies in a definite charge sector, the Krylov construction and $\rho_{\mathrm{DE}}$ are restricted to that block, which considerably reduces the numerical cost. For chaotic spectra the off-diagonal terms in Eq.~\eqref{kcomplexityeq} dephase and $K(t)$ relaxes to the diagonal-ensemble value; the infinite-time average,
\begin{equation}
    \bar{K}=\lim_{T\to\infty}\frac{1}{T}\int_{0}^{T}\mathrm{d}t\,\langle\hat{N}(t)\rangle=\mathrm{Tr}(\rho_{\mathrm{DE}}\hat{N}),
\end{equation}
then coincides with the saturation value \cite{Alishahiha:2024rwm}. Integrable models, constrained by additional conservation laws, fail to explore the Krylov chain uniformly: their long-time dynamics retains persistent oscillations and, whenever an effective plateau forms, it lies below the diagonal-ensemble value,
\begin{equation}
    K_{s}\approx\mathrm{Tr}(\rho_{\mathrm{DE}}\hat{N})\ \ (\text{chaotic}),\qquad K_{s}<\mathrm{Tr}(\rho_{\mathrm{DE}}\hat{N})\ \ (\text{integrable}).
\end{equation}
Fig.~\ref{SPComYP0graph} illustrates this systematics directly: on the random and path graphs the complexity rises quickly, overshoots $\bar{K}$, and then settles close to it, whereas on the complete and cycle graphs $K(t)$ keeps oscillating around the mean with large amplitude over the entire simulation window.

Complementary diagnostics---the inverse participation ratio in Krylov space, the mean ratio of consecutive level spacings, and the SFF---sharpen this picture further and, combined with $K(t)$ and $\mathcal{S}_{\mathrm{vN}}$, allow a precise discrimination between integrable and chaotic regimes, and between local and nonlocal interactions \cite{Alet:2018,Cotler:2016fpe}.

In summary, Krylov complexity---together with the spectral and entanglement diagnostics discussed above---provides a coherent and quantitatively precise framework for nonequilibrium quantum dynamics. It ties operator scrambling, the ETH, and entanglement spreading into a single picture, and it serves as the principal tool in our comparison of local and nonlocal Ising models.

\section{Spectral diagnostics of chaos, the SFF, and their connection to entanglement growth and operator complexity}\label{sec:Spectral}

The statistical arrangement of the many-body energy levels is one of the sharpest fingerprints of quantum chaos \cite{Guhr:1997ve,Camargo:2023eev}. Consider a closed system whose discrete spectrum $\{E_{n}\}$ is ordered as $E_{1}<E_{2}<\cdots$; the most elementary diagnostic is then the distribution of consecutive spacings $s_{n}=E_{n+1}-E_{n}$. Since the raw spacings inherit the global density of states, the spectrum must first be unfolded so that the local mean spacing equals unity; only the residual fluctuations are universal and can be compared across systems. RMT predicts that a time-reversal-invariant chaotic system exhibits the Wigner--Dyson statistics of the GOE, whose spacing distribution is captured to high accuracy by the Wigner surmise
\begin{equation}
    P_{\mathrm{WD}}(s)=\frac{\pi}{2}\,s\,\exp\!\Big(-\frac{\pi}{4}s^{2}\Big),
\end{equation}
whereas a generic integrable system displays uncorrelated levels and hence Poisson statistics,
\begin{equation}
    P_{\mathrm{P}}(s)=e^{-s}.
\end{equation}
The linear rise of $P_{\mathrm{WD}}(s)$ at small $s$---level repulsion---is the local spectral hallmark of chaos.

Because unfolding is numerically delicate, it is customary to work instead with the ratio of adjacent gaps,
\begin{equation}
    r_{n}=\frac{\min(s_{n},s_{n-1})}{\max(s_{n},s_{n-1})},\qquad \langle r\rangle=\frac{1}{N-2}\sum_{n=2}^{N-1}r_{n},
\end{equation}
a dimensionless quantity that is by construction insensitive to the local spectral density. The ensemble averages, $\langle r\rangle\approx0.386$ for Poisson and $\langle r\rangle\approx0.536$ for the GOE \cite{Atas:2012qv}, make $\langle r\rangle$ a convenient practical chaos indicator. Its drift toward the GOE value also heralds the validity of the ETH, since both properties share a common origin in the absence of extensive conserved quantities.

\begin{figure}[tbp]
\centering
\includegraphics[width=.495\textwidth,origin=c]{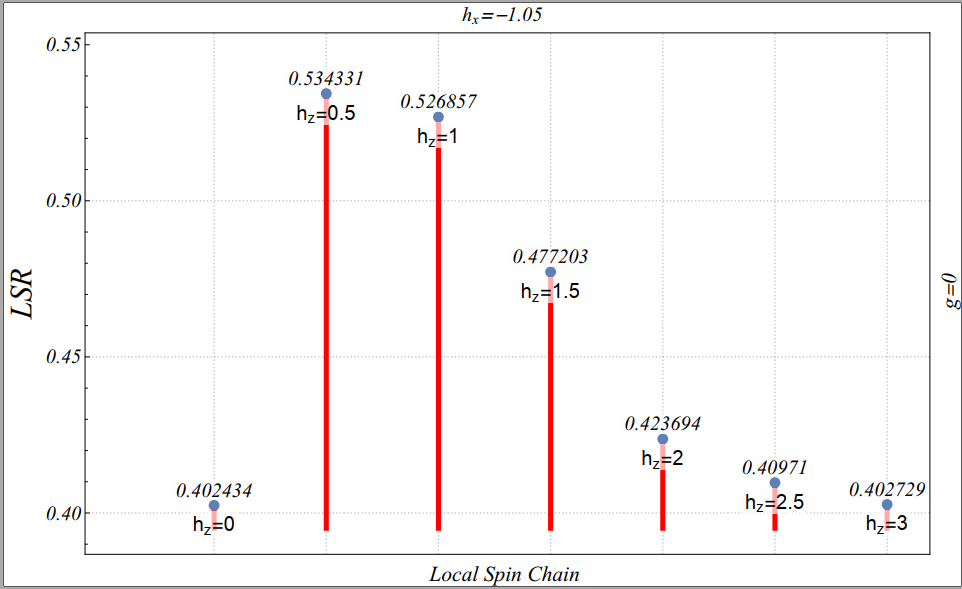}
\includegraphics[width=.495\textwidth,origin=c]{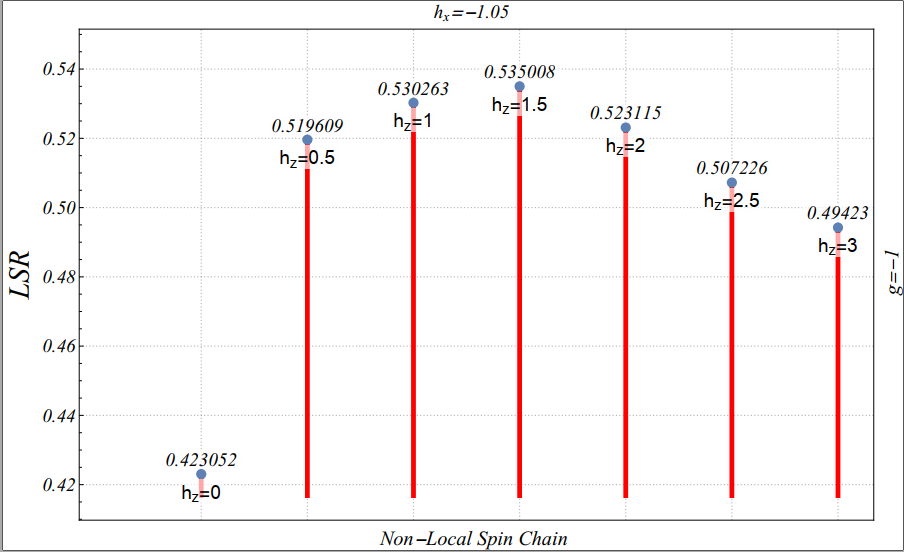}
\includegraphics[width=.495\textwidth,origin=c]{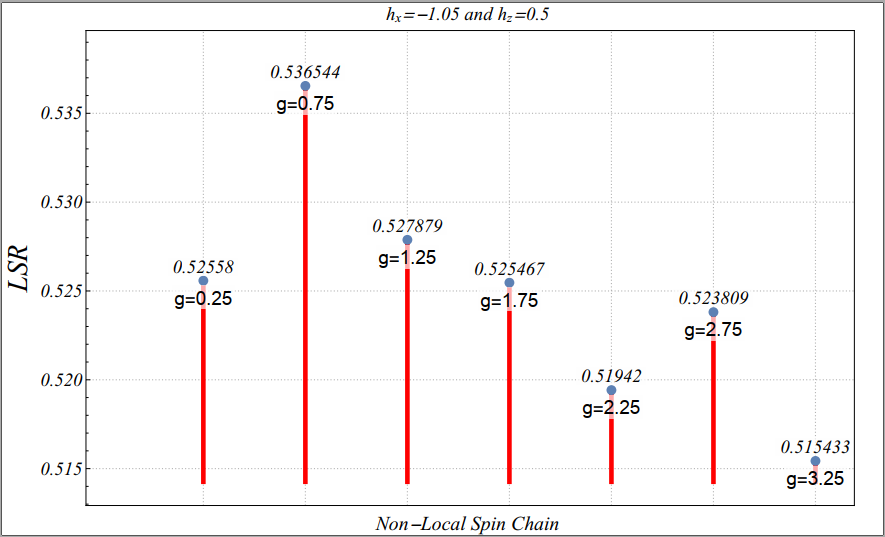}
\caption{Adjacent gap ratios (denoted LSR in the panels) for the Ising model with mixed fields and nonlocal interactions; the numerical value of each ratio is annotated above the corresponding bar. Top left: local chain with $h_{x}=-1.05$ and $g=0$ for $h_{z}=0,\dots,3$; top right: nonlocal chain with $g=-1$ and the same values of $h_{z}$; bottom: nonlocal chain at fixed $h_{z}=0.5$ with $g=0.25,\dots,3.25$. Values close to $0.53$ ($0.39$) indicate GOE (Poisson) statistics.}
\label{LSR}
\end{figure}

In the graph-based Ising models studied here, integrability-breaking perturbations---a longitudinal field or nonlocal couplings---push $\langle r\rangle$ from the Poissonian value toward the GOE one, i.e., they drive a spectral crossover from the integrable to the chaotic regime accompanied by the onset of eigenstate thermalization \cite{Oganesyan:2007,Deutsch:2018ulr}. Fig.~\ref{LSR} quantifies this crossover: for the nonlocal chain, already at $|g|\ge0.25$ the ratios cluster around the GOE value, establishing chaotic dynamics; for the local model with $h_{x}=-1.05$, the case $h_{z}=0.5$ is chaotic ($\langle r\rangle\approx0.53$) whereas $h_{z}=0$ remains integrable ($\langle r\rangle\approx0.40$). These spectral findings are fully consistent with the Krylov-complexity analysis of Sec.~\ref{sec:Krylov}, in which the same parameter regimes exhibit rapid complexity growth and operator scrambling.

\begin{figure}[tbp]
\centering
\includegraphics[width=.495\textwidth,origin=c]{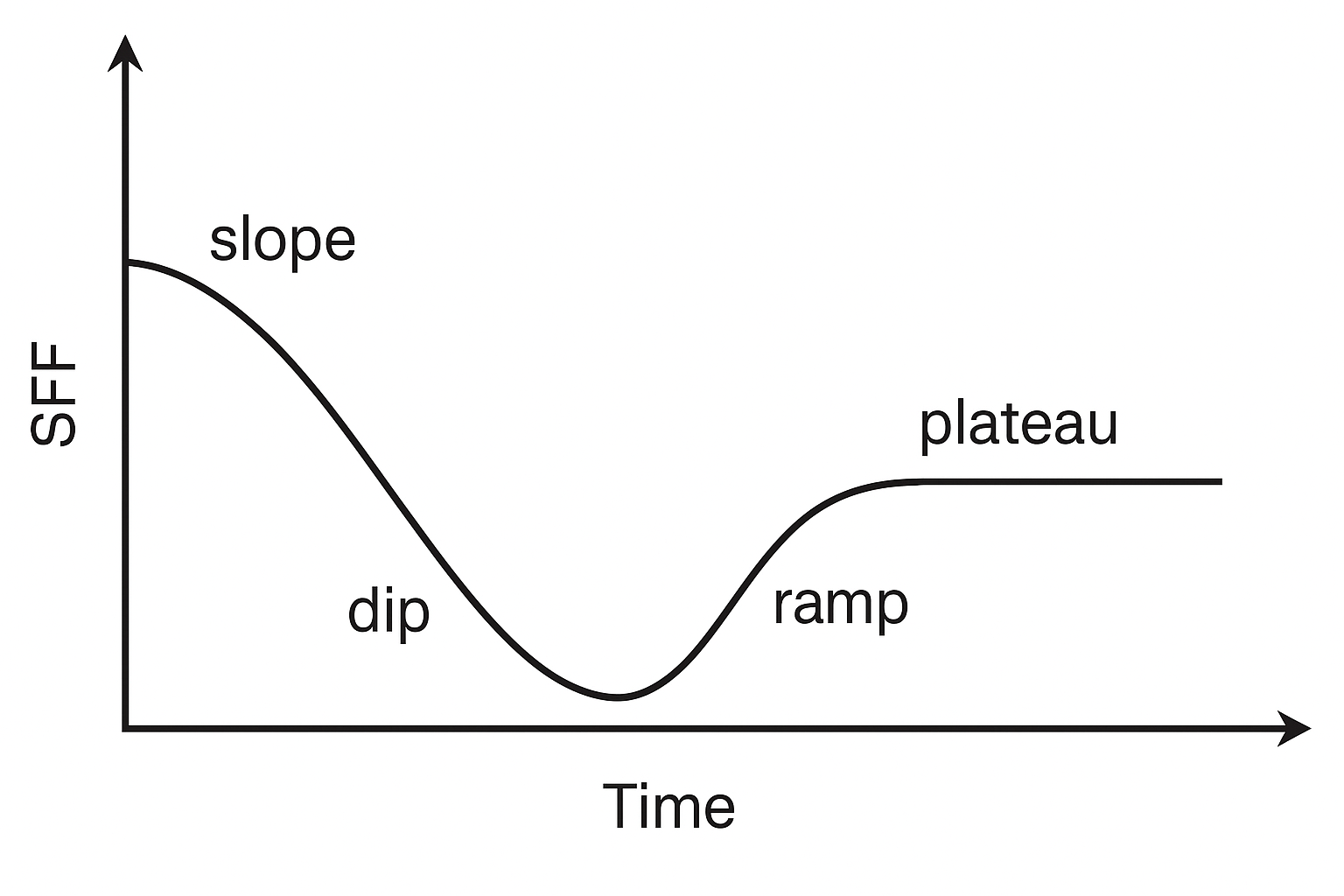}
\caption{Schematic diagram of the SFF for a chaotic system, exhibiting the four characteristic regimes: slope, dip, ramp, and plateau.}
\label{SFF}
\end{figure}

Level-spacing statistics probe only the shortest spectral scales; long-range energy correlations require a dynamical probe, and the SFF is the premier diagnostic of this type. It is defined by
\begin{equation}
    \mathrm{SFF}(\beta,t)=|Z(\beta+it)|^{2}=Z(\beta+it)\,Z(\beta-it),\qquad Z(\beta)=\mathrm{Tr}\,e^{-\beta\hat{H}},
\end{equation}
i.e., as the analytically continued partition function, which weights pairs of energy levels and is therefore sensitive to the global spectral rigidity underlying quantum thermalization. (For a single realization of the Hamiltonian the SFF carries strong sample-to-sample fluctuations; the canonical structure described below refers to its smoothed or ensemble-averaged form.) In a chaotic system the time evolution separates into four regimes:
\begin{enumerate}
    \item \textit{Slope}: the early-time decay, governed by the smooth part of the density of states;
    \item \textit{Dip}: the correlation hole that precedes the ramp and reflects intermediate-range correlations;
    \item \textit{Ramp}: a linear increase generated by the long-range spectral correlations and rigidity characteristic of Wigner--Dyson statistics, and closely tied to the ETH;
    \item \textit{Plateau}: saturation at the Heisenberg time $t_{H}$, the inverse of the mean level spacing, marking the onset of quantum recurrences.
\end{enumerate}
The slope--dip--ramp--plateau sequence is the canonical RMT signature \cite{Cotler:2016fpe,Guhr:1997ve}, sketched in Fig.~\ref{SFF}; in integrable systems, by contrast, the ramp is absent or strongly suppressed and the approach to saturation is qualitatively different, reflecting Poissonian statistics and the lack of complete scrambling.

\begin{figure}[tbp]
\centering
\includegraphics[width=.495\textwidth,origin=c]{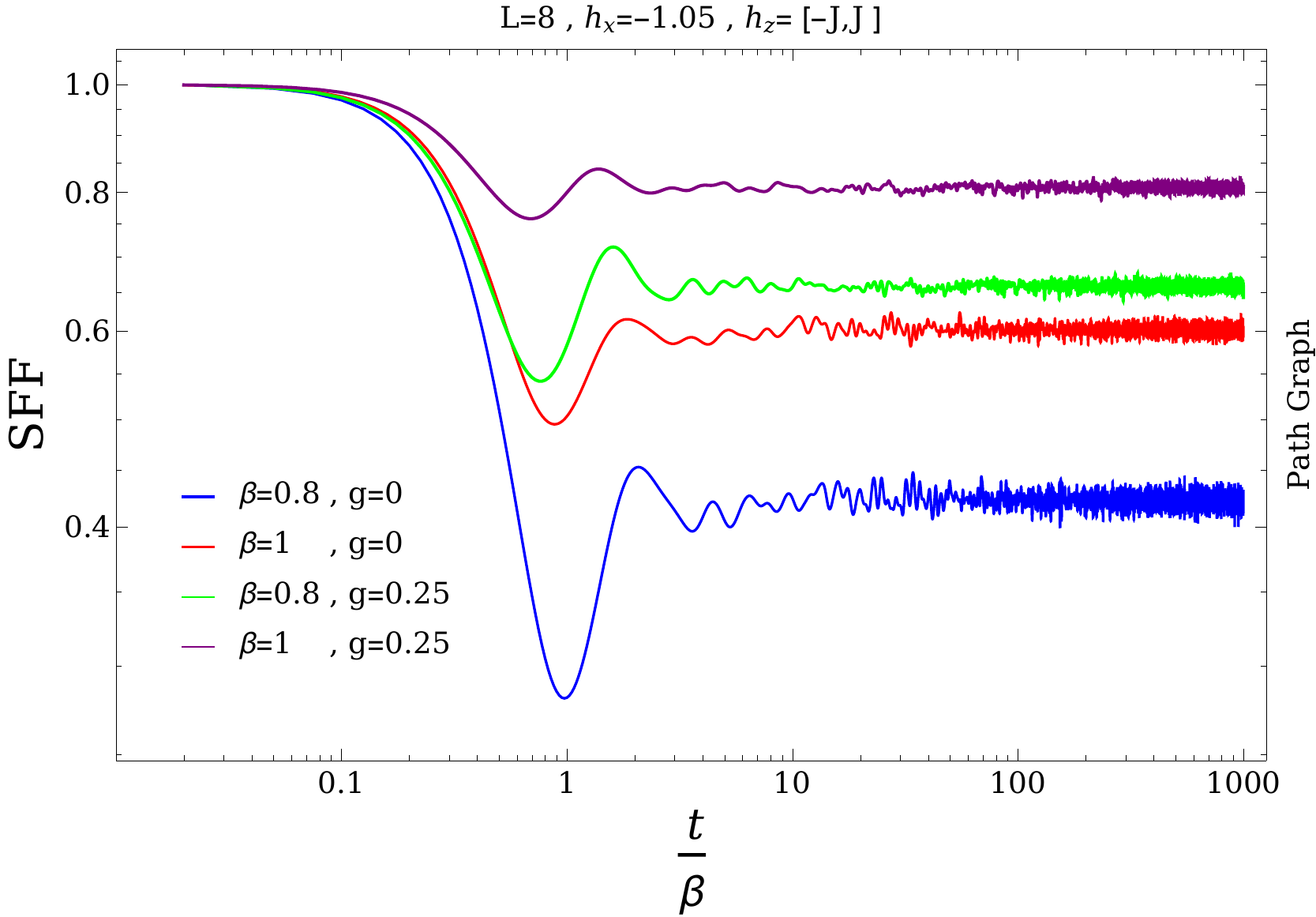}
\includegraphics[width=.495\textwidth,origin=c]{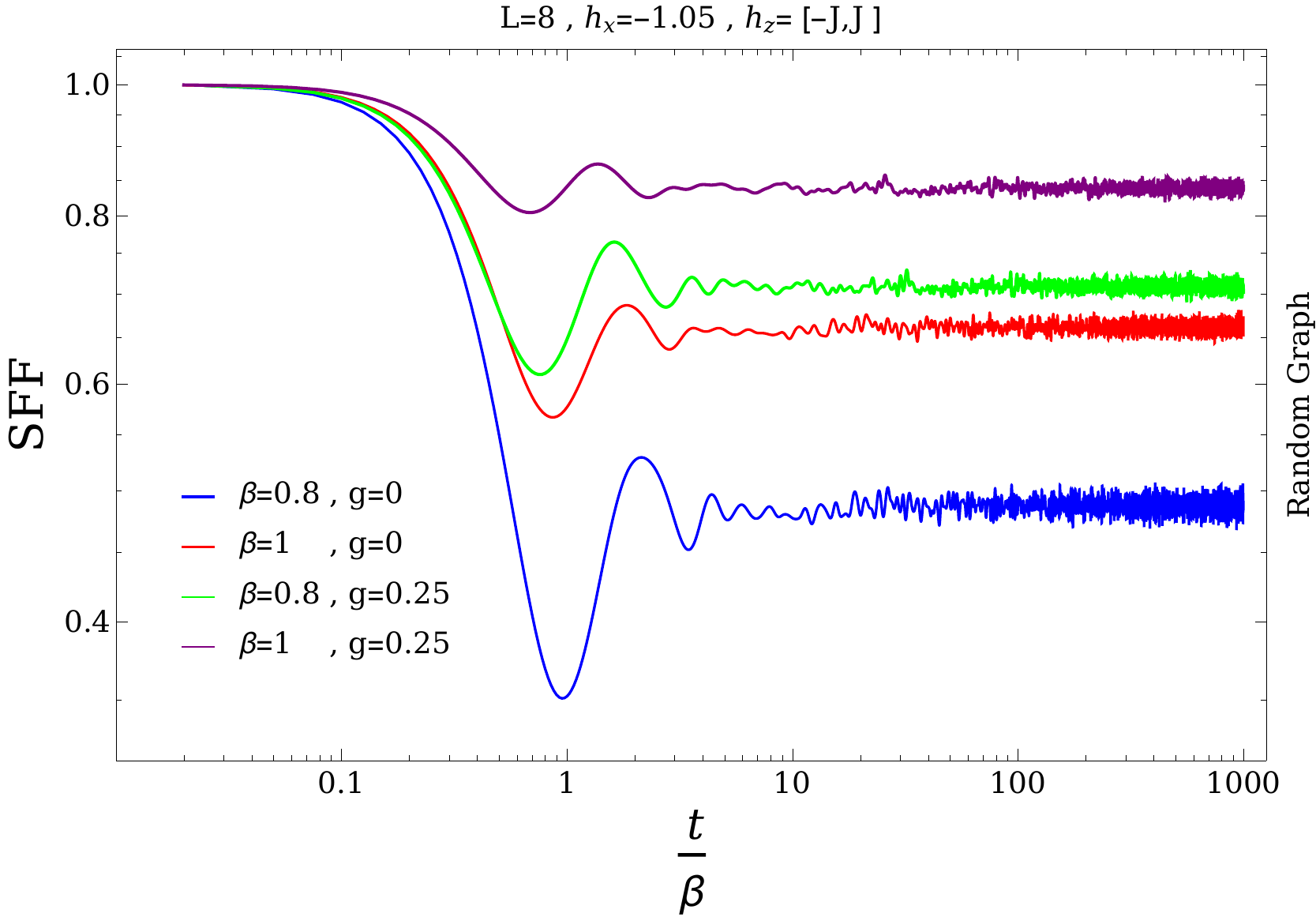}
\caption{SFF of the Ising model with $L=8$ and $h_{x}=-1.05$ on the path graph (left) and the random graph (right), plotted as a function of $t/\beta$ for $\beta=0.8,\,1$ and $g=0,\,0.25$ as indicated in the legends. For $g=0.25$ the curves display a dip followed by a ramp-like rise and saturation, whereas for $g=0$ the dip is deeper and the correlator relaxes to a lower plateau with a much weaker ramp.}
\label{Sffgraph}
\end{figure}

Two spectral time scales organize this structure. The Heisenberg time, $t_{H}\sim2\pi/\overline{\Delta E}$ with $\overline{\Delta E}$ the mean level spacing in the relevant window, sets the time at which the discreteness of the spectrum is resolved and the plateau forms. The Thouless time $t_{\mathrm{Th}}$ \cite{Thouless:1977} marks the onset of universal RMT behavior, i.e., the beginning of the ramp, and in the ETH context is often identified with the thermalization time. In local systems $t_{\mathrm{Th}}$ is tied to transport and operator spreading: it grows polynomially with system size while remaining parametrically below $t_{H}$. In nonlocal systems with rapid entanglement growth (fast scramblers) it can instead grow only logarithmically with the number of degrees of freedom---double-logarithmically in the Hilbert-space dimension---so that the ramp and plateau are reached appreciably earlier \cite{Hayden:2007cs,Cotler:2016fpe}. Fig.~\ref{Sffgraph} illustrates this behavior for the Ising model on the path and random graphs: at $g=0.25$ the curves exhibit a dip followed by a ramp-like rise toward saturation, whereas at $g=0$ the dip is deeper and the system relaxes to a lower plateau with only a weak ramp.

\begin{figure}[tbp]
\centering
\includegraphics[width=.495\textwidth,origin=c]{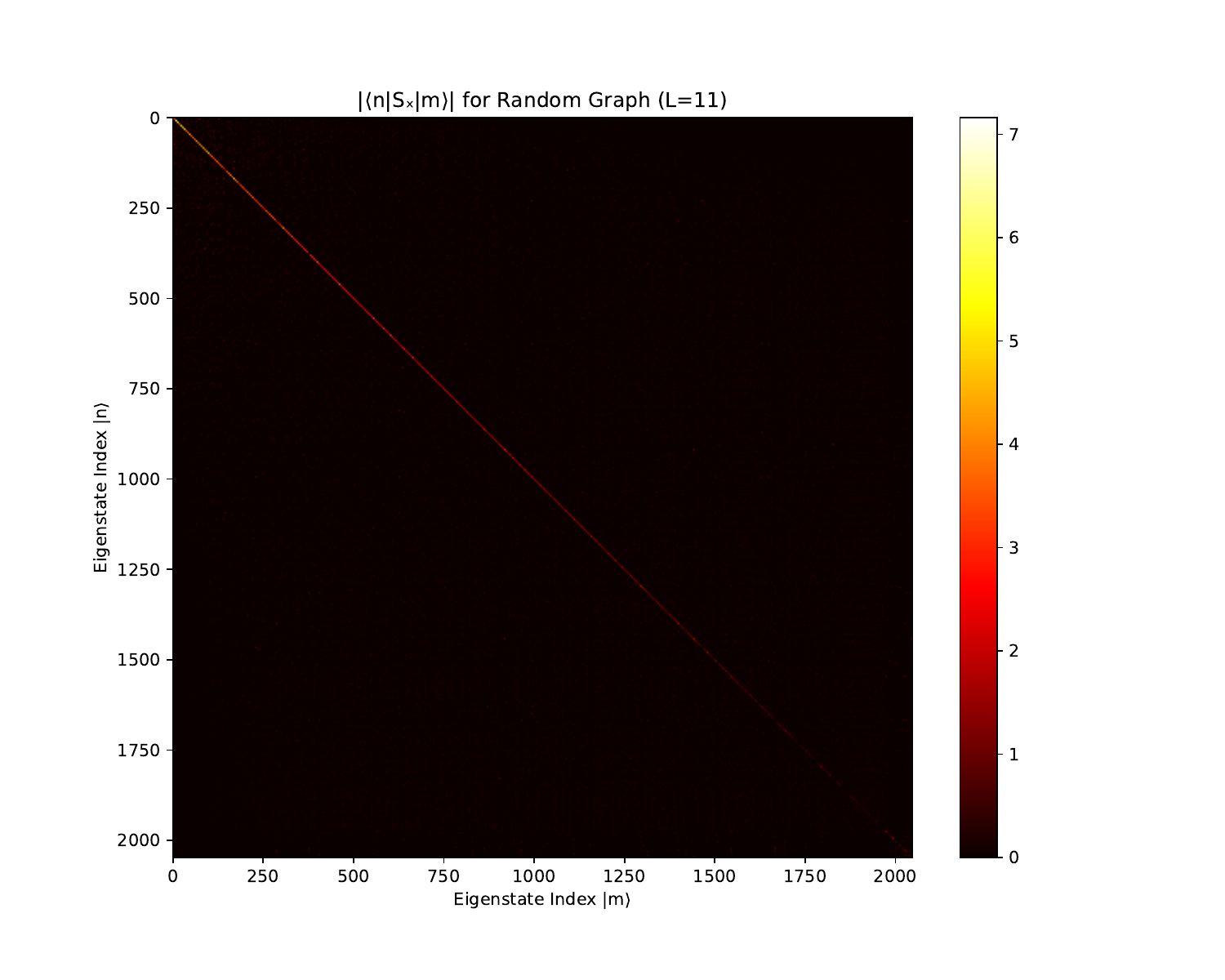}
\includegraphics[width=.495\textwidth,origin=c]{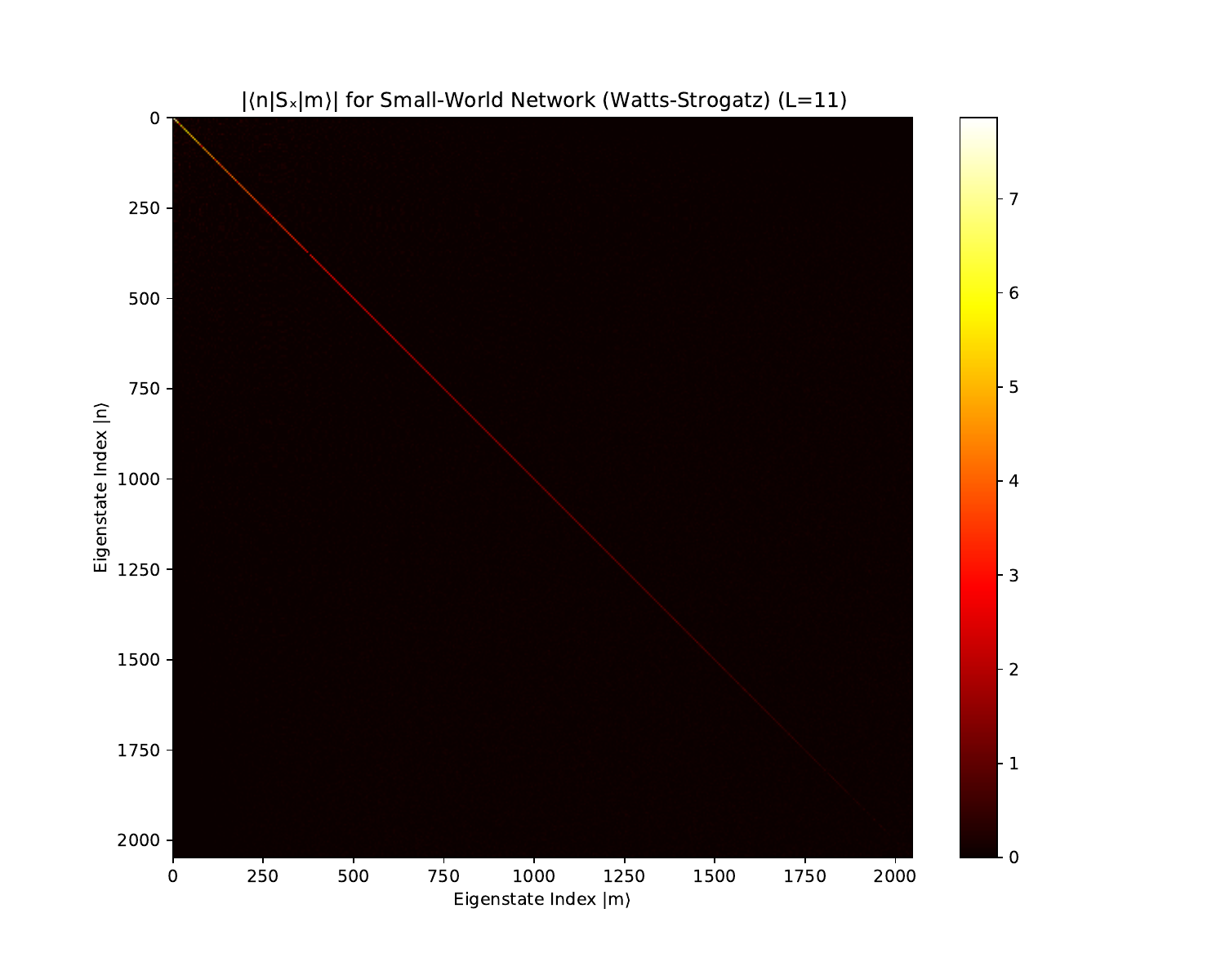}
\caption{Matrix elements $\langle m|S_{x}|n\rangle$ of the spin operator $S_{x}$ in the energy eigenbasis $\{|n\rangle\}$ of the Ising Hamiltonian, plotted against the eigenstate indices $m$ and $n$. Left: random graph; right: Watts--Strogatz small-world network; parameters $h_{x}=0.5$, $h_{z}=1$, $L=11$. The smooth main diagonal realizes the ETH ansatz $\bar{O}(E_{n})$, while the exponentially suppressed off-diagonal elements (dark background) are consistent with the scaling $e^{-S/2}f(E,\omega)R_{mn}$. The nearly identical structure in both topologies shows that ETH is satisfied independently of the graph structure.}
\label{heatmap}
\end{figure}

We now make the connection between spectral measures, entanglement growth, and operator complexity explicit.
\begin{itemize}
    \item[(A)] The ramp and plateau of the SFF indicate that the spectrum is effectively randomized in the RMT sense (``spectral scrambling''). This regime typically coincides with rapid entanglement growth: whenever $t_{\mathrm{Th}}$ is short, the von Neumann entropy $\mathcal{S}_{\mathrm{vN}}$ rises quickly and saturates close to its thermal (Page) value, so $t_{\mathrm{Th}}$ may be regarded as the onset time of effective thermalization and of rapid entropy production \cite{Garrison:2015lva}.
    \item[(B)] Empirically, the entropy saturation time $t_{\mathrm{sat}}$ tracks $t_{\mathrm{Th}}$ while being bounded from above by $t_{H}$: for $t_{\mathrm{Th}}\ll t_{H}$ one finds $t_{\mathrm{sat}}$ close to (or slightly above) $t_{\mathrm{Th}}$, whereas near integrability $t_{\mathrm{Th}}$ becomes large or ill defined and $t_{\mathrm{sat}}$ grows accordingly \cite{Cotler:2016fpe,Martin:2008}. Physically, complete operator spreading and the uniform redistribution of information among subsystems require sufficiently strong long-range spectral correlations.
    \item[(C)] The entanglement spectrum provides a complementary mirror of the dynamics: the eigenvalue spectrum of $\rho_{A}$ not only fixes the entropy values, but its own level-spacing statistics diagnose whether the subsystem is effectively thermalized; Wigner--Dyson statistics of the reduced spectrum are consistent with a developed SFF ramp and exponential OTOC growth \cite{Li:2008,Peschel:2011mcx,Pirmoradian:2025nxw}.
    \item[(D)] Underlying all of these relations is the ETH itself: although the evolution is deterministic and unitary, nonintegrable systems thermalize without any external bath, because individual energy eigenstates reproduce thermal expectation values governed by smooth functions of energy---in marked contrast to classical statistical mechanics, where thermalization requires ergodic exploration of the accessible microstates \cite{Deutsch:1991,Srednicki:1994mfb,DAlessio:2015qtq,Deutsch:2018ulr}.
\end{itemize}
Mathematically, the matrix elements of an observable $\hat{O}$ in the energy eigenbasis take the form
\begin{equation}
    O_{mn}=\langle m|\hat{O}|n\rangle=\bar{O}(E)\,\delta_{mn}+e^{-S(E)/2}f(E,\omega)R_{mn},
\end{equation}
where $\bar{O}(E)$ is the thermal expectation value at energy $E$, $S(E)$ the corresponding thermodynamic entropy, $f(E,\omega)$ a smooth function of the mean energy $E=(E_{m}+E_{n})/2$ and the difference $\omega=E_{m}-E_{n}$, and $R_{mn}$ pseudo-random variables of zero mean and unit variance obeying $R_{mn}=R_{nm}^{*}$ by Hermiticity. Fig.~\ref{heatmap} verifies this structure directly for the spin operator $S_{x}$ in the eigenbasis of the Ising Hamiltonian on two network topologies (random graph and Watts--Strogatz small-world network; $h_{x}=0.5$, $h_{z}=1$, $L=11$): the diagonal entries (the bright main diagonal) vary smoothly with the eigenstate index, as required of $\bar{O}(E)$, while the off-diagonal entries are exponentially suppressed and appear as a dark background. The near-identical patterns on the two topologies indicate that the ETH holds independently of the underlying graph structure.

The suppression factor $e^{-S(E)/2}$ guarantees that off-diagonal matrix elements vanish exponentially with system size, so that long-time averages of observables converge to thermal values despite the unitary evolution. In integrable systems, by contrast, the extensive family of conserved quantities endows the eigenstates with an ordered, non-chaotic structure and thermalization to the Gibbs ensemble fails; relaxation instead proceeds to a GGE \cite{Rigol:2016itf,Ilievski:2015jhc}. A distinct route to ETH violation occurs in disordered systems exhibiting many-body localization (MBL), where ergodicity is broken by localization rather than by conserved quantities \cite{Nandkishore:2014kca,Abanin:2017}.

The ETH ansatz also makes explicit that the off-diagonal components---the very terms responsible for information spreading and OTOC growth---are controlled by the spectral function $f_{O}(\bar{E},\omega)$ and hence by the spectral structure of the Hamiltonian; conversely, the same spectral correlations that produce the ramp bound dynamical rates such as the Lyapunov exponent $\lambda_{L}$ \cite{Maldacena:2015waa} and thereby constrain the growth of entropy and Krylov complexity \cite{Deutsch:2018ulr,Cotler:2016fpe}. Together, these relations close the conceptual loop connecting the SFF, the OTOC, and entropy growth.

Finally, we emphasize that all of the foregoing requires numerical care in finite systems: the ramp may not fully develop, and spectral time scales can be distorted by boundary effects and statistical fluctuations. Any comparison with RMT must therefore be accompanied by a finite-size scaling analysis before results are extrapolated to the thermodynamic limit \cite{Beugeling:2014lnz,Smith:2015dpe}.

In summary, spectral measures---from level-spacing statistics and the $r$-ratio to the SFF and multi-level correlation functions---carry substantial diagnostic and predictive power for nonequilibrium dynamics. They not only discriminate integrable from chaotic regimes but also indicate the time scales of entanglement spreading, scrambling onset, and operator-complexity growth, and, through the ETH, tie the energy spectrum, operator matrix elements, and thermalization into a single testable framework for the local and nonlocal Ising models studied here.

\section{Conclusion}\label{sec:Conclusion}

In this work, we have established a comprehensive framework for investigating the crossover from integrability to quantum chaos by extending the local quantum Ising model with nonlocal interactions and analyzing its dynamics across diverse graph topologies. Our numerical and spectral analyses demonstrate that strengthening the nonlocal couplings $g$ and applying a longitudinal field $h_{z}$ drive a sharp transition in the system's spectral properties, shifting the level-spacing statistics from a Poissonian to a Wigner--Dyson distribution. This spectral crossover is accompanied by an accelerated growth of the von Neumann entropy and a corresponding reduction in saturation times, firmly establishing the onset of eigenstate thermalization in accordance with the ETH.

Complementing the spectral perspective, information-theoretic diagnostics reveal that bipartite and tripartite mutual information (TMI) serve as robust markers of the underlying dynamical regime. In particular, the onset of significantly negative values of the TMI diagnoses the highly delocalized, irretrievable spreading of information characteristic of fast scrambling, whereas near-zero values indicate residual integrability. Consistent with these measures, the OTOCs exhibit an initial exponential decay governed by a quantum Lyapunov exponent $\lambda_{L}$, which is systematically enhanced by increasing $g$ and $h_{z}$.

From an operator-growth perspective, Krylov complexity and the associated Lanczos coefficients provide a microscopic mechanism for operator spreading. The linear growth of the Lanczos coefficients $b_n$ with the Krylov index---a hallmark of chaotic dynamics---directly drives the rapid delocalization of the operator wavefunction and the consequent exponential early-time rise in Krylov complexity. The tight quantitative link between the Lanczos structure and operator growth further corroborates the deep connection between operator spreading and quantum thermalization. Furthermore, the SFF and the adjacent-level spacing ratio $\langle r \rangle$ provide independent, yet fully consistent, corroboration of the chaotic transition. The emergence of the canonical slope--dip--ramp--plateau sequence in the SFF, alongside the shift of $\langle r \rangle$ toward the GOE value, unequivocally signals the buildup of long-range spectral rigidity inherent to the chaotic phase.

Ultimately, our results highlight that quantum chaos in these systems is not merely a spectral artifact, but a multifaceted dynamical process emerging from the synergistic interplay of local and nonlocal correlations, interaction topology, and operator dynamics. The convergence of entropy time scales, the Lyapunov exponent, and Lanczos growth patterns within a single chaotic regime offers a comprehensive diagnostic profile that is far more robust than any single indicator. Moreover, our findings underscore the role of network topology as a structural encoder that can steer the system's dynamics toward either integrability or chaos. This tunability has direct implications for the design of controllable quantum systems and points to viable experimental pathways on platforms such as Rydberg-atom arrays, trapped-ion simulators, and programmable photonic networks.

\subsection*{Future Research Directions}
To advance this research frontier, several pivotal directions are proposed. The primary avenue involves developing analytical frameworks that elucidate the role of graph topology in quantum chaos through rigorous derivations. It is essential to establish quantitative relations between structural graph features—such as spectral dimension and connectivity—and chaotic indicators, including the Lyapunov exponent and the tripartite mutual information. Such a formulation would enable the prediction of chaos transition regimes based solely on topological properties, bypassing computationally expensive numerical simulations.\\
The second avenue entails extending the numerical analysis to larger scales. Simulations must be performed for larger system sizes and a broader variety of graph topologies to disentangle finite-size effects from the intrinsic characteristics of chaos. Finite-size scaling analysis should be employed to identify size-independent critical points and delineate precise boundaries between chaotic and integrable regimes in the thermodynamic limit. Moreover, a systematic comparison among regular, random, small-world, and scale-free graphs is required to construct a comprehensive atlas of dynamical behavior as a function of topology.\\
The third avenue concerns the investigation of disorder effects within the topology. The introduction of disorder or inhomogeneity in the bonds or fields can shift the boundaries of chaos and may lead to the emergence of phases analogous to MBL. This line of inquiry can elucidate how topological disorder modifies the balance between integrability and chaos, and specifically, how it impacts the validity of the ETH and the propagation of quantum information.\\
On the other hand, the fourth direction focuses on experimental realization and validation. The design of protocols for implementing specific topologies in experimental platforms—such as the simulation of small-world graphs in Rydberg-atom arrays or the construction of photonic networks with scale-free architectures—would enable the direct observation of the effect of topology on chaos and thermalization dynamics. Such experiments, in addition to providing theoretical validation, would pave the way for the practical exploitation of these phenomena in quantum technologies.\\
The fifth avenue concerns methodological improvement. Lanczos computations and Krylov complexity measures should be standardized to ensure numerical stability and accuracy, particularly in the presence of finite-size effects. The development of error-resilient algorithms and the release of reference codes and datasets will assist the community in achieving reproducible findings. Furthermore, machine-learning tools may be employed to automate the classification of topologies according to their degree of chaos and adherence to the ETH. Such tools would accelerate the analysis pipeline and yield initial predictions that can guide future research directions.\\
Finally, the sixth avenue should be devoted to interdisciplinary applications. The results of this study can be directly employed in the design of quantum information–processing systems. Regular graphs may serve as platforms for enhanced control and chaos suppression, whereas complex and heterogeneous graphs can be exploited to enhance security and to render information effectively irreversible via scrambling in quantum cryptographic protocols. Furthermore, connecting these findings to holographic theories and Lyapunov bounds may open a new chapter in linking networked quantum systems with gravitational–holographic descriptions.\\
Overall, the continuation of this research program should pursue three concurrent thrusts: (1) deepening theoretical and numerical analyses to elucidate the relationship between topology, chaos, and thermalization; (2) designing and implementing experimental protocols for the direct observation of these phenomena; and (3) exploiting the obtained results in practical architectures for quantum information processing and quantum security. The achievement of these objectives will advance our understanding of quantum chaos from a conceptual level to a practical and engineering one.\\
\subsection*{Acknowledgments}
We gratefully acknowledge Mohsen Alishahiha and M.Javad Vasli for their discussions. The work of  Reza Pirmoradian is based on research funded by the Iran National Science Foundation
(INSF) under Project  No. 4026389. Lastly, we recognize the use of AI tools in assisting with text editing and refinement.


\begin{thebibliography}{99}

\bibitem{PhysRev.65.117}
L.~Onsager,
``Crystal Statistics. I. A Two-Dimensional Model with an Order-Disorder Transition,''
Phys. Rev. \textbf{65}, no.~3--4, 117--149 (1944)
doi:10.1103/PhysRev.65.117

\bibitem{Ising:1925em}
E.~Ising,
``Contribution to the Theory of Ferromagnetism,''
Z. Phys. \textbf{31}, 253-258 (1925)
doi:10.1007/BF02980577

\bibitem{Bethe:1931hc}
H.~Bethe,
``On the theory of metals. 1. Eigenvalues and eigenfunctions for the linear atomic chain,''
Z. Phys. \textbf{71}, 205-226 (1931)
doi:10.1007/BF01341708

\bibitem{Jordan:1928wi}
P.~Jordan and E.~P.~Wigner,
``About the Pauli exclusion principle,''
Z. Phys. \textbf{47}, 631-651 (1928)
doi:10.1007/BF01331938

\bibitem{Eisert:2014jea}
J.~Eisert, M.~Friesdorf and C.~Gogolin,
``Quantum many-body systems out of equilibrium,''
Nature Phys. \textbf{11}, 124 (2015)
doi:10.1038/nphys3215
[arXiv:1408.5148 [quant-ph]].

\bibitem{Calabrese:2006rx}
P.~Calabrese and J.~L.~Cardy,
``Time-dependence of correlation functions following a quantum quench,''
Phys. Rev. Lett. \textbf{96}, 136801 (2006)
doi:10.1103/PhysRevLett.96.136801
[arXiv:cond-mat/0601225 [cond-mat.stat-mech]].

\bibitem{DAlessio:2015qtq}
L.~D'Alessio, Y.~Kafri, A.~Polkovnikov and M.~Rigol,
``From quantum chaos and eigenstate thermalization to statistical mechanics and thermodynamics,''
Adv. Phys. \textbf{65}, no.3, 239-362 (2016)
doi:10.1080/00018732.2016.1198134
[arXiv:1509.06411 [cond-mat.stat-mech]].

\bibitem{RevModPhys.83.863}
A.~Polkovnikov, K.~Sengupta, A.~Silva and M.~Vengalattore,
``Colloquium: Nonequilibrium dynamics of closed interacting quantum systems,''
Rev. Mod. Phys. \textbf{83}, no.~3, 863--883 (2011)
doi:10.1103/RevModPhys.83.863

\bibitem{Deutsch:1991}
J.~M.~Deutsch,
``Quantum statistical mechanics in a closed system,''
Phys. Rev. A \textbf{43}, no.~4, 2046--2049 (1991)
doi:10.1103/PhysRevA.43.2046

\bibitem{Srednicki:1994mfb}
M.~Srednicki,
``Chaos and Quantum Thermalization,''
Phys. Rev. E \textbf{50}, 888
doi:10.1103/PhysRevE.50.888
[arXiv:cond-mat/9403051 [cond-mat]].

\bibitem{Calabrese:2007rg}
P.~Calabrese and J.~Cardy,
``Quantum Quenches in Extended Systems,''
J. Stat. Mech. \textbf{0706}, P06008 (2007)
doi:10.1088/1742-5468/2007/06/P06008
[arXiv:0704.1880 [cond-mat.stat-mech]].

\bibitem{Rigol:2008}
M.~Rigol, V.~Dunjko and M.~Olshanii,
``Thermalization and its mechanism for generic isolated quantum systems,''
Nature \textbf{452}, no.~7189, 854--858 (2008)
doi:10.1038/nature06838
[arXiv:0708.1324 [cond-mat.stat-mech]].

\bibitem{Rigol:2009}
M.~Rigol,
``Breakdown of Thermalization in Finite One-Dimensional Systems,''
Phys. Rev. Lett. \textbf{103}, no.~10, 100403 (2009)
doi:10.1103/PhysRevLett.103.100403
[arXiv:0904.3746 [cond-mat.stat-mech]].

\bibitem{Blatt:2012chk}
R.~Blatt and C.~F.~Roos,
``Quantum simulations with trapped ions,''
Nature Phys. \textbf{8}, 277-284 (2012)
doi:10.1038/nphys2252

\bibitem{Bloch:2012uep}
I.~Bloch, J.~Dalibard and S.~Nascimb{\`e}ne,
``Quantum simulations with ultracold quantum gases,''
Nature Phys. \textbf{8}, 267-276 (2012)
doi:10.1038/nphys2259

\bibitem{Cirac:2012}
J.~I.~Cirac and P.~Zoller,
``Goals and opportunities in quantum simulation,''
Nat. Phys. \textbf{8}, no.~4, 264--266 (2012)
doi:10.1038/nphys2275

\bibitem{Grabarits:2024snh}
A.~Grabarits, G.~Sammartino and A.~del Campo,
``Annealing dynamics of regular rotor networks: Universality and its breakdown,''
Phys. Rev. Res. \textbf{7}, no.1, 013123 (2025)
doi:10.1103/PhysRevResearch.7.013123
[arXiv:2407.09316 [quant-ph]].

\bibitem{Pirmoradian:2023uvt}
R.~Pirmoradian and M.~R.~Tanhayi,
``Symmetry-resolved entanglement entropy for local and non-local QFTs,''
Eur. Phys. J. C \textbf{84}, no.8, 849 (2024)
doi:10.1140/epjc/s10052-024-13212-8
[arXiv:2311.00494 [hep-th]].

\bibitem{Tasaki:1998}
H.~Tasaki,
``From Quantum Dynamics to the Canonical Distribution: General Picture and a Rigorous Example,''
Phys. Rev. Lett. \textbf{80}, no.~7, 1373--1376 (1998)
doi:10.1103/PhysRevLett.80.1373

\bibitem{Goldstein:2005aib}
S.~Goldstein, J.~L.~Lebowitz, R.~Tumulka and N.~Zanghi,
``Canonical Typicality,''
Phys. Rev. Lett. \textbf{96}, 050403 (2006)
doi:10.1103/PhysRevLett.96.050403
[arXiv:cond-mat/0511091 [cond-mat.stat-mech]].

\bibitem{Rigol:2007ti}
M.~Rigol, V.~Dunjko, V.~Yurovsky and M.~Olshanii,
``Relaxation in a Completely Integrable Many-Body Quantum System: An Ab Initio Study of the Dynamics of the 1D Bose Gas,''
Phys. Rev. Lett. \textbf{98}, 050405 (2007)
doi:10.1103/PhysRevLett.98.050405
[arXiv:cond-mat/0604556 [cond-mat.stat-mech]].

\bibitem{Huse:2014tqa}
D.~A.~Huse, R.~Nandkishore and V.~Oganesyan,
``Phenomenology of fully many-body-localized systems,''
Phys. Rev. B \textbf{90}, no.17, 174202 (2014)
doi:10.1103/PhysRevB.90.174202
[arXiv:1408.4297 [cond-mat.stat-mech]].

\bibitem{Nandkishore:2014kca}
R.~Nandkishore and D.~A.~Huse,
``Many body localization and thermalization in quantum statistical mechanics,''
Ann. Rev. Condensed Matter Phys. \textbf{6}, 15-38 (2015)
doi:10.1146/annurev-conmatphys-031214-014726
[arXiv:1404.0686 [cond-mat.stat-mech]].

\bibitem{Pfeuty:1970qrn}
P.~Pfeuty,
``The one-dimensional Ising model with a transverse field,''
Annals Phys. \textbf{57}, no.1, 79-90 (1970)
doi:10.1016/0003-4916(70)90270-8

\bibitem{Lieb:1972wy}
E.~H.~Lieb and D.~W.~Robinson,
``The finite group velocity of quantum spin systems,''
Commun. Math. Phys. \textbf{28}, 251-257 (1972)
doi:10.1007/BF01645779

\bibitem{Nachtergaele:2006}
B.~Nachtergaele and R.~Sims,
``Lieb-Robinson Bounds and the Exponential Clustering Theorem,''
Commun. Math. Phys. \textbf{265}, no.~1, 119--130 (2006)
doi:10.1007/s00220-006-1556-1
[arXiv:math-ph/0506030].

\bibitem{Hastings:2005pr}
M.~B.~Hastings and T.~Koma,
``Spectral gap and exponential decay of correlations,''
Commun. Math. Phys. \textbf{265}, 781-804 (2006)
doi:10.1007/s00220-006-0030-4
[arXiv:math-ph/0507008 [math-ph]].

\bibitem{Barankov:2008qq}
R.~Barankov and A.~Polkovnikov,
``Microscopic diagonal entropy and its connection to basic thermodynamic relations,''
Annals Phys. \textbf{326}, 486-499 (2011)
doi:10.1016/j.aop.2010.08.004
[arXiv:0806.2862 [cond-mat.stat-mech]].

\bibitem{Abanin:2017}
D.~A.~Abanin and Z.~Papi\'c,
``Recent progress in many-body localization,''
Ann. Phys. \textbf{529}, no.~7, 1700169 (2017)
doi:10.1002/andp.201700169
[arXiv:1705.09103 [cond-mat.dis-nn]].

\bibitem{Bennett:1995tk}
C.~H.~Bennett, H.~J.~Bernstein, S.~Popescu and B.~Schumacher,
``Concentrating partial entanglement by local operations,''
Phys. Rev. A \textbf{53}, 2046-2052 (1996)
doi:10.1103/PhysRevA.53.2046
[arXiv:quant-ph/9511030 [quant-ph]].

\bibitem{Page:1993df}
D.~N.~Page,
``Average entropy of a subsystem,''
Phys. Rev. Lett. \textbf{71}, 1291-1294 (1993)
doi:10.1103/PhysRevLett.71.1291
[arXiv:gr-qc/9305007 [gr-qc]].

\bibitem{Affleck:1986bv}
I.~Affleck,
``Universal Term in the Free Energy at a Critical Point and the Conformal Anomaly,''
Phys. Rev. Lett. \textbf{56}, 746-748 (1986)
doi:10.1103/PhysRevLett.56.746

\bibitem{Shenker:2013pqa}
S.~H.~Shenker and D.~Stanford,
``Black holes and the butterfly effect,''
JHEP \textbf{03}, 067 (2014)
doi:10.1007/JHEP03(2014)067
[arXiv:1306.0622 [hep-th]].

\bibitem{Maldacena:2015waa}
J.~Maldacena, S.~H.~Shenker and D.~Stanford,
``A bound on chaos,''
JHEP \textbf{08}, 106 (2016)
doi:10.1007/JHEP08(2016)106
[arXiv:1503.01409 [hep-th]].

\bibitem{Pirmoradian:2021wvo}
R.~Pirmoradian and M.~R.~Tanhayi, 
`Non-local probes of entanglement in the scale-invariant gravity,'' Int. J. Geom. Meth. Mod. Phys. \textbf{18}, no.12, 2150197 (2021) doi:10.1142/S0219887821501978 [arXiv:2103.02998 [hep-th]].

\bibitem{Hayden:2007cs}
P.~Hayden and J.~Preskill,
``Black holes as mirrors: Quantum information in random subsystems,''
JHEP \textbf{09}, 120 (2007)
doi:10.1088/1126-6708/2007/09/120
[arXiv:0708.4025 [hep-th]].

\bibitem{Altland:1997}
A.~Altland and M.~R.~Zirnbauer,
``Nonstandard symmetry classes in mesoscopic normal-superconducting hybrid structures,''
Phys. Rev. B \textbf{55}, no.~2, 1142--1161 (1997)
doi:10.1103/PhysRevB.55.1142

\bibitem{Verbaarschot:2000dy}
J.~J.~M.~Verbaarschot and T.~Wettig,
``Random matrix theory and chiral symmetry in QCD,''
Ann. Rev. Nucl. Part. Sci. \textbf{50}, 343-410 (2000)
doi:10.1146/annurev.nucl.50.1.343
[arXiv:hep-ph/0003017 [hep-ph]].

\bibitem{Parker:2018yvk}
D.~E.~Parker, X.~Cao, A.~Avdoshkin, T.~Scaffidi and E.~Altman,
``A Universal Operator Growth Hypothesis,''
Phys. Rev. X \textbf{9}, no.4, 041017 (2019)
doi:10.1103/PhysRevX.9.041017
[arXiv:1812.08657 [cond-mat.stat-mech]].

\bibitem{Balasubramanian:2022tpr}
V.~Balasubramanian, P.~Caputa, J.~M.~Magan and Q.~Wu,
``Quantum chaos and the complexity of spread of states,''
Phys. Rev. D \textbf{106}, no.4, 046007 (2022)
doi:10.1103/PhysRevD.106.046007
[arXiv:2202.06957 [hep-th]].

\bibitem{Alishahiha:2024rwm}
M.~Alishahiha and M.~J.~Vasli,
``Thermalization in Krylov basis,''
Eur. Phys. J. C \textbf{85}, no.1, 39 (2025)
doi:10.1140/epjc/s10052-025-13757-2
[arXiv:2403.06655 [quant-ph]].

\bibitem{Vidal:2003lvx}
G.~Vidal,
``Efficient simulation of one-dimensional quantum many-body systems,''
Phys. Rev. Lett. \textbf{93}, 040502 (2004)
doi:10.1103/PhysRevLett.93.040502
[arXiv:quant-ph/0310089 [quant-ph]].

\bibitem{Schollwock:2011}
U.~Schollw\"ock,
``The density-matrix renormalization group in the age of matrix product states,''
Ann. Phys. \textbf{326}, no.~1, 96--192 (2011)
doi:10.1016/j.aop.2010.09.012
[arXiv:1008.3477 [cond-mat.str-el]].

\bibitem{Essler:2004ht}
F.~H.~L.~Essler and R.~M.~Konik,
``Applications of massive integrable quantum field theories to problems in condensed matter physics,''
doi:10.1142/9789812775344{\_}0020
[arXiv:cond-mat/0412421 [cond-mat.str-el]].

\bibitem{Gogolin:2015gts}
C.~Gogolin and J.~Eisert,
``Equilibration, thermalisation, and the emergence of statistical mechanics in closed quantum systems,''
Rept. Prog. Phys. \textbf{79}, no.5, 056001 (2016)
doi:10.1088/0034-4885/79/5/056001
[arXiv:1503.07538 [quant-ph]].

\bibitem{Sachdev:1993}
S.~Sachdev and J.~Ye,
``Gapless spin-fluid ground state in a random quantum Heisenberg magnet,''
Phys. Rev. Lett. \textbf{70}, no.~21, 3339--3342 (1993)
doi:10.1103/PhysRevLett.70.3339
[arXiv:cond-mat/9212030].

\bibitem{Jurcevic:2014}
P.~Jurcevic, B.~P.~Lanyon, P.~Hauke, C.~Hempel, P.~Zoller, R.~Blatt and C.~F.~Roos,
``Quasiparticle engineering and entanglement propagation in a quantum many-body system,''
Nature \textbf{511}, no.~7508, 202--205 (2014)
doi:10.1038/nature13461

\bibitem{Pachos:2004tqe}
J.~K.~Pachos and M.~B.~Plenio,
``Three-Spin Interactions in Optical Lattices and Criticality in Cluster Hamiltonians,''
Phys. Rev. Lett. \textbf{93}, no.5, 056402 (2004)
doi:10.1103/PhysRevLett.93.056402
[arXiv:quant-ph/0401106 [quant-ph]].

\bibitem{Oganesyan:2007}
V.~Oganesyan and D.~A.~Huse,
``Localization of interacting fermions at high temperature,''
Phys. Rev. B \textbf{75}, no.~15, 155111 (2007)
doi:10.1103/PhysRevB.75.155111
[arXiv:cond-mat/0610854 [cond-mat.str-el]].

\bibitem{Haake:2010fgh}
F.~Haake,
``Quantum Signatures of Chaos,''
Springer, 2010,
ISBN 978-3-642-26330-9, 978-3-642-05428-0
doi:10.1007/978-3-642-05428-0

\bibitem{Grabarits:2024yhi}
A.~Grabarits, K.~R.~Swain, M.~S.~Heydari, P.~Chandarana, F.~J.~G{\'o}mez-Ruiz and A.~del Campo,
``Quantum chaos in random Ising networks,''
Phys. Rev. Res. \textbf{7}, no.1, 013146 (2025)
doi:10.1103/PhysRevResearch.7.013146
[arXiv:2405.14376 [cond-mat.dis-nn]].

\bibitem{Rigol:2016itf}
M.~Rigol and L.~Vidmar,
``Generalized Gibbs ensemble in integrable lattice models,''
J. Phys. A \textbf{2016}, no.6, 064007 (2016)
doi:10.1088/1742-5468/2016/06/064007
[arXiv:1604.03990 [cond-mat.stat-mech]].

\bibitem{Ilievski:2015jhc}
E.~Ilievski, J.~De Nardis, B.~Wouters, J.~S.~Caux, F.~H.~L.~Essler and T.~Prosen,
``Complete Generalized Gibbs Ensembles in an Interacting Theory,''
Phys. Rev. Lett. \textbf{115}, no.15, 157201 (2015)
doi:10.1103/PhysRevLett.115.157201
[arXiv:1507.02993 [quant-ph]].

\bibitem{Eisler:2007}
V.~Eisler and I.~Peschel,
``Evolution of entanglement after a local quench,''
J. Stat. Mech.: Theory Exp. \textbf{2007}, no.~06, P06005 (2007)
doi:10.1088/1742-5468/2007/06/P06005
[arXiv:cond-mat/0703379 [cond-mat.stat-mech]].

\bibitem{Wigner:1951}
E.~P.~Wigner,
``On the statistical distribution of the widths and spacings of nuclear resonance levels,''
Math. Proc. Cambridge Philos. Soc. \textbf{47}, no.~4, 790--798 (1951)
doi:10.1017/S0305004100027237

\bibitem{Islam:2015mom}
R.~Islam, R.~Ma, P.~M.~Preiss, M.~E.~Tai, A.~Lukin, M.~Rispoli and M.~Greiner,
``Measuring entanglement entropy through the interference of quantum many-body twins,''
doi:10.1038/nature15750
[arXiv:1509.01160 [cond-mat.quant-gas]].

\bibitem{Bernien:2017ubn}
H.~Bernien, S.~Schwartz, A.~Keesling, H.~Levine, A.~Omran, H.~Pichler, S.~Choi, A.~S.~Zibrov, M.~Endres and M.~Greiner, \textit{et al.}
``Probing many-body dynamics on a 51-atom quantum simulator,''
Nature \textbf{551}, 579-584 (2017)
doi:10.1038/nature24622
[arXiv:1707.04344 [quant-ph]].

\bibitem{Deutsch:2018ulr}
J.~M.~Deutsch,
``Eigenstate thermalization hypothesis,''
Rept. Prog. Phys. \textbf{81}, no.8, 082001 (2018)
doi:10.1088/1361-6633/aac9f1
[arXiv:1805.01616 [quant-ph]].

\bibitem{Reimann:2008}
P.~Reimann,
``Foundation of Statistical Mechanics under Experimentally Realistic Conditions,''
Phys. Rev. Lett. \textbf{101}, no.~19, 190403 (2008)
doi:10.1103/PhysRevLett.101.190403
[arXiv:0810.3092 [cond-mat.stat-mech]].

\bibitem{Hosur:2015ylk}
P.~Hosur, X.~L.~Qi, D.~A.~Roberts and B.~Yoshida,
``Chaos in quantum channels,''
JHEP \textbf{02}, 004 (2016)
doi:10.1007/JHEP02(2016)004
[arXiv:1511.04021 [hep-th]].

\bibitem{Nahum:2017yvy}
A.~Nahum, S.~Vijay and J.~Haah,
``Operator Spreading in Random Unitary Circuits,''
Phys. Rev. X \textbf{8}, no.2, 021014 (2018)
doi:10.1103/PhysRevX.8.021014
[arXiv:1705.08975 [cond-mat.str-el]].

\bibitem{Swingle:2016var}
B.~Swingle, G.~Bentsen, M.~Schleier-Smith and P.~Hayden,
``Measuring the scrambling of quantum information,''
Phys. Rev. A \textbf{94}, no.4, 040302 (2016)
doi:10.1103/PhysRevA.94.040302
[arXiv:1602.06271 [quant-ph]].

\bibitem{Garcia-Mata:2022voo}
I.~Garc{\'\i}a-Mata, R.~A.~Jalabert and D.~A.~Wisniacki,
``Out-of-time-order correlators and quantum chaos,''
Scholarpedia \textbf{18}, 55237 (2023)
doi:10.4249/scholarpedia.55237
[arXiv:2209.07965 [quant-ph]].

\bibitem{Nandy:2024evd}
P.~Nandy, A.~S.~Matsoukas-Roubeas, P.~Mart{\'\i}nez-Azcona, A.~Dymarsky and A.~del Campo,
``Quantum dynamics in Krylov space: Methods and applications,''
Phys. Rept. \textbf{1125-1128}, 1-82 (2025)
doi:10.1016/j.physrep.2025.05.001
[arXiv:2405.09628 [quant-ph]].

\bibitem{Hastings:2007iok}
M.~B.~Hastings,
``An area law for one-dimensional quantum systems,''
J. Stat. Mech. \textbf{0708}, P08024 (2007)
doi:10.1088/1742-5468/2007/08/P08024
[arXiv:0705.2024 [quant-ph]].

\bibitem{Ghasemi:2021jiy}
M.~Ghasemi, A.~Naseh and R.~Pirmoradian,
``Odd entanglement entropy and logarithmic negativity for thermofield double states,''
JHEP \textbf{10}, 128 (2021)
doi:10.1007/JHEP10(2021)128
[arXiv:2106.15451 [hep-th]].

\bibitem{Kim:2013etb}
H.~Kim and D.~A.~Huse,
``Ballistic Spreading of Entanglement in a Diffusive Nonintegrable System,''
Phys. Rev. Lett. \textbf{111}, no.12, 127205 (2013)
doi:10.1103/PhysRevLett.111.127205
[arXiv:1306.4306 [quant-ph]].

\bibitem{Cardy:2015xaa}
J.~Cardy,
``Quantum Quenches to a Critical Point in One Dimension: some further results,''
J. Stat. Mech. \textbf{1602}, no.2, 023103 (2016)
doi:10.1088/1742-5468/2016/02/023103
[arXiv:1507.07266 [cond-mat.stat-mech]].

\bibitem{Bertini:2016tmj}
B.~Bertini, M.~Collura, J.~De Nardis and M.~Fagotti,
``Transport in Out-of-Equilibrium $XXZ$ Chains: Exact Profiles of Charges and Currents,''
Phys. Rev. Lett. \textbf{117}, no.20, 207201 (2016)
doi:10.1103/PhysRevLett.117.207201
[arXiv:1605.09790 [cond-mat.stat-mech]].

\bibitem{Daley:2012xhf}
A.~J.~Daley, H.~Pichler, J.~Schachenmayer and P.~Zoller,
``Measuring Entanglement Growth in Quench Dynamics of Bosons in an Optical Lattice,''
Phys. Rev. Lett. \textbf{109}, no.2, 020505 (2012)
doi:10.1103/PhysRevLett.109.020505
[arXiv:1205.1521 [cond-mat.quant-gas]].

\bibitem{Pirmoradian:2025xxl}
R.~Pirmoradian, E.~Sadoogh, M.~Teymouri, N.~Abolqasemi-Azad, M.~R.~Lahooti and Z.~Mohammad-Ali,
``Investigation of quantum chaos in local and non-local Ising models,''
[arXiv:2512.21713 [quant-ph]].

\bibitem{Jaiswal:2024fotoc}
N.~Jaiswal, M.~Gautam, A.~Gill and T.~Sarkar,
``FOTOC complexity in the Lipkin--Meshkov--Glick model and its variant,''
Eur.\ Phys.\ J.\ B \textbf{97}, 5 (2024)
doi:10.1140/epjb/s10051-023-00646-4.
\bibitem{Roberts:2014isa}
D.~A.~Roberts, D.~Stanford and L.~Susskind,
``Localized shocks,''
JHEP \textbf{03}, 051 (2015)
doi:10.1007/JHEP03(2015)051
[arXiv:1409.8180 [hep-th]].

\bibitem{Khemani:2018sdn}
V.~Khemani, D.~A.~Huse and A.~Nahum,
``Velocity-dependent Lyapunov exponents in many-body quantum, semiclassical, and classical chaos,''
Phys. Rev. B \textbf{98}, no.14, 144304 (2018)
doi:10.1103/PhysRevB.98.144304
[arXiv:1803.05902 [cond-mat.stat-mech]].

\bibitem{Yao:2016ayk}
N.~Y.~Yao, F.~Grusdt, B.~Swingle, M.~D.~Lukin, D.~M.~Stamper-Kurn, J.~E.~Moore and E.~A.~Demler,
``Interferometric Approach to Probing Fast Scrambling,''
[arXiv:1607.01801 [quant-ph]].

\bibitem{Khorasani:2023usq}
F.~Khorasani, R.~Pirmoradian and M.~R.~Tanhayi,
``Position dependence of Nielsen complexity for the thermofield double state,''
Phys. Lett. B \textbf{851}, 138585 (2024)
doi:10.1016/j.physletb.2024.138585
[arXiv:2308.15836 [quant-ph]].

\bibitem{Roberts:2015}
D.~A.~Roberts and D.~Stanford,
``Diagnosing Chaos Using Four-Point Functions in Two-Dimensional Conformal Field Theory,''
Phys. Rev. Lett. \textbf{115}, no.~13, 131603 (2015)
doi:10.1103/PhysRevLett.115.131603

\bibitem{Swingle:2016jdj}
B.~Swingle and D.~Chowdhury,
``Slow scrambling in disordered quantum systems,''
Phys. Rev. B \textbf{95}, no.6, 060201 (2017)
doi:10.1103/PhysRevB.95.060201
[arXiv:1608.03280 [cond-mat.str-el]].

\bibitem{Fan:2017}
R.~Fan, P.~Zhang, H.~Shen and H.~Zhai,
``Out-of-time-order correlation for many-body localization,''
Science Bulletin \textbf{62}, no.~10, 707--711 (2017)
doi:10.1016/j.scib.2017.04.011
[arXiv:1608.01914 [cond-mat.quant-gas]].

\bibitem{Bhattacharya:2022gbz}
A.~Bhattacharya, P.~Nandy, P.~P.~Nath and H.~Sahu,
``Operator growth and Krylov construction in dissipative open quantum systems,''
JHEP \textbf{12}, 081 (2022)
doi:10.1007/JHEP12(2022)081
[arXiv:2207.05347 [quant-ph]].

\bibitem{Alishahiha:2022nhe}
M.~Alishahiha,
``On quantum complexity,''
Phys. Lett. B \textbf{842}, 137979 (2023)
doi:10.1016/j.physletb.2023.137979
[arXiv:2209.14689 [hep-th]].

\bibitem{Roberts:2016hpo}
D.~A.~Roberts and B.~Yoshida,
``Chaos and complexity by design,''
JHEP \textbf{04}, 121 (2017)
doi:10.1007/JHEP04(2017)121
[arXiv:1610.04903 [quant-ph]].

\bibitem{Doroudiani:2019llj}
M.~Doroudiani, A.~Naseh and R.~Pirmoradian,
``Complexity for Charged Thermofield Double States,''
JHEP \textbf{01}, 120 (2020)
doi:10.1007/JHEP01(2020)120
[arXiv:1910.08806 [hep-th]].

\bibitem{Pirmoradian:2020}
R.~Pirmoradian and M.~R.~Tanhayi,
``On the Complexity of a Charged Quantum Oscillator,''
J. Korean Phys. Soc. \textbf{77}, no.~2, 102--106 (2020).
doi:10.3938/jkps.77.102
[arXiv:1911.08886 [physics.gen-ph]].

\bibitem{Brown:2017jil}
A.~R.~Brown and L.~Susskind,
``Second law of quantum complexity,''
Phys. Rev. D \textbf{97}, no.8, 086015 (2018)
doi:10.1103/PhysRevD.97.086015
[arXiv:1701.01107 [hep-th]].

\bibitem{Bhattacharya:2023zqt}
A.~Bhattacharya, P.~Nandy, P.~P.~Nath and H.~Sahu,
``On Krylov complexity in open systems: an approach via bi-Lanczos algorithm,''
JHEP \textbf{12}, 066 (2023)
doi:10.1007/JHEP12(2023)066
[arXiv:2303.04175 [quant-ph]].

\bibitem{Caputa:2024vrn}
P.~Caputa, H.~S.~Jeong, S.~Liu, J.~F.~Pedraza and L.~C.~Qu,
``Krylov complexity of density matrix operators,''
JHEP \textbf{05}, 337 (2024)
doi:10.1007/JHEP05(2024)337
[arXiv:2402.09522 [hep-th]].

\bibitem{Vasli:2023syq}
M.~J.~Vasli, K.~Babaei Velni, M.~R.~Mohammadi Mozaffar, A.~Mollabashi and M.~Alishahiha,
``Krylov complexity in Lifshitz-type scalar field theories,''
Eur. Phys. J. C \textbf{84}, no.3, 235 (2024)
doi:10.1140/epjc/s10052-024-12609-9
[arXiv:2307.08307 [hep-th]].

\bibitem{Baggioli:2024wbz}
M.~Baggioli, K.~B.~Huh, H.~S.~Jeong, K.~Y.~Kim and J.~F.~Pedraza,
``Krylov complexity as an order parameter for quantum chaotic-integrable transitions,''
Phys. Rev. Res. \textbf{7}, no.2, 023028 (2025)
doi:10.1103/PhysRevResearch.7.023028
[arXiv:2407.17054 [hep-th]].

\bibitem{Camargo:2024deu}
H.~A.~Camargo, K.~B.~Huh, V.~Jahnke, H.~S.~Jeong, K.~Y.~Kim and M.~Nishida,
``Spread and spectral complexity in quantum spin chains: from integrability to chaos,''
JHEP \textbf{08}, 241 (2024)
doi:10.1007/JHEP08(2024)241
[arXiv:2405.11254 [hep-th]].

\bibitem{Khorasani:2021zus} 
F.~Khorasani, M.~R.~Tanhayi and R.~Pirmoradian, 
`Ultimate limits to computation: anharmonic oscillator,'' Eur. Phys. J. Plus \textbf{137}, no.6, 699 (2022) doi:10.1140/epjp/s13360-022-02900-7 [arXiv:2103.03124 [quant-ph]].

\bibitem{Rakovszky:2017qit}
T.~Rakovszky, F.~Pollmann and C.~W.~von Keyserlingk,
``Diffusive hydrodynamics of out-of-time-ordered correlators with charge conservation,''
Phys. Rev. X \textbf{8}, no.3, 031058 (2018)
doi:10.1103/PhysRevX.8.031058
[arXiv:1710.09827 [cond-mat.stat-mech]].

\bibitem{Cotler:2017jue}
J.~Cotler, N.~Hunter-Jones, J.~Liu and B.~Yoshida,
``Chaos, Complexity, and Random Matrices,''
JHEP \textbf{11}, 048 (2017)
doi:10.1007/JHEP11(2017)048
[arXiv:1706.05400 [hep-th]].

\bibitem{Alet:2018}
F.~Alet and N.~Laflorencie,
``Many-body localization: An introduction and selected topics,''
C. R. Physique \textbf{19}, no.~6, 498--525 (2018)
doi:10.1016/j.crhy.2018.03.003
[arXiv:1711.03145 [cond-mat.str-el]].

\bibitem{Cotler:2016fpe}
J.~S.~Cotler, G.~Gur-Ari, M.~Hanada, J.~Polchinski, P.~Saad, S.~H.~Shenker, D.~Stanford, A.~Streicher and M.~Tezuka,
``Black Holes and Random Matrices,''
JHEP \textbf{05}, 118 (2017)
[erratum: JHEP \textbf{09}, 002 (2018)]
doi:10.1007/JHEP05(2017)118
[arXiv:1611.04650 [hep-th]].

\bibitem{Camargo:2023eev}
H.~A.~Camargo, V.~Jahnke, H.~S.~Jeong, K.~Y.~Kim and M.~Nishida,
``Spectral and Krylov complexity in billiard systems,''
Phys. Rev. D \textbf{109}, no.4, 046017 (2024)
doi:10.1103/PhysRevD.109.046017
[arXiv:2306.11632 [hep-th]].

\bibitem{Guhr:1997ve}
T.~Guhr, A.~Muller-Groeling and H.~A.~Weidenmuller,
``Random matrix theories in quantum physics: Common concepts,''
Phys. Rept. \textbf{299}, 189-425 (1998)
doi:10.1016/S0370-1573(97)00088-4
[arXiv:cond-mat/9707301 [cond-mat]].

\bibitem{Atas:2012qv}
Y.~Y.~Atas, E.~Bogomolny, O.~Giraud and G.~Roux,
``Distribution of the ratio of consecutive level spacings in random matrix ensembles,''
Phys. Rev. E \textbf{87}, 032104 (2013)
doi:10.1103/PhysRevE.87.032104
[arXiv:1212.5241 [math-ph]].

\bibitem{Thouless:1977}
D.~J.~Thouless,
``Maximum Metallic Resistance in Thin Wires,''
Phys. Rev. Lett. \textbf{39}, no.~18, 1167--1169 (1977)
doi:10.1103/PhysRevLett.39.1167

\bibitem{Garrison:2015lva}
J.~R.~Garrison and T.~Grover,
``Does a single eigenstate encode the full Hamiltonian?,''
Phys. Rev. X \textbf{8}, 021026 (2018)
doi:10.1103/PhysRevX.8.021026
[arXiv:1503.00729 [cond-mat.str-el]].

\bibitem{Martin:2008}
J.~Martin, O.~Giraud and B.~Georgeot,
``Multifractality and intermediate statistics in quantum maps,''
Phys. Rev. E \textbf{77}, no.~3, 035201 (2008)
doi:10.1103/PhysRevE.77.035201
\bibitem{Pirmoradian:2025nxw}
R.~Pirmoradian, M.~H.~Bek-Khoshnevis, S.~Ebadi and M.~R.~Tanhayi,
``Entanglement Structure of Nonlocal Field Theories,''
[arXiv:2511.10505 [quant-ph]].

\bibitem{Li:2008}
H.~Li and F.~D.~M.~Haldane,
``Entanglement Spectrum as a Generalization of Entanglement Entropy: Identification of Topological Order in Non-Abelian Fractional Quantum Hall Effect States,''
Phys. Rev. Lett. \textbf{101}, no.~1, 010504 (2008)
doi:10.1103/PhysRevLett.101.010504
[arXiv:0805.0332 [cond-mat.mes-hall]].

\bibitem{Peschel:2011mcx}
I.~Peschel,
``Special Review: Entanglement in Solvable Many-Particle Models,''
Braz. J. Phys. \textbf{42}, no.3, 267-291 (2012)
doi:10.1007/s13538-012-0074-1
[arXiv:1109.0159 [cond-mat.stat-mech]].

\bibitem{Beugeling:2014lnz}
W.~Beugeling, R.~Moessner and M.~Haque,
``Finite-size scaling of eigenstate thermalization,''
Phys. Rev. E \textbf{89}, no.4, 042112 (2014)
doi:10.1103/PhysRevE.89.042112
[arXiv:1308.2862 [cond-mat.stat-mech]].

\bibitem{Smith:2015dpe}
J.~Smith, A.~Lee, P.~Richerme, B.~Neyenhuis, P.~W.~Hess, P.~Hauke, M.~Heyl, D.~A.~Huse and C.~Monroe,
``Many-body localization in a quantum simulator with programmable random disorder,''
Nature Phys. \textbf{12}, no.10, 907-911 (2016)
doi:10.1038/nphys3783
[arXiv:1508.07026 [quant-ph]].





\end{thebibliography}
\end{document}